\documentclass{article}
\usepackage{graphicx}
\usepackage{multirow}
\usepackage{threeparttable}
\usepackage{mathrsfs}
\usepackage{hyperref}

\usepackage{amsmath} \usepackage{amsfonts,euscript,amssymb}
\usepackage{graphicx} \usepackage{subcaption} \usepackage[font=scriptsize]{caption}
\usepackage{epsf}
\usepackage{subcaption}
\usepackage{tabularx}

\usepackage{mathtools}
\usepackage{calc}   

\usepackage{color}

\makeatletter
\newcommand{\vast}{\bBigg@{4}}
\makeatother

\definecolor{Darkgreen}{rgb}{0,0.45,0}
\newcommand{\dt}[1]{\textcolor{black}{#1}}

\newcommand{\eequ}{\end{equation}}
\newcommand{\bequ}{\begin{equation}}
\newcommand{\eequd}{\end{eqnarray*}}
\newcommand{\bequd}{\begin{eqnarray*}}

\def\I{\mathcal{I}}

\def\J{\mathcal{J}}

\def\Sb{\textbf{S}}

\def\Sect{\mathcal{S}}

\def\A{\mathcal{A}}

\def\K{\mathcal{K}}

\def\Ncal{\mathcal{N}}

\def\P{\mathcal{P}}

\def\Bila{\mathbf{B}}

\def\normal{\mathbf{n}}

\def\bary{\mathbf{b}}

\def\P{\mathcal{P}}

\def\D{\mathcal{D}}

\def\F{\mathcal{F}}

\def\K{\mathcal{K}}

\def\I{\mathcal{I}}

\def\R{\mathbb{R}}

\def\N{\mathbb{N}}

\def\Bila{\mathbf{B}}

\newcommand {\nor} [1]{\parallel #1 \parallel}

\begin{document}

\title{Multiscale Modelling of Fibres Dynamics and Cell Adhesion within Moving Boundary Cancer Invasion\thanks{The two authors contributed equally to the paper, the order being purely alphabetical.}}



\author{Robyn Shuttleworth\footnote{Division of Mathematics, University of Dundee, Dundee, DD1 4HN, Scotland, UK, Email: r.shuttleworth@dundee.ac.uk
}\ and  Dumitru Trucu\footnote{Division of Mathematics, University of Dundee, Dundee, DD1 4HN, Scotland, UK, Email: trucu@maths.dundee.ac.uk} \footnote{Corresponding Author: Dumitru Trucu}}


%
\date{27 September 2018}

\maketitle

\begin{abstract}
Cancer cell invasion, recognised as one of the hallmarks of cancer, is a complex process involving the secretion of matrix-degrading enzymes that have the ability to degrade the surrounding extracellular matrix (ECM). Combined with cell proliferation, migration, and changes in cell-cell and cell-matrix adhesion, a tumour is able to spread into the surrounding tissue, and in this context, we highlight the multiscale character of this process enabled through complex feedback links between the cell-scale molecular \dt{activity} and tissue-scale \dt{cell-population dynamics}. 

In order to gain a deeper understanding of the tumour invasion process, we \dt{pay special attention to the interacting dynamics between the cancer cell population and various constituents of the} the surrounding \dt{tumour} microenvironment. \dt{To that end, we consider the key role that ECM plays within the human body tissue}, providing not only structure and support to surrounding cells, but also acting as a platform \dt{for} cells communication \dt{and} spatial movement. There are several other vital structures within the ECM, however we are going to focus primarily on fibrous proteins, such as fibronectin. These fibres are key players in the function of healthy cells, contributing to many essential processes such as cell migration and proliferation. They also play a crucial role in tumour progression with the ability to anchor cells to other components of the ECM. 

In this work we consider the two-scale dynamic cross-talk between cancer cells and a two component ECM (consisting of both a fibre and a non-fibre phase). To that end, we incorporate the interlinked \dt{two-scale} dynamics of cells-ECM interactions within the tumour support that contributes simultaneously both to cell-adhesion and to the dynamic rearrangement and restructuring of the ECM fibres. Furthermore, this is embedded within a multiscale moving boundary approach for the invading cancer cell population, in the presence of cell-adhesion at the tissue scale and cell-scale fibre redistribution activity and leading edge matrix degrading enzyme molecular proteolytic processes. The overall modelling framework will be accompanied by computational results that will explore the impact on cancer invasion patterns of different levels of cell adhesion in conjunction with the continuous ECM fibres \dt{rearrangement}.


\end{abstract}

\section{Introduction}

Cancer invasion of the human body is a complex, multiscale phenomenon that incorporates both molecular and cellular interactions as well as interconnections within tissues. Recognised as one of the hallmarks of cancer \cite{Hanahan2000}, cancer invasion is a process \dt{that takes advantage of important changes in the} behaviour of \dt{many molecular activities typical for} healthy cell, \dt{such as the abnormal} secretion of proteolytic enzymes that \dt{lead to} the degradation of its surrounding environment that ultimately translate in further tumour progression. Changes in cell adhesion properties also contribute to the success of tumour invasion. 

Led by the \dt{proteolytic processes induced by the cancer} cells \dt{from its} outer \dt{proliferating rim}, the tumour locally invades neighbouring sites \dt{via an up-regulated} cell-matrix adhesion \cite{Berrier2007} \dt{concomitant with a loss in cell-cell adhesion}.  \dt{This local invasion marks} the first \dt{of a cascade of stages, that ultimately result} in the process of cells escaping the primary tumour and creating metastases at distant sites in the body. Without treatment, metastasised tumours can lead to organ failure and eventual death in around 90\% of patients \cite{Chaffer}.

\dt{Acknowledge for the essential role that ECM plays in many vital processes, such as in embryogenesis \cite{rozario_2010} and wound healing \cite{xue_2015}, this plays a crucial part also in cancer invasion.} The success of tumour invasion is greatly influenced by the extracellular matrix (ECM), a key biological structure formed from an interlocking network of proteins including collagen and elastin, which provide necessary structure and elasticity, as well as proteoglycans that aid the secretion of growth factors. However, during cancer invasion, the over-secretion of proteolytic enzymes, such as the urokinase-type plasminogen activator (uPA) and matrix metallo-proteinases (MMPs) \cite{Parsons1997} by the cancer cells, \dt{followed by interactions} of these enzymes with the ECM components results in the degradation and remodelling of the ECM \cite{Lu2011,Pickup2014}, largely contributing to \dt{further tumour progression.} 

The invasive capabilities of a tumour gather their strength from \dt{a cascade of processes enabled by the cancer cells, which, besides abnormal secretion of matrix degrading enzymes, includes also} enhanced proliferation and altered cellular adhesion abilities. Both cell-cell and cell-matrix adhesion play important roles in tumour progression, \dt{and changes to either of these contribute directly} to the overall pattern of invasion. Certain proteins found in the ECM, for example \dt{collagen and} fibronectin, aid in the binding of cells to the matrix through \dt{the} cell-matrix adhesion\dt{,} process {which is} regulated by a family of specific molecules on the cell surface known as calcium independent cell adhesion molecules (CAMs), or integrins \cite{Humphries2006}. \dt{While collagen is a main component of the ECM, being one of the most common protein found in the human body, fibronectin plays a crucial role during cell adhesion having the ability to anchor cells to collagen and other components of the ECM. Thus, while collagen provides structure and rigidity to the ECM, fibronectin contributes to cell migration, growth and proliferation, both ensuring the normal functionality of healthy cells and being of crucial importance in cancer progression.}

On the other hand, calcium dependent CAMs on the cell surface naturally mediate cell-cell adhesion. Adhesion is dependent on the cell signalling pathways that are formed due to interactions between $\text{Ca}^{2+}$ ions and the distribution of calcium sensing receptors in the ECM. Specifically, the \dt{molecular} subfamily \dt{of} E-cadherins is responsible for binding with the intra-cellular proteins known as catenins, typically $\beta$-catenin, forming the E-cadherin/catenin complex. The recruitment of cadherins and  $\beta$-catenin to the cell cytoskeleton is effectuated by intracellular calcium signalling \cite{Ko_2001}. Evidence suggests that activation of calcium sensing receptors results in an increase of E-cadherins \dt{which} in turn increases the binding of $\beta$-catenin \cite{Hills2012}. However, any alteration to the function of $\beta$-catenin will result in the loss of the ability of E-cadherin to initiate cell-cell adhesion \cite{Wijnhoven2000}.

As tumour malignancy increases, normal fibroblasts are subverted to promote tumour growth, known as cancer-associated fibroblasts (CAFs) \cite{kalluri_2016,Shiga_2015}. CAFs proliferate at a much higher rate than normal fibroblasts in healthy tissue \cite{Erdogan3799}.  \dt{Biological} evidence that CAFs induce tumour growth, metastasis, angiogenesis and resistance to chemotherapeutic treatments \cite{tao_2017}. Unlike normal fibroblasts, CAFs are specific to tumour cells and their microenvironment and possess the ability to change the structure and influence functions within the ECM \cite{Jolly_2016}. Many \textit{in vitro} experiments have shown that CAFs rearrange both collagen fibres and fibronectin, enabling a smooth invasion of the cancer cells \cite{Erdogan3799,fang2014,gopal2017,Ioachim2002}. The ability to reorganise fibrous proteins in the microenvironment is aided by the high secretion of collagen types I and II and fibronectin by the fibroblasts \cite{cirri_2011}. \dt{For that reason, we choose to give here special consideration to a two-component ECM in the context of cancer invasion, and, to that end, to regard the ECM as consisting of both a fibre and a non-fibre phase.}   
 
\dt{Despite increasingly abundant \textit{in vivo} and \textit{in vitro} investigations and modelling for cancer invasion from a variety of standpoints}, only a snippet of the interactions between the cancer cells and components of the extracellular matrix and surrounding tissues could be so far depicted and understood. \dt{However, alongside all these biological research efforts, the past 25 years have witnessed increasing focus on the mathematical modelling of cancer invasion, \cite{Andasari2011,Anderson2005,Anderson2000,Chaplain_et_al_2011,Chaplain2006a,Gerisch2008,Peng2016,Ramis-Conde_et_al_2000,szymanska_08,Dumitru_et_al_2013}, addressing various processes of cancer cells and their interactions with the surrounding environment through a variety of approaches ranging from discrete, local and non-local continuous models to hybrid and multiscale models. Among these models, we note here the ones concerning the secretion and transport of proteolytic enzymes such as uPAs and MMPs, with direct impact upon the degradation of ECM \cite{Andasari2011,Chaplain2005,Peng2016,Dumitru_et_al_2013} as well as those exploring the direct effects of chemotaxis, proliferation and adhesion on tumour invasion \cite{Bitsouni_2017,Chauviere_2007,Domschke_et_al_2014,Gerisch2008,Painter2008,Ramis-Conde_et_al_2000}, all these aspects being of direct interest for us in the current investigation.} 

There are several models which have previously focused on the components of the surrounding microenvironment of tumours and how these contribute to invasion \cite{Perumpanani_et_al_1998,Scianna_Presiosi_2012}. A model describing the mesenchymal motion of cells in a fibre network and \dt{suggesting that the cells will preferentially follow the direction of the fibres} was \dt{proposed in} \cite{Hillen2006}. Chemotactic and haptotactic effects between cells and the fibrous environment of the ECM \dt{where considered in \cite{Chauviere_2007} and explored} two scenarios, \dt{namely that either} cancer cells will try to gather in to high density regions of fibres, or they will try to avoid these regions altogether. 

\dt{Finally, as the invasion process is genuinely multiscale, with its dynamics ranging from molecular sub-cellular and cellular-scale to intercellular- and tissue- scale, the multiscale modelling of cancer invasion have witnessed major advances over the past 15 years \cite{anderson_07,Peng2016,Ramis-Conde_et_al_2000,Dumitru_et_al_2013}}. \dt{However, while recognised by most previous works that a combination of \dt{information from} different scales \dt{would pave the way for} a \dt{better understanding} of cancer invasion, the naturally interlinked multiscale dynamics of this process was for the first time addressed in a genuinely spatially multiscale fashion in \cite{Dumitru_et_al_2013}, where a novel multiscale moving boundary model was developed by exploring the double feedback link between tissue-scale tumour dynamics  and the tumour invasive edge cell-scale matrix-degrading enzymes (MDEs) activity. In that multiscale model, while the tissue-scale macro-dynamics of cancer cells induces the source for the leading edge cell-scale molecular micro-dynamics of MDEs, in turn, through its proteolytic activity, this molecular micro-dynamics causes significant changes in the structure of the ECM in the peritumoural region that ultimately translate in a tissue-scale relocation of the tumour boundary. Later on,} that model was adapted \dt{in \cite{Peng2016}} to capture the \dt{influence within the tumour invasion process of the proteolytic dynamics of urokinase-plasminogen activator (uPA) system}, \dt{exploring various scenarios for} ECM degradation and proliferation of cancer cells. \dt{More recently, a further extension of that modelling was developed in \cite{shutt_chapter}, where the dynamics of cell adhesion within a two cell population heterogeneous context was explored by adopting the non-local modelling proposed in \cite{Domschke_et_al_2014,Gerisch_Chaplain_2008} as the macro-scale part of the multiscale platform introduced in \cite{Dumitru_et_al_2013}.} 

\dt{Building on the modeling platform introduced in \cite{Dumitru_et_al_2013} and extended in \cite{shutt_chapter}}, in this paper we \dt{will pay a special attention} \dt{to} the complicated structure of the ECM, \dt{and to that end we will propose a novel multiscale-moving boundary} model to \dt{account upon the multiscale dynamics of} a two-component ECM, \dt{ considered here to consist of both a} fibre and a non-fibre \dt{phase}. \dt{T}his way we \dt{will highlight the significance of} the fibrous structure \dt{of the invading tumour} and \dt{explore not only the influence of }these fibres \dt{within} the macroscopic cancer cell dynamics, but also \dt{capture} their microscopic rearrangement. \dt{This new two-scale fibres dynamics will be considered in the context of the multiscale moving boundary invasion process as formulated in \cite{Dumitru_et_al_2013}, leading this way to a novel multiscale moving boundary framework, with two simultaneous but different in nature micro-dynamic processes that are each connected through two double feedback loops to a shared tissue-scale cancer macro-dynamics}.

\section{The \dt{Novel} Multiscale Modell\dt{ing Framework}}
\label{framework}

\dt{Building on the multiscale moving boundary framework initially introduced in \cite{Dumitru_et_al_2013}, in the following we describe the novel modelling platform for cancer invasion. Besides the underlying tumour invasive edge two-scale proteolytic activity of the matrix degrading enzymes considered in \cite{Dumitru_et_al_2013}, the new modelling framework will now incorporate and explore the multiscale ECM fibre dynamics within the bulk of the invading tumour, accounting in a double feedback loop for their microscopic rearrangement as well as their macro-scale effect on cancer cell movement.}

\dt{Let us denote the support of the locally invading tumour by $\Omega(t)$ and assume that this evolves within a maximal environmental tissue cube $Y\in \R^{N}$, with $N=2,3$, which is centred at the origin of the space. In this context, at any tissue-scale spatio-temporal node $(x,t)\in \Omega(t)\times [0,T]$, we consider the tumour as being a dynamic mixture consisting of a cancer cell distribution $c(x,t)$ combined with a cumulative extracellular matrix density $v(x,t):=F(x,t)+l(x,t)$ whose multiphase configuration $(F,l)$ will be detailed in Sections \ref{mmECM}-\ref{mfomECM}.}

\subsection{The multiscale moving boundary perspective}\label{multMovingBoundaryPerspect}
\begin{figure}[h]
    \centering
    {\includegraphics[scale=0.6]{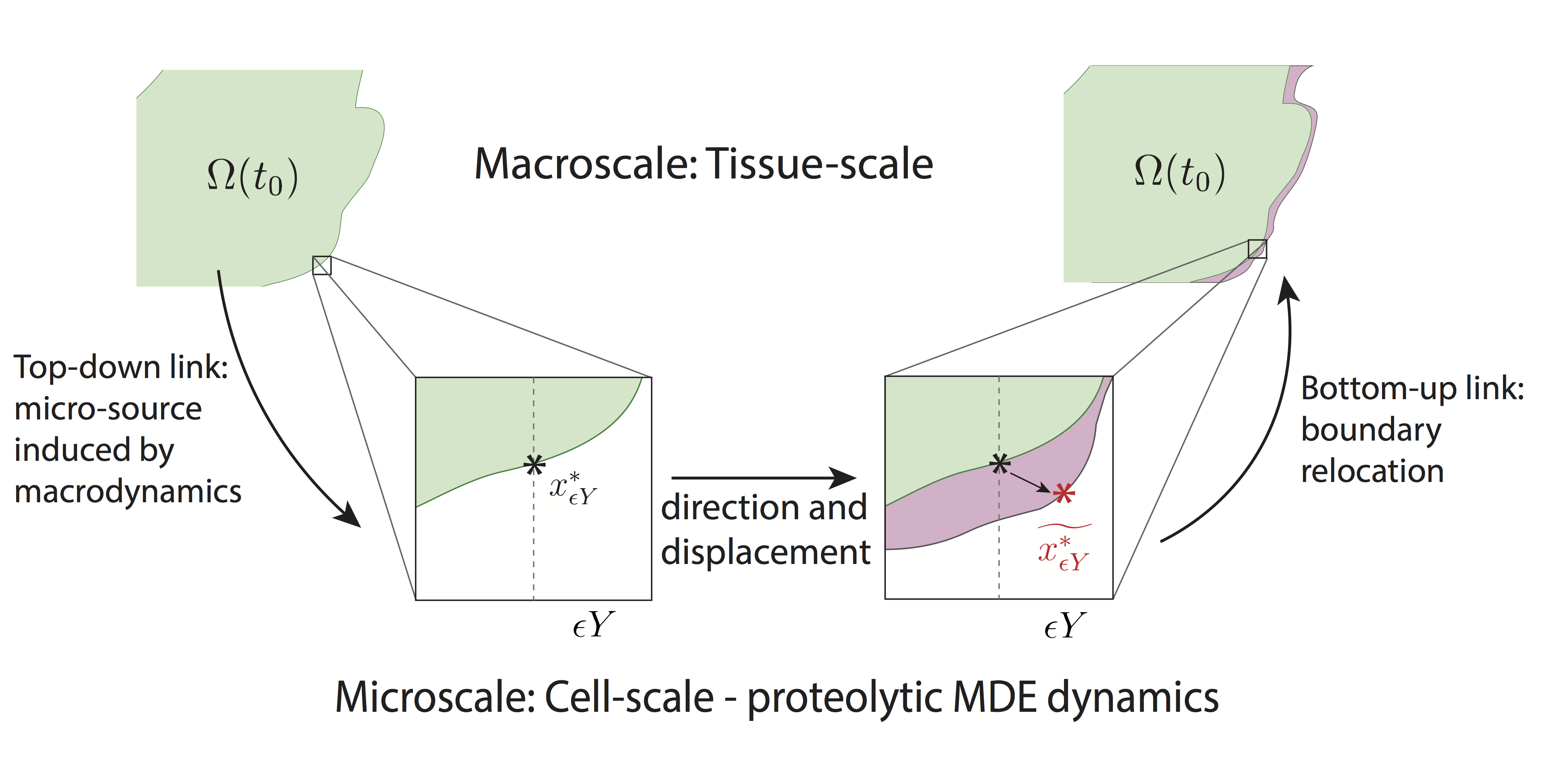} }
\caption{\emph{Schematic showing the interactions between the macro- and \dt{the proteolytic MDEs} micro- scale \dt{dynamics} and the role this plays in boundary reallocation.}}
    \label{fig:reallo}
\end{figure}

\dt{While postponing for the moment the precise details of the macro-dynamics (leaving this to be fully introduced and explored in Section \ref{tumourMacroDynamicSection}), since the novel modelling platform that we develop here builds on the initial multiscale moving boundary framework proposed in \cite{Dumitru_et_al_2013}, let us briefly revisit this here. In this context, let us express for the moment the tissue-scale tumour macro-dynamics on the the evolving $\Omega(t)$ in the form of the following pseudo-differential operator equation
\bequ\label{genericTumourMacroEq}
\mathcal{T}(c,F,l)={\bf 0} 
\eequ
where $\mathcal{T}(\cdot,\cdot,\cdot)$ denotes an appropriately derived reaction-diffusion-taxis operator. As detailed in \cite{Dumitru_et_al_2013}, the key multiscale role played by the tumour invasive proteolytic enzymes processes in cancer invasion is captured here in a multiscale moving boundary approach where the link between the tumour macro-dynamics \eqref{genericTumourMacroEq} and the cell-scale leading edge proteolytic molecular micro-dynamics is captured via a double feedback loop. This double feedback loop is realised via a \emph{top-down} and a \emph{bottom-up} link, as illustrated schematically in Figure \ref{fig:reallo} and detailed below.  
}
\dt{\paragraph{The top-down link.} As discussed previously, cancer invasion is a multiscale process in which the matrix-degrading enzymes (MDEs), such as matrix metallo-proteinases (MMP) which are secreted by the cancer cells from within the outer proliferation rim of the tumour, are responsible for the degradation of the peritumoural ECM, enabling further tumour expansion.  Thus, adopting the terminology and framework developed in \cite{Dumitru_et_al_2013}, this tumour invasive edge molecular micro-dynamics, which occurs within a cell-scale neighbourhood of the tumour interface $\partial \Omega(t)$, can be explored on an appropriately constructed bundle of  $\epsilon-$size \emph{half-way overlapping micro-domains} $\{\epsilon Y\}_{\epsilon Y \in \P(t)}$ satisfying special topological requirements that ensure that each $\epsilon Y$ \emph{``sits on the interface"} and captures relevant parts of both inside and outside regions of the tumour (as briefly detailed in Appendix \ref{movingBoundaryFrameworkAPP_01sept2018} and illustrated in schematic Figure \ref{fig:fullmodelintro}), where the proteolytic activity takes place. This allows us to decouple this leading-edge proteolytic activity in a bundle of corresponding MDE micro-processes occurring on each $\epsilon Y$. In this context, a source of MDEs arises at each $z\in \epsilon Y \cap \Omega(t_{0})$ as a collective contribution of all the cells that (subject to macro-dynamics \eqref{genericTumourMacroEq}) arrive within the outer proliferating rim at a spatial distance from $z$ smaller than a certain radius $\gamma>0$ (representing the maximal thickness of the outer proliferating rim).  Thus, the source of MDEs that is this way induced by the macro-dynamics at the micro-scale on each $\epsilon Y$ realises a significant \emph{top-down link} that can be mathematically expressed as
\begin{align} \label{eq:sourceMDEs}
\begin{split}
1. \quad &g_{\epsilon Y}(z,\tau) = \frac{\int\limits_{\textbf{B}(z,\gamma)\cap\Omega(t_0)} \alpha c (x,t_0 + \tau) dx}{\lambda (\textbf{B}(z,\gamma)\cap\Omega(t_0))}, \quad  z \in \epsilon Y \cap \Omega(t_0), \\[0.7cm]
2. \quad &g_{\epsilon Y}(z,\tau) = 0,\quad  z \in \epsilon Y \setminus \big( \Omega(t_0)+\{ z \in Y|  \ ||z||_2 < \zeta\}), 
\end{split}
\end{align}
where $\Bila(z,\gamma):=\{\xi\in Y\,|\, \nor{z-\xi}_{_{\infty}}\leq \zeta\}$ and $\alpha$ is an MDE secreting rate for the cancer cell population.
In the presence of this source, a cross interface MDEs transport takes place. As in this paper we only consider the micro-dynamics of a single class of MDEs, such as MMPs, this simply results in a diffusion type transport over the entire $\epsilon Y$ micro-domain, and so denoting the MDE molecular density by $m(z,\tau)$, this can be mathematically formulated as
\bequ\label{microdyn_01Sept18_eq1}
\frac{\partial m}{\partial \tau} = D_{m} \Delta m + g_{\epsilon Y} (z,\tau), \quad \quad z \in \epsilon Y, \ \tau \in [0, \Delta t],
\eequ
}

\paragraph{The bottom-up link.} During the micro-dynamics \eqref{microdyn_01Sept18_eq1}, the MDEs \dt{transported across the interface in the peritumoural region} interact with ECM \dt{distribution that they meet in the immediate tumour proximity outside the cancer region within} each \dt{boundary micro-domain} $\epsilon Y$. \dt{On each microdomain $\epsilon Y$, provided that a sufficient amount of MDEs have been transported across the cancer invading edge enclosed in this microdomain, it is the pattern of the front of the advancing spatial distribution of MDEs that characterises the way in which the ECM is locally degraded. As introduced and described in \cite{Dumitru_et_al_2013}}, within each $\epsilon Y$, the pattern of degradation of ECM \dt{caused by the significant levels of the advancing front of} MDEs \dt{give rise to a direction $\eta_{\epsilon Y}$ and displacement magnitude $\xi_{\epsilon Y}$ (detailed in Appendix \ref{movingBoundaryFrameworkAPP_01sept2018}),} \dt{which determine the cancer boundary movement characteristics represented back at macro-scale through the movement of the appropriately defined boundary mid-points} $x^*_{\epsilon Y}$ to \dt{their} new spatial positions $\widetilde{x^*_{\epsilon Y}}$, see Figure \ref{fig:reallo}. Thus, \dt{over a given time perspective $[t_{0}, t_{0}+\Delta t]$}, the \textit{bottom-up link} of the interaction between the \dt{proteolytic tumour invasive edge micro-dynamics} and macro-scale is \dt{realised through the macro-scale boundary movement characteristics that are provided by the micro-scale MDEs activity, leading to the expansion of the tumour boundary $\Omega(t_{0})$ to an enhanced domain $\Omega(t_{0}+\Delta t)$ where the multiscale dynamics is continued.}

\subsection{\dt{The multiscale and multi-component} structure of the \dt{ECM}}\label{mmECM}
\dt{To gain a deeper understanding of the invasion process, in this work we pay special attention to the ECM structure within the overall multiscale dynamics.} While in previous \dt{multiscale approaches (such as those proposed in} \cite{Peng2016,Dumitru_et_al_2013}) the ECM has been \dt{considered} as a \emph{``well mixed" matrix distribution}, with no individual components taken in to consideration, \dt{in the following we account for the structure of the ECM by regarding this as a two-component media.} \dt{The first ECM component that we distinguish accounts for all significant ECM fibres such as collagen fibres or fibronectin fibrils. This will be denoted by $F(x,t)$ and will simply be referred to as the \emph{fibres} component. Finally, the second ECM component that we distinguish consists of the rest of ECM constituents bundled together. This will be referred to as the \emph{non-fibres} component and will be denoted by $l(x,t)$. }
\begin{figure}[h!]
\centering
\includegraphics[scale=0.4]{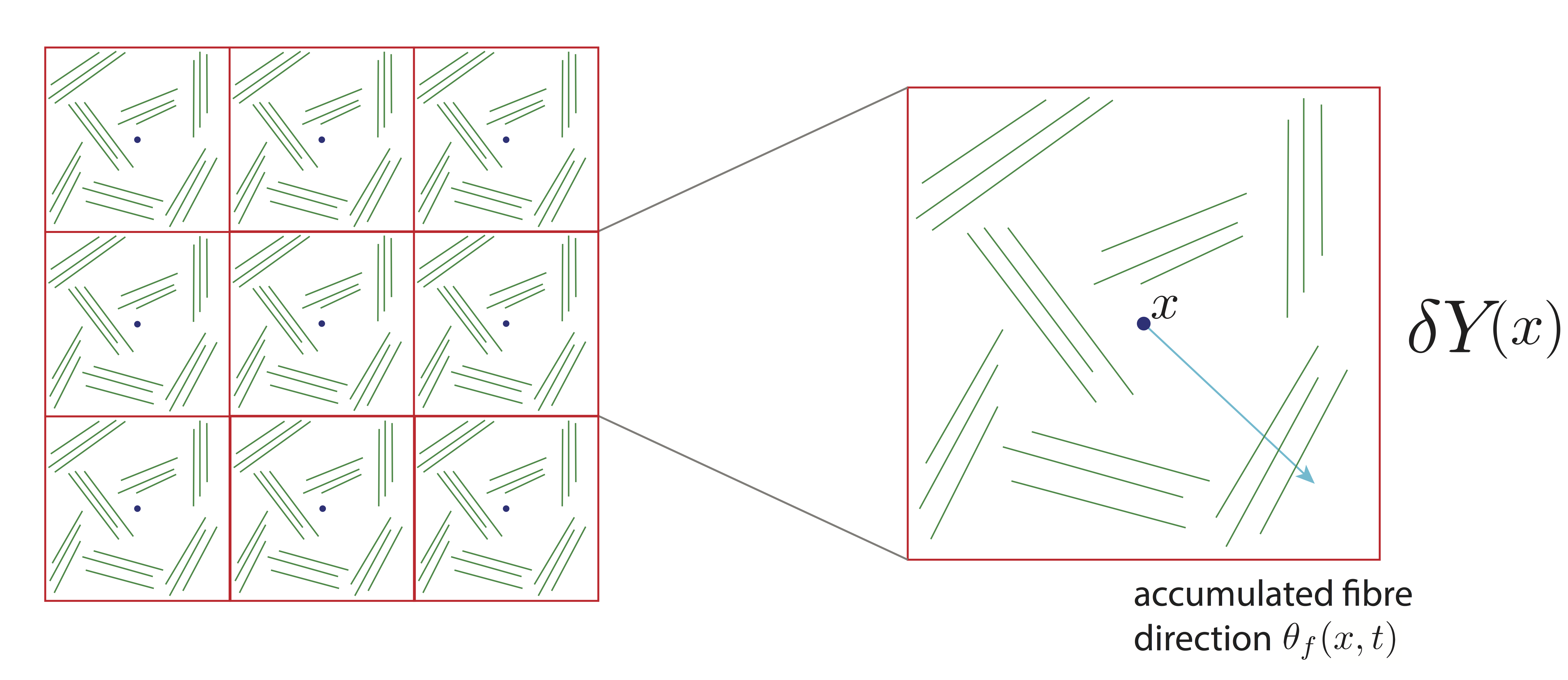}
\caption{\emph{Schematic showing copies of $\delta Y$cube on the grid with \dt{micro-}fibres distribution in green and their \dt{induced macroscopic} direction $\theta_{f}(x,t)$ in pale blue.}}
\label{fig:fibexample}
\end{figure}

\dt{While from the tissue-scale (macro-scale) stand point the fibres are regarded as a continuous distribution at any $x\in Y$, from the cell-scale (micro-scale) point of view, a specific micro-structure can be in fact distinguished as a mass-distribution of the \emph{ECM micro-fibres} $f(z,t)$ that are spatially distributed on a small micro-domain of micro-scale size $\delta>0$ centred at any macroscopic point $x\in Y$, namely on $\delta Y(x):=\delta Y+x$.}
\begin{figure}[ht!]
    \centering
    \begin{subfigure}{.4\linewidth}
        \includegraphics[scale=0.26]{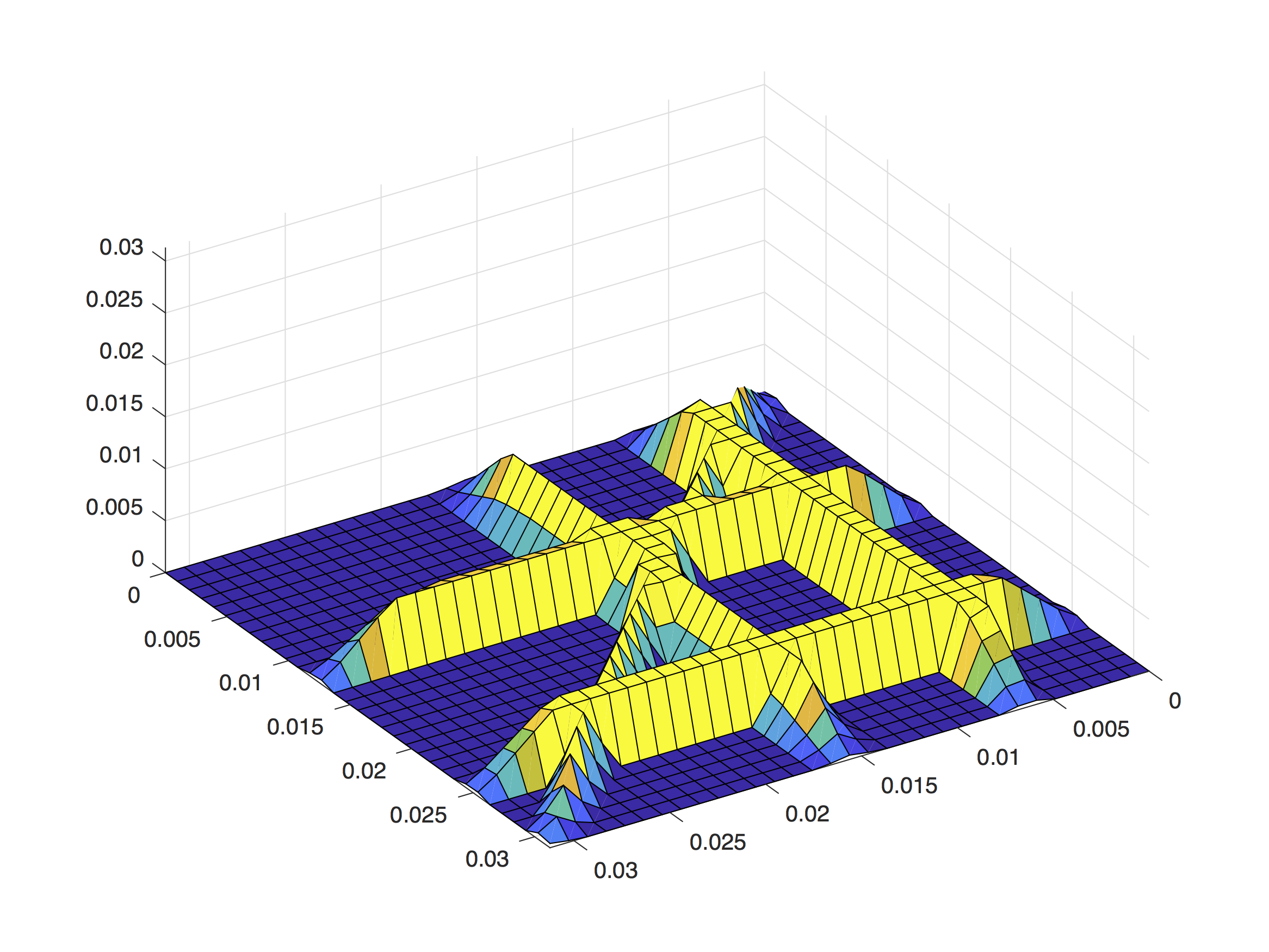}
        \caption{\emph{\dt{graph} of \dt{micro-fibre distribution on} $\delta Y(x)$}}
    \end{subfigure}
    \hskip2em
    \begin{subfigure}{.4\linewidth}
        \includegraphics[scale=0.12]{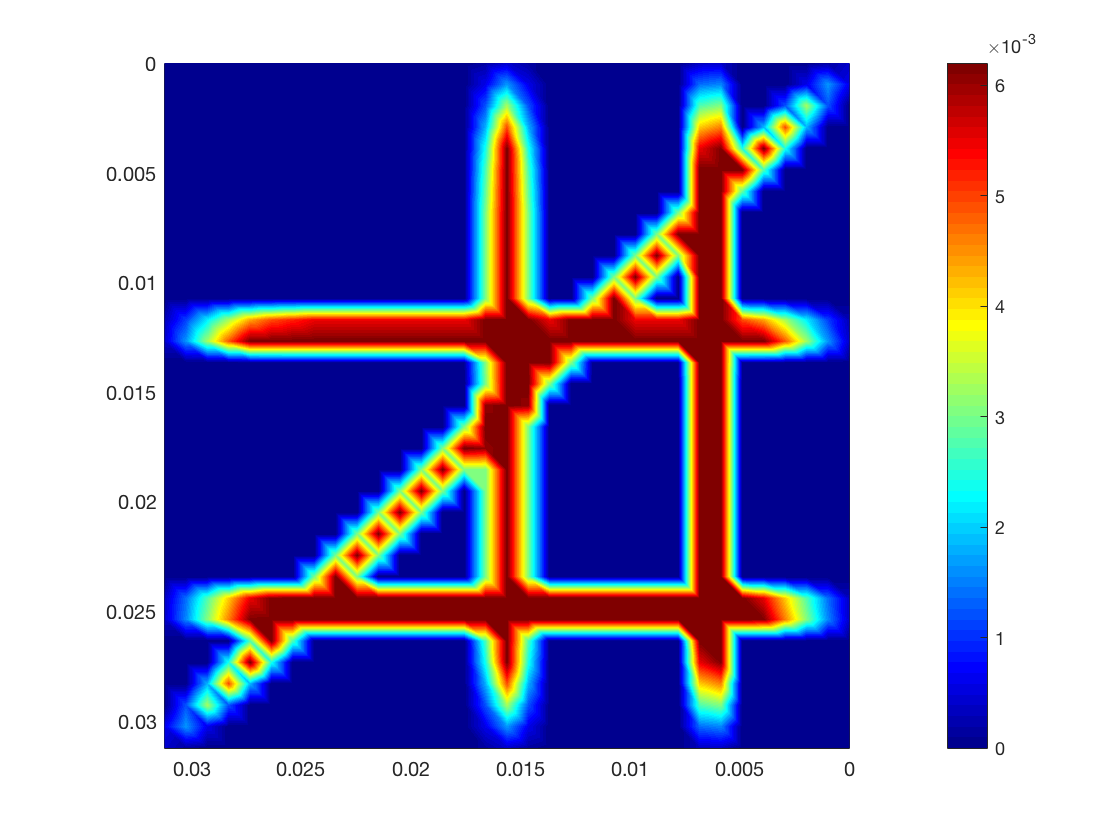}
        \caption{\emph{2D \dt{color} plot of \dt{the micro-fibre distribution on} $\delta Y(x)$}}
    \end{subfigure}
    \caption{\emph{\dt{Micro-fibre distribution on $\delta Y(x)$}.}}
    \label{fig:fibre}
\end{figure}
\dt{In this context, as we will detail below, the microscopic mass-distribution of ECM micro-fibres will be able to supply important macro-scale fibre characteristics, both in terms of their associated \emph{macroscopic fibre orientation} $\theta_{_{f}}(x,t)$ and \emph{magnitude} $F(x,t)$, which will be introduced in Section \ref{mfomECM}.}
\dt{Figure \ref{fig:fibexample} illustrates such micro-fibres distribution in micro-domains $\delta Y(x)$, $x\in \Omega(t)$. A concrete example of such micro-scale fibres pattern is then proposed in Figure \ref{fig:fibre}, this being given as 
\bequ\label{ECMfibreInitialConditions}
f(z,t):=\sum\limits_{j\in J}\psi_{h_{j}}(z)(\chi_{_{(\delta-2\gamma)Y(x)}}\ast \psi_{\gamma})(z)
\eequ
where $\{\psi_{h_{j}}\}_{j\in J}$ are smooth compact support functions of the form
\bequ
\begin{array}{l}
\psi_{h_{j}}: \delta Y(x) \rightarrow \mathbb{R}\\[0.3cm]
\textrm{which, at every $z:=(z_{1},z_{2})\in\delta Y(x)$, are given by:}\\[0.2cm]
\psi_{h_{j}}(z_1,z_2):=
\left\{
\begin{array}{ll}
C_{h_{j}} e^{-{\frac{1}{r^2-(h_{j}(z_2) - z_1)^2}}}, &\quad \text{if} \ z_1 \in [h_{j}(z_2)-r, h_{j}(z_2)+r], \\[0.2cm]
0, &\quad \text{if} \ z_1 \not\in [h_{j}(z_2)-r, h_{j}(z_2)+r].
\end{array}
\right.
\end{array}
\label{eq:fib}
\eequ
with $r>0$ being the width of the micro-fibres and $C_{\dt{h_{j}}}$ \dt{being constants} that determine the maximum hight of $\psi_{h_{j}}$ along the smooth paths $\{h_{j}\}_{j\in J}$ in $\delta Y(x)$ that are given in Appendix \ref{microFibresAPP_01sept2018}. Finally, $\chi_{_{(\delta-\gamma)Y(x)}}(\cdot)$ represents the characteristic function of the cubic micro-domain $(\delta-\gamma)Y(x)$ centred at $x$ and of size $(\delta-2\gamma)$, with $\gamma>0$ a small enough radius, while $\psi_{\gamma}$ is the usual mollifier defined in Appendix \ref{mollifierUsed} that is smoothing out this characteristic function to a smooth compact support function on $\delta Y$}.  

\dt{Furthermore, as we will discuss in the following, while the fibre micro-structure will be dynamically rearranged at micro-scale by the incoming flux of cancer cell population, their \emph{``on the fly"} updated revolving orientation $\theta_{f}(x,t)$ and magnitude $F(x,t)$ will be involved in the dynamics at macro-scale.} 

\subsection{\dt{Macroscale fibres orientation and magnitude induced by the ECM micro-fibres spatial distribution at microscale: derivation and well-posedness}}\label{mfomECM}
\dt{On every micro-domain $\delta Y(x)$ centred at a macro-point $x\in \Omega(t)$, at a given time instance $t\in[0,T]$, the spatial distribution of the micro-fibres $f(z,t)$ on $\delta Y(x)$ naturally provides a cumulative revolving orientation of these with respect to the barycentre $x$, and to derive this we proceed as follows.} 

\dt{Considering an arbitrary dyadic decomposition $\{\D_j\}_{j\in \J_{n}}$ of size $\delta 2^{-n}$ for the micro-domain $\delta Y(x)$, let us denote  by $z_{j}$ the barycentre of each dyadic cube $\D_{j}$. Then, for any $j\in \J_{n}$, the mass of micro-fibres distributed on $\D_{j}$ will influence the overall revolving fibre orientation on $\delta Y(x)$ through its contribution in direction of the position vector $\overrightarrow{x \, z_{j}}:=z_{j}-x$ in accordance with its weight relative standing with respect to the micro-fibre mass distributed on all other $\D_{j}$ covering $\delta Y(x)$. Therefore, the overall \emph{revolving micro-fibres orientation on $\delta Y(x)$ associated with the dyadic decomposition  $\{\D_j\}_{j\in \J_{n}}$} is given by:
\bequ
\begin{array}{lll}
\theta^{n}_{_{f,\delta Y(x)}}(x,t):&=&\sum\limits_{j \in \J_{n}} \frac{\int_{\D_{j}} f(\zeta,t) d\zeta}{\sum\limits_{j \in \J_{n}} \int\limits_{\D_{j}} f(\zeta,t) \ d\zeta } \overrightarrow{x \,z_{j}} \\[0.3cm]
&=&\sum\limits_{j \in \J_{n}} \frac{\int\limits_{\D_{j}} f(\zeta,t)  d\zeta}{ \int\limits_{\delta Y(x)} f(\zeta,t) d\zeta } \overrightarrow{x \,z_{j}} \\[0.3cm]
&=&\frac{\sum\limits_{j \in \J_{n}}\big(\frac{1}{\lambda(\D_{j})}\int\limits_{\D_{j}} f(\zeta,t)  d\zeta\big)\lambda(\D_{j})\overrightarrow{x \,z_{j}}}{ \int\limits_{\delta Y(x)} f(\zeta,t)  d\zeta } \\[0.3cm]
&=&\frac{\int\limits_{\delta Y(x)}\big[\sum\limits_{j \in \J_{n}}\big(\frac{1}{\lambda(\D_{j})}\int\limits_{\D_{j}} f(\zeta,t)  d\zeta\big)\chi_{_{\D_{j}}}(z)\overrightarrow{x \,z_{j}}\big]dz}{ \int\limits_{\delta Y(x)} f(\zeta,t)  d\zeta }\\[0.3cm] 
&=&\frac{\int\limits_{\delta Y(x)}\big[\sum\limits_{j \in \J_{n}}\big(\frac{1}{\lambda(\D_{j})}\int\limits_{\D_{l}} f(\zeta,t)  d\zeta\big)\chi_{_{\D_{j}}}(z)(z_{j}-x)\big]dz}{ \int\limits_{\delta Y(x)} f(\zeta,t)  d\zeta }
\label{eq1_03Aug2018}
\end{array}
\eequ
where $\lambda(\cdot)$ is the usual Lebesgue measure and $\chi_{_{\D_{l}}}(\cdot)$ is the characteristic function of the dyadic cube $D_{l}$. 
Thus, for any $n\in \N^{*}$ denoting the numerator function
\vspace{-0.3cm}
\bequ\label{eq2_03Aug2018}
g_{n}(z):=\sum\limits_{j \in \J_{n}}\bigg(\frac{1}{\lambda(\D_{j})}\int\limits_{\D_{j}} f(\zeta,t)  d\zeta\bigg)\chi_{_{\D_{j}}}(z)(z_{j}-x),
\eequ
let's observe immediately that $\{g_{n}\}_{n\in \N^{*}}$ is actually a sequence of vector-valued simple functions that is convergent to $f(z,t)(z-x)$ and that its associated sequence of integrals converges to the Bochner Intergal of $f(z,t)(z-x)$ on $\delta Y(x)$ with respect to $\lambda(\cdot)$ \cite{yosida1980}, namely
\bequ\label{eq3_03Aug2018}
\int\limits_{\delta Y(x)}f(z,t)(z-x)dz:=\lim_{n \to \infty }\int\limits_{\delta Y(x)} g_{n}(z)dz.
\eequ 
Hence, from \eqref{eq1_03Aug2018}-\eqref{eq3_03Aug2018}, we obtain that the sequence of revolving $\{\theta_{_{f}}^{n}(x,t)\}_{n\in \N^{*}}$ fibres orientations associated to the entire family of dyadic decompositions $\{\{D_j\}_{j\in \J_{n}}\}_{n\in \N^{*}}$ is convergent to the unique \emph{revolving barycentral micro-fibres orientation on $\delta Y(x)$} denoted by $\theta_{_{f,\delta Y(x)}}(x,t)$ and given by
\bequ\label{eq3_04Aug2018}
\begin{array}{lll}
\theta_{_{f,\delta Y(x)}}(x,t)&:=&\lim\limits_{n \to \infty }\theta^{n}_{_{f,\delta Y(x)}}(x,t)\\[0.3cm]
&=&\lim\limits_{n \to \infty }\frac{\int\limits_{\delta Y(x)}\big[\sum\limits_{j \in \J_{n}}\big(\frac{1}{\lambda(\D_{j})}\int\limits_{\D_{j}} f(\zeta,t)  d\zeta\big)(z_{j}-x)\chi_{_{\D_{j}}}(z)\big]dz}{ \int\limits_{\delta Y(x)} f(\zeta,t)  d\zeta }\\[0.3cm]
&=&\frac{\lim\limits_{n \to \infty }\int\limits_{\delta Y(x)} g_{n}(z)dz}{\int\limits_{\delta Y(x)} f(\zeta,t)  d\zeta}\\[0.3cm]
&=&\frac{\int\limits_{\delta Y(x)}f(z,t)(z-x)dz}{\int\limits_{\delta Y(x)} f(\zeta,t)  d\zeta}\\[0.3cm]
&=&\frac{\int\limits_{\delta Y(x)}f(z,t)(z-x)dz}{\int\limits_{\delta Y(x)} f(z,t)  dz},
\end{array}
\eequ
which is actually precisely the \emph{Bochner-mean-value} of the position vectors function $\delta Y(x)\ni z\mapsto z-x\in\R^{N}$ with respect to the measure $f(x,t)\lambda(\cdot)$ that is induced by the micro-fibres distribution.
\begin{figure}[h!]
\centering
\includegraphics[scale=0.4]{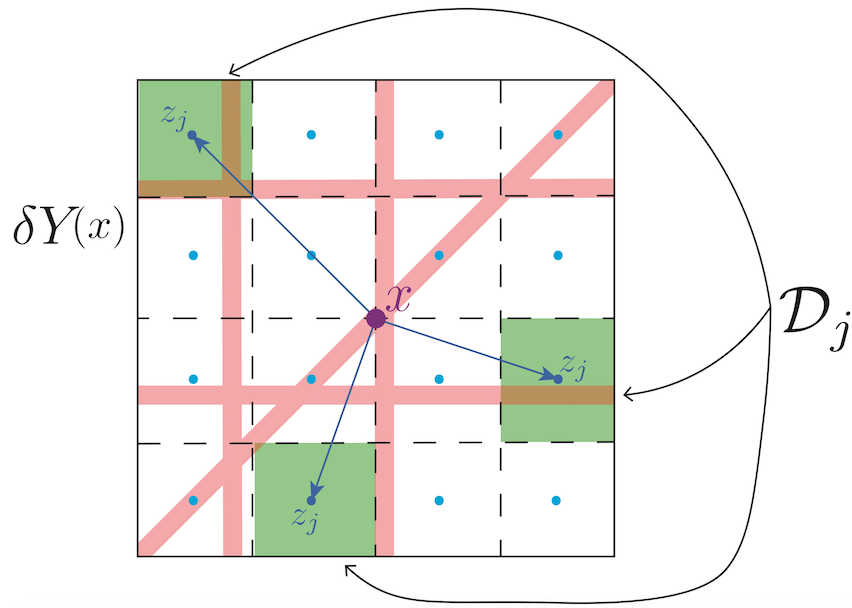}
\caption{\emph{Schematic of the fibre micro-domain $\delta Y(x)$ decomposed into dyadic cubes shown in green, $\D_{j}$, with associated barycentre $z_{j}$ in cyan. The position vectors are shown in dark blue.}}
\label{fig:fibmicro}
\end{figure}
Therefore, denoting by $\theta_{_{f}}(x,t)$ \emph{the macroscopic fibres orientation} at $(x,t)$ induced by the \emph{revolving barycentral micro-fibres orientation} on $\delta Y(x)$, we have that this is given by
\bequ\label{macroFibEq1}
\theta_{_{f}}(x,t):=\frac{1}{\lambda(\delta Y(x))}\int\limits_{\delta Y(x)} f(z,t)  dz\cdot \frac{\theta_{_{f,\delta Y(x)}}(x,t)}{\nor{\theta_{_{f,\delta Y(x)}}(x,t)}_{_{2}} }
\eequ 
Finally, the macroscopic representation of the ECM fibres distributed at $(x,t)$ is denoted by $F(x,t)$ and is given by the Euclidean magnitude of $\theta_{_{f}}(x,t)$, namely:
\bequ\label{macroFibEq2}
F(x,t):=\nor{\theta_{_{f}}(x,t)}_{_{2}},
\eequ
and so using now \eqref{macroFibEq1}, from \eqref{macroFibEq2} we obtain that 
\bequ\label{macroFibEq3}
\begin{array}{lll}
F(x,t):&=&\nor{\theta_{_{f}}(x,t)}_{_{2}}\\
&=&\nor{\frac{1}{\lambda(\delta Y(x))}\int\limits_{\delta Y(x)} f(z,t)  dz\cdot \frac{\theta_{_{f,\delta Y(x)}}(x,t)}{\nor{\theta_{_{f,\delta Y(x)}}(x,t)}_{_{2}} }}_{_{2}}\\
&=&\frac{1}{\lambda(\delta Y(x))}\int\limits_{\delta Y(x)} f(z,t)  dz \cdot \frac{\nor{\theta_{_{f,\delta Y(x)}}(x,t)}_{_{2}}}{\nor{\theta_{_{f,\delta Y(x)}}(x,t)}_{_{2}}}\\
&=& \frac{1}{\lambda(\delta Y(x))}\int\limits_{\delta Y(x)} f(z,t)  dz,
\end{array}
\eequ
which is precisely the mean-value of the micro-fibres distributed on $\delta Y(x)$.} \dt{Therefore, \emph{the macroscopic fibres orientation} at $(x,t)$ induced by the \emph{revolving barycentral micro-fibres orientation on $\delta Y(x)$} has its magnitude given by the mean-value of the micro-fibres on $\delta Y(x)$, and since in \eqref{eq1_03Aug2018}-\eqref{eq3_04Aug2018} we have ensured the well-posedness of $\theta_{_{f,\delta Y(x)}}(x,t)$, from \eqref{macroFibEq1}-\eqref{macroFibEq3}, we obtain that $\theta_{_{f}}(x,t)$ and $F(x,t)$ are also well-posed.}

\dt{With all these preparations, we are now in the position to describe the tumour macro-dynamics, which will be detailed in full in the next section.}

\dt{
\subsection{Tumour macro-dynamics}\label{tumourMacroDynamicSection}
To explore mathematically the macro-scale coupled dynamics exercised by the cancer cells mixed with the ECM, for notation convenience, let's first gather the macroscopic distributions of cancer and the two ECM phases considered here in the three-dimensional vector 
\[
\textbf{u}(x,t):=(c(x,t),F(x,t),l(x,t))^T,
\]
and let's denote tumour's volume fraction of occupied space by
\bequ\label{volFraction_eq17Sept2018}
\rho(\textbf{u}) \equiv \rho(\textbf{u}(x,t)) := \vartheta_v (F(x,t)+l(x,t)) + \vartheta_c c(x,t),
\eequ
with $\vartheta_v $ representing physical space occupied by the fibre and non-fibre phases of the ECM taken together and $\vartheta_c $ being the fraction of physical space occupied by the cancer cell population $c$.}

\dt{
Therefore, focusing first upon the cancer cell population, per unit time, under the presence of a proliferation law, its spatial dynamics is not only due to random motility (approximated mathematically by diffusion), but this is also crucially influenced by a combination of cell adhesion processes that include cell-cell adhesion and cell-matrix adhesion, with cell-matrix adhesion exhibiting distinctive characteristics in relation to the two ECM phases (namely: the fibres and non-fibres components).
Hence, assuming here a logistic proliferation law, the dynamics of the cancer cell population can be mathematically represented as 
\bequ 
\frac{\partial c}{\partial t} = \nabla \cdot [D_{1} \nabla c - c \mathcal{A}(t,x,\textbf{u}(\cdot, t), \theta_{_{f}}(\cdot, t))] +\mu_{1}c(1-\rho(\textbf{u})), 
\label{eq:c1} 
\eequ
where: $D_1$ and $\mu_1$ are non-negative diffusion and proliferation rates, respectively, while $\mathcal{A}(t,x,\textbf{u}(\cdot,t), \theta_{_{f}}(\cdot,t))$ represents a non-local constitutive flux term accounting for the critically important cell-adhesion processes that influence directly the spatial tumour movement, whose precise form will be explored as follows. }

\begin{figure}[h!]
\centering
\includegraphics[scale=0.7]{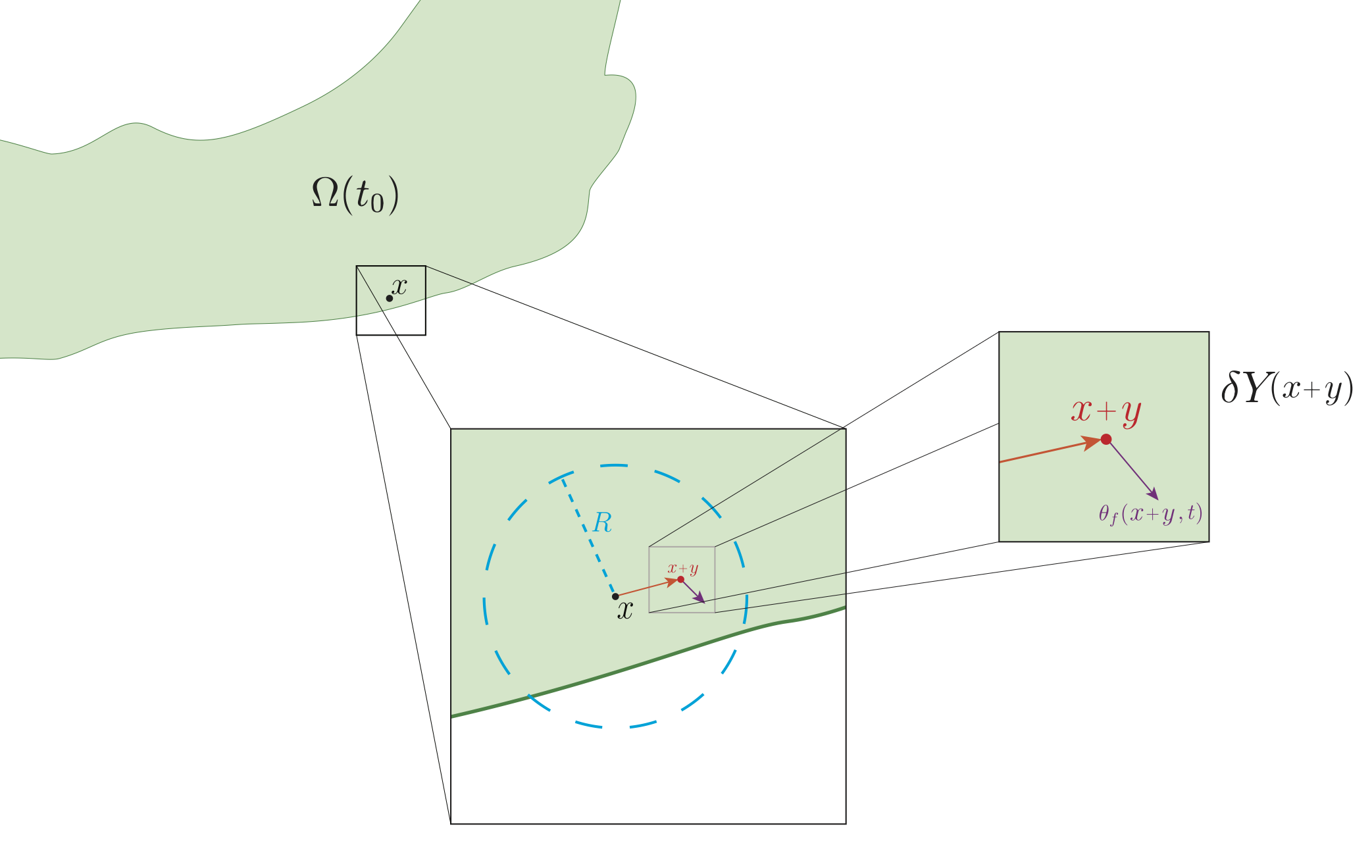}
\caption{\emph{Schematic to describe the process of the new cell-fibre adhesion term which includes the distribution of fibres. It shows the ball $B(x,R)$ centred at $x$ and of radius $R$, the point $x+y$ with the direction vector $n(y)$ in orange, and the fibre orientation $\theta_{f}(x+y,t)$ in purple.}}
\label{Fig:c-m_adhesion0}
\end{figure}

\dt{While generally adopting a similar perspective to the one in \cite{Armstrong_et_al_2006,Domschke_et_al_2014,Gerisch_Chaplain_2008} concerning cell-cell adhesion and cell-ECM-non-fibres substrate, here we move beyond the context considered in those works by accounting for the crucial role played by the cell-fibres adhesive interaction. Thus, within a sensing radius $R$, at a given time $t$ and spatial location $x$, the adhesive flux associated to the cancer cells distributed at $(x,t)$ will account for not only the adhesive interactions with the other cancer cells and ECM non-fibres phase distributed on $\Bila(x, R)$, but this will also appropriately consider and cumulate the adhesive interaction arising between cancer cells and the oriented ECM fibres, resulting in the following novel non-local adhesion flux term: 
}
\bequ\label{adhesiveTermExpression}
\begin{split}
\mathcal{A}(t,x,\textbf{u}(\cdot, t), \theta_{_{f}}(\cdot, t))=\frac{1}{R} \int_{B(0,R)} \mathcal{K}(\nor{\!y\!}_{2}) & \big(n(y) (\textbf{S}_{_{cc}} c(x\dt{+y},t) + \textbf{S}_{_{c\dt{l}}} \dt{l}(x\dt{+y},t)) \\
&+ \hat{n}(y) \ \textbf{S}_{_{cF}} \dt{F}(x\dt{+y},t) \big)(1-\rho(\textbf{u}))^{+}
\end{split}
\eequ
\dt{While the influence on adhesive interactions of the distance from the spatial location $x$ is accounted for through the radial kernel $\K(\cdot)$ detailed in Appendix \ref{kernelAppendix}, $n(\cdot)$ represents the usual unit radial vector given by}
\bequ
n(y):=
\left\{
\begin{array}{l}
y/||y||_2 \quad\text{if} \  y \in \Bila(0,R)\setminus\{0\}, \\[0.2cm]
(0,0) \quad \quad \text{if} \ y=(0,0),
\end{array}
\right.
\eequ
\dt{along which we consider the cell-cell and cell-ECM-non-fibres adhesion bonds established between the cancer cells distributed at $x$ and the cells and non-fibre ECM phase distributed at $x+y$ with strengths $\Sb_{_{cc}}$ and $\Sb_{_{cl}}$, respectively. Specifically, while $\Sb_{_{cl}}$ is considered here to be constant, as biological evidence discussed in \cite{Gu2014,Hofer2000} suggests that, in direct correlation to collagen levels, it is the high level of extracellular Ca$^{+2}$ ions rather than the sole production and presence of intracellular Ca$^{+2}$ that enables strong and stable adhesive bonds between cells, hence directly impacting the strength of cell-cell adhesion, we therefore assume that $\Sb_{_{cc}}$ is dependent on the collagen density, smoothly ranging between $0$ and a Ca$^{+2}$-saturation level $S_{_{max}}$, this being taken here of the form
\[
\Sb_{_{cc}}(x,t):=S_{_{max}}e^{\big({1-\frac{1}{1-(1-l(x,t))^2}}\big)}.
\]}
\dt{Finally, the last term in \eqref{adhesiveTermExpression} considers the crucially important adhesive interactions between the cancer cells  distributed at $x$ and the oriented fibres distributed on $\Bila(x, R)$. In this context, while the strength of this interaction is proportional to the macro-scale amount of fibres $F(\cdot, t)$ distributed at $x+y$, and, as illustrated in Figure \ref{Fig:c-m_adhesion0}, the orientation $\theta_{_{f}}(\cdot, t)$ of these fibres biases the direction of these adhesive interactions in the direction of the vector $\hat{n}(\cdot)$ defined by
}
\bequ
\hat{n}(y):=
\left\{
\begin{array}{l}
\frac{y+\theta_f(x+y)}{||y+\theta_f\dt{(x+y)}||_2} \quad\text{if} \  (y+\theta_f(x+y))  \neq (0,0) \\[0.2cm]
(0,0) \in \mathbb{R}^2 \quad \quad  \text{otherwise}.
\end{array}
\right.
\eequ

\dt{Further, per unit time, the fibres distribution are simply degraded by the cancer cells at macroscale, and so the dynamics of their macroscopic dynamics is simply governed by 
\bequ
\frac{d F}{d t} = -\gamma_{1} c F 
\label{eq:f} 
\eequ
where $\gamma_{1}$ describes the rate of degradation.} This macroscopic degradation of fibres is feed back to the micro-fibres $f(\cdot,t)$ on the micro-domains $\delta Y(x)$ as a factor which will lower their microscopic height accordingly. \dt{To complete the description of the macroscopic system,  the non-fibre ECM is degraded and remodelled by the cancer cells, and so its dynamics can be mathematically formulated as}
\bequ
\frac{d l}{d t} = -\gamma_{\dt{2}} c l + \omega(1-\rho(\textbf{u}))^{+} 
\label{eq:l}
\eequ
where $\gamma_{\dt{2}}$ describes the rate of degradation and $\omega$ denotes the rate of remodelling, \dt{while} the \textit{volume filling term} $(1-\rho(\textbf{u}))^{+}:=\max(0,(1-\rho(\textbf{u}))$ prevents the overcrowding of physical space.

\subsection{Microscopic fibre rearrangement induced by the macro-dynamics}

As the \dt{cancer} cells invade, they push the fibres in the direction they are travelling, thereby influencing the ECM fibres \dt{by their} own directive movement. \dt{Thus, in addition to the macroscale fibre degradation (explored in \eqref{eq:f}), during the tumour dynamics, at any instance in time $t$ and spatial location $x\in \Omega(t)$, the cancer cell population are also pushing and realigning the fibres, causing a microscale spatial rearrangement of the micro-fibres distributed on $\delta Y(x)$.  Specifically, this micro-fibres rearrangement is triggered by the macro-scale spatial flux of migratory cancer cells, namely by} 
 \bequ
 \mathcal{F}(x,t)=D_1 \nabla c(x,t)-c(x,t)\mathcal{A}(t,x,\textbf{u}(\cdot,t),\theta_{_{f}}(\cdot,t)).
 \label{spatial_flux}
 \eequ
\dt{Naturally magnified in accordance to the amount of cancer cells distributed at $(x,t)$ relative to the overall macro-scale amount of cancer cells and fibres that they meet at $(x,t)$, expressed here throught the weight
\[
\omega(x,t)=\frac{c(x,t)}{c(x,t)+F(x,t)},
\]
the spatial flux of migratory cells $ \mathcal{F}(x,t)$ gets balanced in a weighted manner by the macroscopic orientation $\theta_{f}(t,x)$, resulting in a \emph{rearrangement flux} vector-valued function given by
\bequ
r(\delta Y(x)\dt{, t}):=\omega(x,t)\mathcal{F}(x,t)+(1-\omega(x,t))\theta_{f}(x,t),
\eequ
which acts uniformly upon the micro-fibres distributed on $\delta Y(x)$, leading to an \emph{on-the-fly} change in the spatial distribution of micro-fibres on $\delta Y(x)$. In this context, denoting the barycentric position vector of any micro-scale position $z\in \delta Y(x)$ by
\[
x_\text{dir}(z)=\overrightarrow{ x \, z},
\]
let's observe that this microscale fibres rearrangement will be exercised provided that the micro-fibres $f(z,t)$ would not have already reached a certain maximum level $f_{\max}$ (when the micro fibre distribution would be very ``stiff'' and the cancer cells would struggle to move through those micro-locations) and that their movement magnitude will be affected by the micro-fibre saturation fraction 
\[
f^{*}=\frac{f(z,t)}{f_\text{max}}
\] 
combined with size of the micro-scale position defect with respect to $r(\delta Y(x)\dt{, t})$ that is given simply by 
\[
||r(\delta Y(x)) - x_\text{dir}(z)||_{2}.
\]
\begin{figure}[ht!]
    \centering
    \hspace{-0.5cm}\begin{subfigure}{.4\linewidth}
        \includegraphics[scale=0.3]{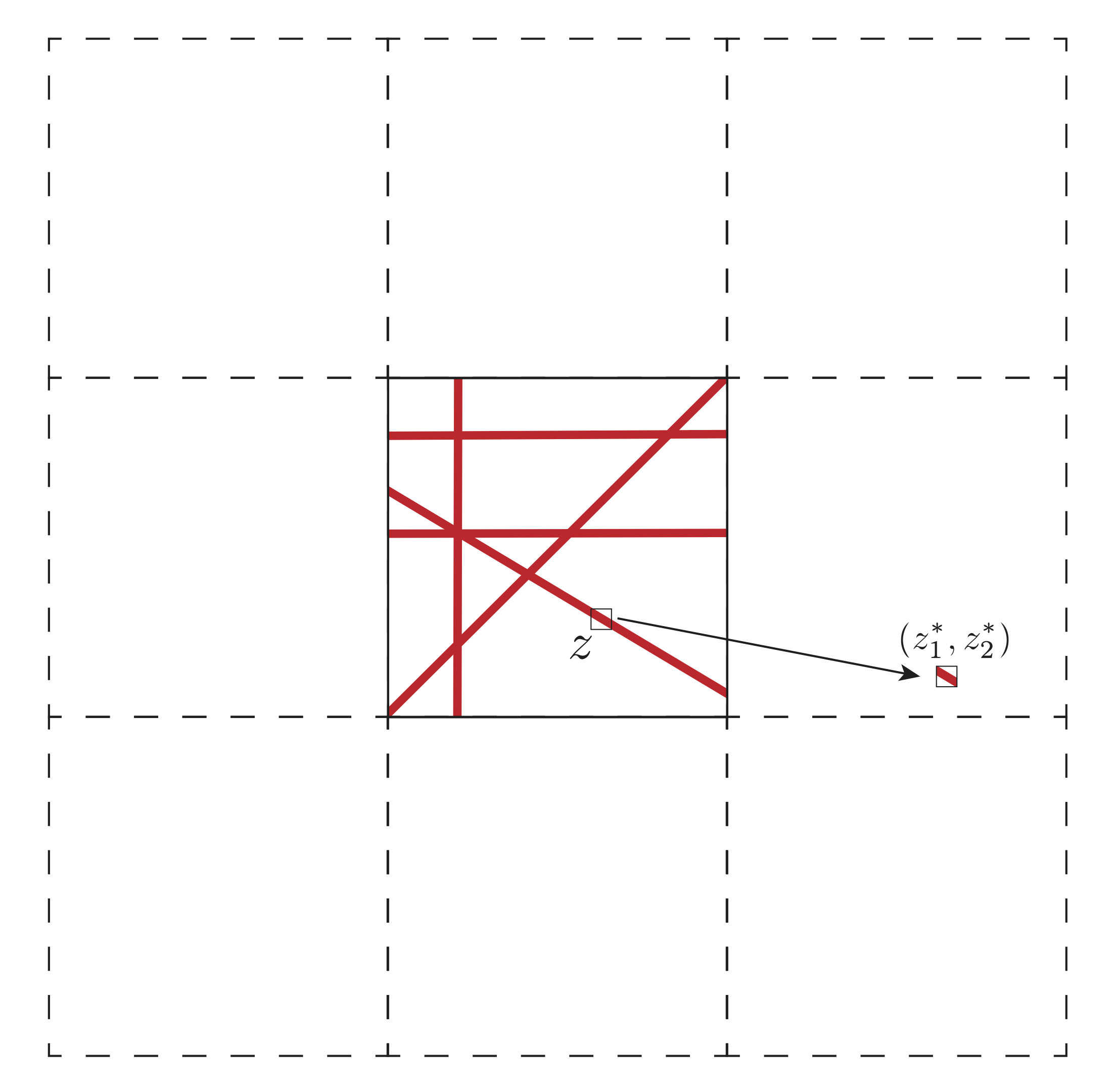}
        \caption{\emph{Relocation micro-fibre distributions in $ \delta Y(x)$ within its neighbouring micro-domains}}
    \end{subfigure}
    \hskip5em
    \begin{subfigure}{.4\linewidth}
        \includegraphics[scale=0.35]{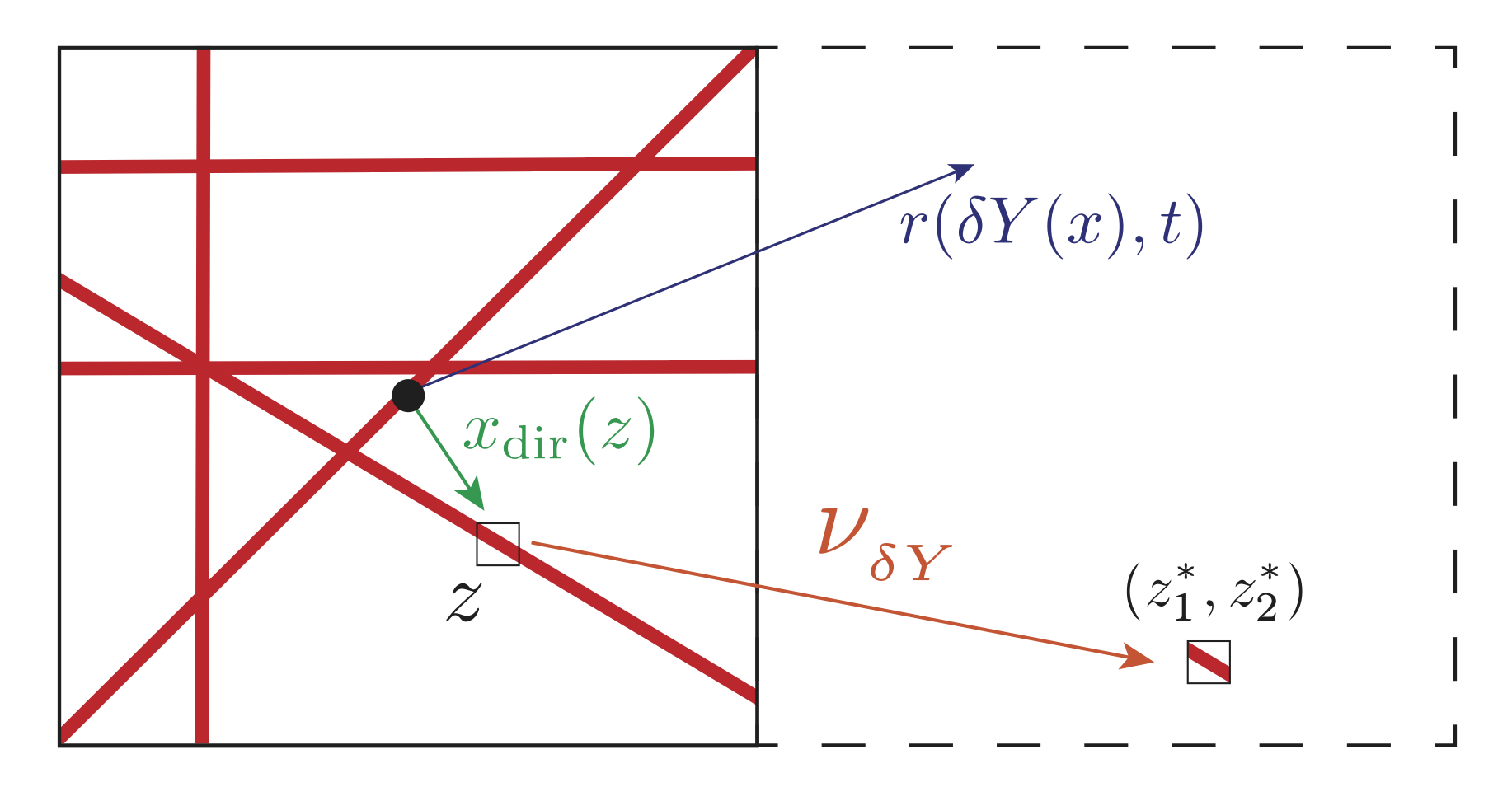}
        \caption{\emph{Enlarged version of $\delta Y$ cube illustrating the components required for relocation}}
    \end{subfigure}
    \caption{\emph{Schematics to describe the process of reallocation of fibre distribution in each $\delta Y$ cube.}}
    \label{fig:grid_move}
\end{figure}
Therefore, under the action of the rearrangement flux $r(\delta Y(x)\dt{, t})$, the micro-fibres distributed at $z$ will attempt to exercise their movement in the direction of the resulting vector  $x_\text{dir}(z)+r(\delta Y(x)\dt{, t})$, and so their relocation to the corresponding position within neighbouring micro-domain will be given by the vector-valued function: }
\begin{equation}
\nu_{_{\delta Y\dt{(x)}}}(z,\dt{t})=\left(x_\text{dir}(z) + r(\delta Y(x)\dt{, t})\right) \cdot \frac{f(z,t)(f_{\text{max}}-f(z,t))}{f^{*}+||r(\delta Y(x)) - x_\text{dir}(z)||_{2}} \cdot \chi_{_{\dt{\{f(\cdot,t)>0\}}}}
\label{eq:fibnu}
\end{equation}
\dt{where} $\chi_{_{\dt{\{f(\cdot,t)>0\}}}}$ represents the \dt{usual} characteristic function of \dt{the micro-fibres support set $\{f(\cdot,t)>0\}:=\{z\in \delta Y(x)|\,f(z,t)>0\}$. Finally, the movement of the micro-fibres distributed at $z$ to the newly attempted location $z^{*}$ given by
\[
z^{*}:=z+\nu_{_{\delta Y(x)}}(z,\dt{t})
\]
is exercised in accordance with the space available at the new position $z^{*}$. Thus, this is explored here through the movement probability
\[
p_{move}:=\max\big(0,\frac{f_\text{max}-f(z^{*},t)}{f_\text{max}}\big)
\]
which enables only an amount of $p_{move}f(z,t)$ of micro-fibres to be transported to position $z^{*}$ (as illustrated in Figure \ref{fig:grid_move}), while the rest of $(1-p_{move})f(z,t)$ remains at $z$. }
\begin{figure}[h!]
\centering
\includegraphics[scale=0.6]{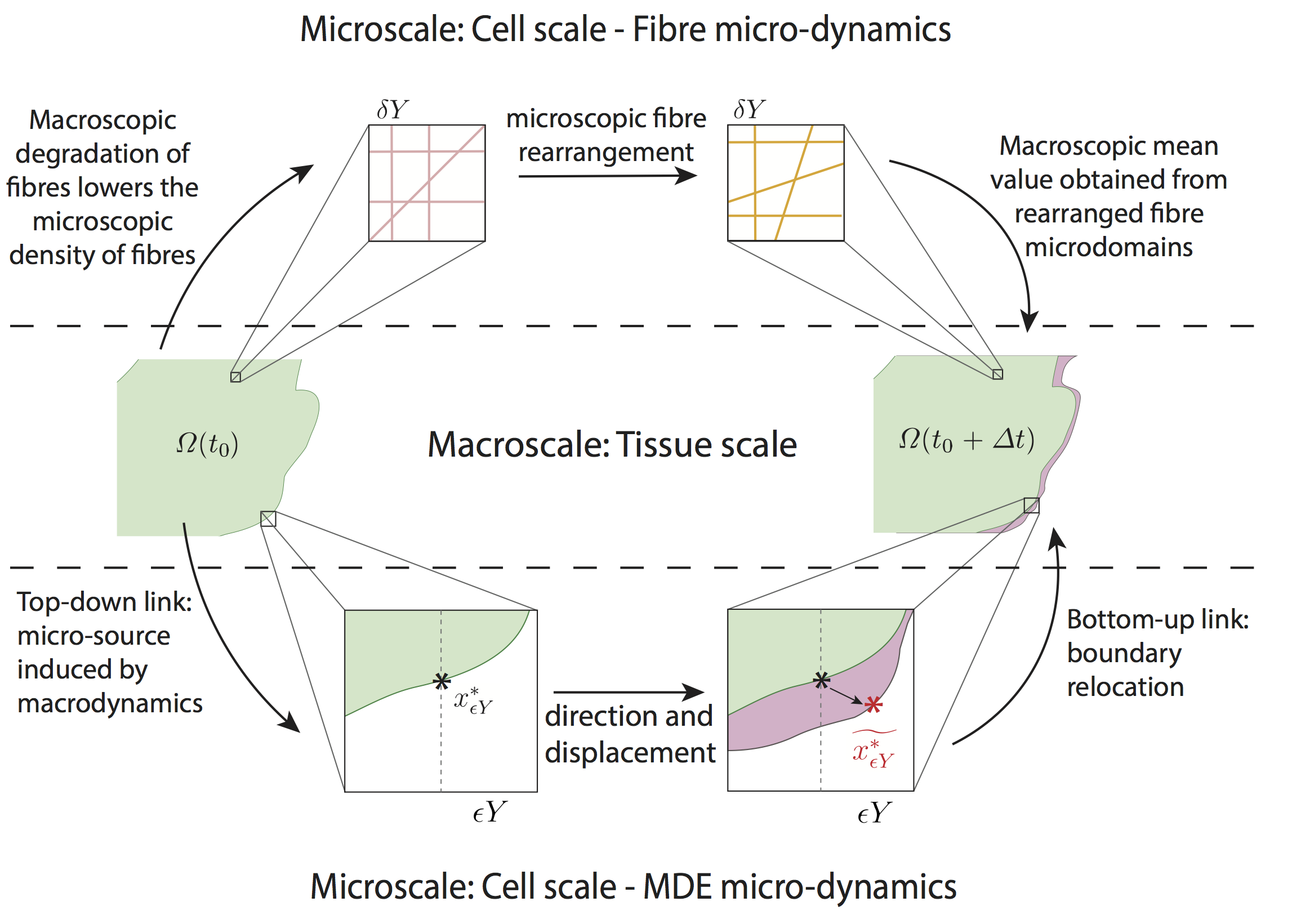}
\caption{\emph{Schematic summary of global multi-scale model. }}
\label{Fig:c-m_adhesion}
\end{figure}
\subsection{\dt{Schematic} summary of global multi-scale model}
In this model, there are two \dt{interconnected} multi-scale systems, \dt{each with their own distinct cell-scale micro-dynamics, but} both \dt{of them sharing} the same macro-scale cancer dynamics at the tissue scale, being linked to this through two double feedback loops, as illustrated in Figure \ref{Fig:c-m_adhesion}. The macro-scale dynamics governs the spatial distribution of both the \dt{invading} cancer cells and the fibrous and non-fibrous \dt{density components} of the ECM, and is given by the following non-local coupled system:
\begin{subequations}
\begin{align}
\begin{split}
\frac{\partial c}{\partial t} &= \nabla \cdot [D_{1} \nabla c - c \mathcal{A}(t,x,\textbf{u}(\cdot,t), \theta_{_{f}}(\cdot, t))] +\mu_{1}c(1-\rho(\textbf{u})),
\end{split}\label{globalMacroSystem-c}\\[2mm]
\begin{split}
\frac{d F}{d t} &= -\gamma_{1} c F ,
\end{split}\label{globalMacroSystem-F}\\[2mm]
\frac{d l}{d t} &= -\gamma_{\dt{2}} c l + \omega(1-\rho(\textbf{u})).
\label{globalMacroSystem-l}
\end{align}\label{globalMacroSystem}
\end{subequations}
Occurring on the \dt{cell}-scale, the \dt{micro-scale part of the first multiscale system} controls the dynamic redistribution of microscopic fibres within the \dt{entire} cancer region. \dt{On each micro-domain $\delta Y(x)$, t}he realignment of the \dt{existing micro-}fibres \dt{is triggered by} the spatial flux of cancer cells from the macro-scale and \dt{this is realised by weighted action of this over} the \dt{oriented} macroscopic fibres \dt{distribution that they meet at $x$}. Once all fibre micro-domains within the cancer region have undergone redistribution, \dt{a} new macroscopic fibre orientation and mean value per each $\delta Y(x)$ is \dt{obtained and that in its turn will have its effect in the important cell-adhesion behaviour that the cancer exhibits at macro-scale}. 
\dt{Finally, in} the second \dt{multi}-scale system, the spatial distribution of cancer cells induces a source of MDEs on the boundary at the micro-scale level. \dt{In return, the leading edge} \dt{proteolytic micro-}dynamics of MDEs instigates a change in the position of the \dt{tissue-scale} tumour boundary that corresponds to the pattern of the peritumoural ECM degradation, enabling this way the invasion process to continue on the expanding domain.

\section{Numerical Approach:  \dt{Key Points of the Implementation}} 
\dt{Building on the numerical multiscale platform initially introduced in \cite{Dumitru_et_al_2013}, the implementation of the novel multiscale moving boundary model that we proposed in this work required a number of \dt{new major computational} steps, which will be detailed in the next three subsections. These include a special treatment for several computational aspects, such as those concerned with: macroscale computations on the expanding tumour;  the macro-scale adhesion term; and a new predictor-corrector scheme for the cancer dynamics equation \eqref{globalMacroSystem-c}}. 

\dt{Finally, the approach for the cross-interface proteolytic micro-dynamics on each tumour boundary micro-domain $\epsilon Y$ follow precisely the steps described in \cite{Dumitru_et_al_2013}, involving a finite element scheme using bilinear shape functions and square elements, reason for which we do not include that here.}

\subsection{Macroscale computations on the expanding tumour domain}
\label{macro-comp}
\dt{While considering a uniform spatial mesh of size $\Delta x= \Delta y = h$ for the maximal cube $Y$, recoded on a square grid $\{(x_{i},x_{j})\}_{i,j=1...M}$, with $M:=length(Y)/h+1$, the actual macroscopic computation will be performed exclusively on the expanding tumour $\Omega(t_{0})$ over every macro-micro time interval $[t_{0}, t_0 + \Delta t]$ as will be detailed in the following. Specifically, to explore this, let us first denote by $\I(\cdot,\cdot):\{1,...,M\}\times \{1,...,M\}\to\{0,1\}$ the \emph{on-grid cancer indicator} function given as usual by
 \bequ\label{movingMeshOp}
 \I(i,j):=
 \left\{
 \begin{array}{lll}
 1 & \quad \text{if} & (x_{i},x_{j}) \in \Omega(t_{0}),\\[0.2cm]
 0 & \quad \text{if} & (x_{i},x_{j}) \not\in \Omega(t_{0}),
 \end{array}
 \right.
 \eequ
Further, let's observe that the \emph{on-grid closest neighbour indicator} functions $I_{x,+1}(\!\cdot,\!\cdot)\,,I_{x,-1}(\!\cdot,\!\cdot),I_{y,+1}(\!\cdot,\!\cdot)\,,I_{y,-1}(\!\cdot,\!\cdot)\!:\{2,...,M-1\}\times \{2,...,M-1\}\to\{0,1\}$, defined by 
\bequ
\begin{array}{lll}
I_{x,\pm1}(i,j) &:=& |\mathcal{I}(i,j)-\mathcal{I}(i,j\pm1)|\cdot \mathcal{I}(i,j), \\[0.2cm]
I_{y,\pm1}(i,j) &:=& |\mathcal{I}(i\pm1,j)-\mathcal{I}(i,j)|\cdot \mathcal{I}(i,j),
\end{array}
\label{eq:IndCbackforw}
\eequ
enable us to detect \emph{on-the-fly} the grid positions immediately outside the cancer boundary as the points of non-zero value along each spatial direction, given by the union of preimages $I^{-1}_{x,-1}(\{1\})\cup I^{-1}_{x,+1}(\{1\})$ and $I^{-1}_{y,-1}(\{1\})\cup I^{-1}_{y,+1}(\{1\})$ for $x-$ and $y-$ direction, respectively.}

\dt{Over each macroscale time perspective $[t_{0}, t_{0}+\Delta t]$ the overall macroscopic scheme for \eqref{globalMacroSystem} involves the method of lines coupled with a novel predictor-corrector method for time marching (whose main steps will be detailed in the next subsection), the discretisation of the spatial operators appearing in the right-hand side of \eqref{globalMacroSystem-c} is based on central differences and midpoint approximations. For this, considering a uniform discretisation $\{t_{p}\}_{p= 0...k}$ of $[t_{0}, t_{0}+\Delta t]$, of time step $\delta t>0$, let's denote by $c^{p}_{i,j},\, \A^{p}_{i,j}, \,F^{p}_{i,j},\, l^{p}_{i,j}  $ the discretised values of $c, \A, F, l$ at $((x_{i}, x_{j}), t_{p})$, respectively. Thus, at any spatial node $(x_{i},x_{j}) \in \Omega(t_{0})$, the no-flux across the moving boundary dynamics is accounted for via the indicators \eqref{movingMeshOp}-\eqref{eq:IndCbackforw} on the expanding spatial mesh, and results into the midpoint approximations
\bequ
\begin{array}{lll}
c^{p}_{i,j\pm\frac{1}{2}}&:=&\frac{c^{p}_{i,j}+[I_{x,\pm 1}(i,j)c^{p}_{i,j}+\I(i,j\pm 1)c^{p}_{i,j\pm 1}]}{2},\\
c^{p}_{i\pm\frac{1}{2},j}&:=&\frac{c^{p}_{i,j}+[I_{y,\pm 1}(i,j)c^{p}_{i,j}+\I(i\pm 1, j)c^{p}_{i\pm 1, j}]}{2},
\end{array}
\eequ
and 
\bequ
\begin{array}{lll}
\A^{p}_{i,j\pm\frac{1}{2}}&:=&\frac{\A^{p}_{i,j}+[I_{x,\pm 1}(i,j)\A^{p}_{i,j}+\I(i,j\pm 1)\A^{p}_{i,j\pm 1}]}{2},\\
\A^{p}_{i\pm\frac{1}{2},j}&:=&\frac{\A^{p}_{i,j}+[I_{y,\pm 1}(i,j)\A^{p}_{i,j}+\I(i\pm 1, j)\A^{p}_{i\pm 1, j}]}{2},
\end{array}
\eequ
while the central differences for $c$ at the virtual nodes $(i,j\pm \frac{1}{2})$ and $(i\pm \frac{1}{2},j)$ are given by
\bequ
\begin{array}{lll}
[c_{x}]^p_{i,j+\frac{1}{2}}& := & \frac{[I_{x,+1}(i,j)c_{i,j}+\I(i,j+1)c_{i,j+1}]-c_{i,j}}{dx}, \\[2mm]
[c_{x}]^p_{i,j-\frac{1}{2}}& := & \frac{c_{i,j}-[I_{x,-1}(i,j)c_{i,j}+\I(i,j-1)c_{i,j-1}]}{dx}, \\ [2mm]
[c_{y}]^p_{i+\frac{1}{2},j}& := & \frac{[I_{y,+1}(i,j)c_{i,j}+\I(i+1,j)c_{i+1,j}]-c_{i,j}}{dy}, \\[2mm]
[c_{y}]^p_{i-\frac{1}{2},j}& := & \frac{c_{i,j}-[I_{y,-1}(i,j)c_{i,j}+\I(i-1,j)c_{i-1,j}]}{dy}.
\end{array}
\eequ
Therefore, the approximation for the term $\nabla \cdot [D_{1}\nabla c - c \mathcal{A}(t,x,\textbf{u}(\cdot, t), \theta_{_{f}}(\cdot,t))]$ in \eqref{globalMacroSystem-c} is obtained as
\begin{align}\label{discreteExplicitSpFlux}
\begin{split}
(\nabla \cdot [D_{1}\nabla c & - c \mathcal{A}(t,x,\textbf{u}(\cdot,t),\theta_{_{f}}(\cdot,t))])^{p}_{i,j}   \\ 
\simeq&\frac{D_{1}([c_{x}]^p_{i,j+\frac{1}{2}} - [c_{x}]^p_{i,j-\frac{1}{2}}) - c^p_{{i,j+\frac{1}{2}}}\cdot \mathcal{A}^p_{{i,j+\frac{1}{2}}} + c^p_{{i,j-\frac{1}{2}}}\cdot \mathcal{A}^p_{{i,j-\frac{1}{2}}}}{\Delta x}\\[0.1cm]
&+\frac{D_{1}([c_{y}]^p_{i+\frac{1}{2},j} - [c_{y}]^p_{i-\frac{1}{2},j}) - c^p_{{i+\frac{1}{2},j}}\cdot \mathcal{A}^p_{{i+\frac{1}{2},j}} + c^p_{{i-\frac{1}{2},j}}\cdot \mathcal{A}^p_{{i-\frac{1}{2},j}}}{\Delta y}, 
\end{split}
\end{align}
and so, denoting by $\F^{p}_{i,j}$ the discretised value of the flux $\F(\cdot,\cdot)$ at the spatio-temporal node $((x_{i},x_{j}),t_{p})$, the spatio-temporal discretisation of $\nabla\cdot \F:=\nabla \cdot [D_{1}\nabla c - c \mathcal{A}(t,x,\textbf{u}(\cdot, t), \theta_{_{f}}(\cdot,t))]$ given in \eqref{discreteExplicitSpFlux} can therefore be equivalently expressed in a compact form as 
\bequ
(\nabla\cdot \F)^{p}_{i,j}\simeq \frac{\F^{p}_{i,j+\frac{1}{2}}-\F^{p}_{i,j-\frac{1}{2}}}{\Delta x}+\frac{\F^{p}_{i+\frac{1}{2},j}-\F^{p}_{i-\frac{1}{2},j}}{\Delta y}
\label{dist_flux}
\eequ
where $\F^{p}_{\!i,j\pm\frac{1}{2}}\!\!\!:=\!\!D_{1}[c_{x}]^p_{i,j\pm\frac{1}{2}}\!\!-c^p_{{i,j\pm\frac{1}{2}}}\!\cdot\! \mathcal{A}^p_{{i,j\pm\frac{1}{2}}} $ and $\F^{p}_{\!i\pm\frac{1}{2},j}\!\!\!:=\!\!D_{1}[c_{y}]^p_{i\pm\frac{1}{2},j}\!\!-c^p_{{i\pm\frac{1}{2},j}}\!\cdot\! \mathcal{A}^p_{{i\pm\frac{1}{2},j}} $.}

\subsection{Adhesive flux computation}
\dt{As already mentioned above}, an important aspect within the macroscopic part of our solver is the numerical approach addressing the adhesive flux $\mathcal{A}(t,x,\textbf{u}(\cdot,t), \theta_{_{f}}(\cdot, t))$, \dt{which} explores the effects of cell-cell, cell-ECM-non-fibre and cell-fibre adhesion of cancer cell population. \dt{Although to a certain extent we adopt a similar approach to the one  that we previously proposed in \cite{shutt_chapter} (for a similar macro-scale invasion context but in the absence of fibre dynamics), the numerical approximation for the non-local term $\mathcal{A}(t,x,\textbf{u}(t,\cdot), \theta_{_{f}}(\cdot, t))$ involves a series of off-grid computations on a new decomposition of the sensing region, developing further the approach introduced \cite{shutt_chapter} and adapting that to the new context of the current macro-model. For completeness, we detail this here as follows}. \dt{Thus, at a given spatio-temporal node $((x_{i},x_{j}), t_{p})$, w}e decompose the sensing region $\dt{\Bila((x_{i}, x_{j}),R)}$ in 
\bequd
\dt{q:=\sum\limits_{i=1}^{s} 2^{m+(i-1)}}\textrm{ annulus radial sectors }\Sect_{1}, \dots, \Sect_{\dt{q},}
\eequd
\begin{figure}[h] 
\centering
\includegraphics[scale=0.2]{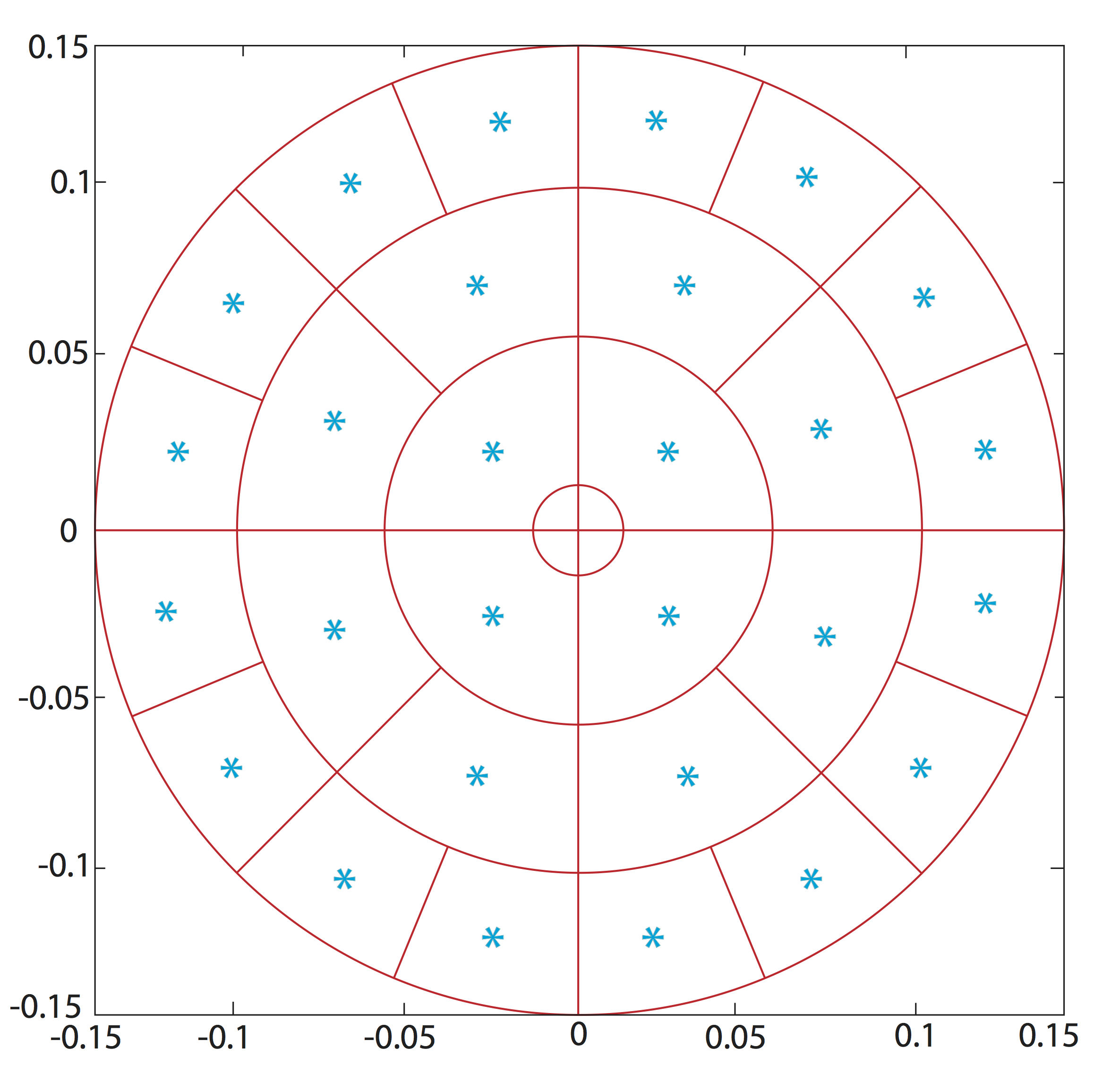}
\caption{\emph{Sensing region $B(x,R)$ approximated by the annulus radial sectors with the barycentre $\bary_{_{\Sect_{\nu}}}$ associated to each sector $\Sect_{\nu}$ highlighted with a blue star.}}
\label{figcircle}
\end{figure}
\dt{which are} obtained by intersecting \dt{each annulus $i\in\{1,...,s\}$} annuli with \dt{a corresponding number of $2^{m+(i-1)}$} uniformly distributed radial sectors of $\dt{\Bila (x_{i},x_{j}),R)}$, as shown in Figure \ref{figcircle}, \dt{while considering the remaining central circle to be of a computationally negligible radius.} Then, using a standard barycentric interpolation approach for approximating the off-grid values, \dt{$\forall \nu\in \{1,\dots,q\}$, on each annulus sector $S_{\nu}$, we calculate the mean-values of all the macro-scale densities of cancer cells $c(\cdot,t_{p})$, ECM non-fibres component $l(\cdot, t_{p})$, macroscopic ECM fibres $F(\cdot, t_{p})$ and their associated directions $\theta_{_{f}}(\cdot,t_{p})$, namely: 
\bequd
\begin{array}{lll}
W^{p}_{\Sect_{\nu}, c}:=\!\!\frac{1}{\lambda(\Sect_{\nu})}\int\limits_{\Sect_{\nu}}c(\xi,t_{p}) \ d\xi,& \qquad & 
W^{p}_{\Sect_{\nu}, l}:=\!\!\frac{1}{\lambda(\Sect_{\nu})}\int\limits_{\Sect_{\nu}}l(\xi,t_{p}) \ d\xi,\\[4mm]
W^{p}_{\Sect_{\nu}, F}:=\!\!\frac{1}{\lambda(\Sect_{\nu})}\int\limits_{\Sect_{k}}F(\xi,t_{p}) \ d\xi, & \text{and} &
W^{p}_{\Sect_{\nu}, \theta_{_{f}}}:=\!\!\frac{1}{\lambda(\Sect_{\nu})}\int\limits_{\Sect_{\nu}}\theta_{f}(\xi,t_{p}) \ d\xi,
\end{array}
\eequd
respectively.} \dt{Further}, \dt{$\forall \nu\in\{1,\dots, q\}$}, denoting by $\bary_{_{\Sect_{\nu}}}$ the barycenter of $\Sect_{\nu}$, \dt{this enable us to evaluate the} unit vector denoted by $\normal_{\nu}$ \dt{that points from the centre of the sensing region to $\bary_{_{\Sect_{\nu}}}$}, i.e., 
\[
\normal_{\nu}:=\frac{\bary_{_{\Sect_{\nu}}}-\dt{(x_{i}, x_{j})}}{\nor{\bary_{_{\Sect_{\nu}}}-\dt{(x_{i}, x_{j})}}_{_{2}}},
\]
\dt{as well as the corresponding macroscopic vector accounting for the influence of the cumulative mean-value direction of the fibres on $\Sect_{\nu}$, namely 
\bequd
\widehat\normal^{p}_{\nu}:=\frac{\normal_{\nu}+W^{p}_{\Sect_{\nu}, \theta_{_{f}}}}{\nor{\normal_{\nu}+W^{p}_{\Sect_{\nu}, \theta_{_{f}}}}_{_{2}}}
\eequd
Thus, finally, the approximation of the adhesion flux $\mathcal{A}(t,x,\textbf{u}(\cdot, t), \theta_{_{f}}(\cdot, t))$ at the spatio-temporal node $((x_{i},x_{j}), t_{p})$, is denoted by $\A^{p}_{i,j}$ and is given by
\bequ
\A^{p}_{i,j}\!\!=\!\!\frac{1}{R}\!\!\!\!\!\!\sum\limits_{\substack{\nu=1\\[0.1cm] \mathmakebox[\widthof{$l$}+8\fboxrule][c]{\bary_{_{\Sect_{\nu}}}}\in\Omega(t_{0})}}^{q}\!\!\!\!\!\!\K(\bary_{_{\Sect_{\nu}}}\!)[\normal_{\nu}\cdot(\textbf{S}_{cc}W^{p}_{\Sect_{\nu}, c} +\textbf{S}_{cl} W^{p}_{\Sect_{\nu}, l})+\widehat\normal^{p}_{\nu}\cdot\textbf{S}_{cF}W^{p}_{\Sect_{\nu}, F}](1-\rho(\mathbf{u}^{p}_{\bary_{_{\Sect_{\nu}}}}\!)\!)^{+}\lambda(\Sect_{\nu})
\eequ
where, $\forall \nu\in\{1,\dots, q\}$, denoting $\mathbf{u}^{p}_{\bary_{_{\Sect_{\nu}}}}:=[W^{p}_{\Sect_{\nu}, c},  W^{p}_{\Sect_{\nu}, l}, W^{p}_{\Sect_{\nu}, F}]^T$, we have that $\rho(\mathbf{u}^{p}_{\bary_{_{\Sect_{\nu}}}}\!)$ is volume fraction defined in \eqref{volFraction_eq17Sept2018} that corresponds to the discrete vector $\mathbf{u}^{p}_{\bary_{_{\Sect_{\nu}}}}$.}

\subsection{\dt{The predictor-corrector step}}
\label{p-c}
\dt{For the time discretisation of equation \eqref{globalMacroSystem-c}, we develop a novel predictor-corrector scheme involving a non-local trapezoidal corrector. For this, let us denote by $H(\cdot,\cdot,\cdot)$ the right-hand side spatial operator of \eqref{globalMacroSystem-c}, which is defined as follows. At any instance in time and any corresponding triplet $(\overline{\F}, \overline{c}, \overline{\textbf{u}})$ of given spatially discretised values for the flux $\F$, the cell population $c$, and the tumour vector $\textbf{u}$, by ignoring for simplicity the time notation we have that $H$ is given by  
\bequ\label{eqOperatorH}
H(\overline{\F}_{i,j},\overline{c}_{i,j},\overline{\textbf{u}}_{i,j}):=(\nabla \cdot \overline{\F})_{i,j} + \mu_{1}\overline{c}_{i,j}(1-\rho(\overline{\textbf{u}}_{i,j})), 
\eequ
where the spatial discretisation $(\nabla \cdot \overline{\F})_{i,j}$ is given here still by \eqref{dist_flux} but applied to the spatial flux $\overline{\F}$, and  $\rho(\overline{\textbf{u}}_{i,j})$ is simply the volume fraction defined in \eqref{volFraction_eq17Sept2018} evaluated for the discrete vector $\overline{\textbf{u}}_{i,j}:=[\overline{c}_{i,j}, \overline{F}_{i,j},\overline{l}_{i,j}]$, $\forall i,j=1...M$. In this context, on the time interval $[t_{p}, t_{p+1}]$, we first predict \emph{on-the-fly} values for $c$ at $t_{p+\frac{1}{2}}$, namely
\bequ\label{predictoPartOne}
\tilde{c}^{p+\frac{1}{2}}_{i,j}=c^{p}_{i,j}+\frac{\delta t}{2}H(\F^{p}_{i,j},c^{p}_{i,j},\textbf{u}^{p}_{i,j}). 
\eequ
where $\textbf{u}^{p}_{i,j}:=[c^{p}_{i,j}, F^{p}_{i,j},l^{p}_{i,j}]$, $\forall i,j=1...M$. 
Further, using these predicted values $\tilde{c}^{p+\frac{1}{2}}$, we calculate the corresponding predicted flux at $t_{p+\frac{1}{2}}$, namely $\tilde{\F}^{p+\frac{1}{2}}$, and then we construct a non-local corrector that involves the average of the flux at the active neighbouring spatial locations 
\bequ\label{activeLocations}
\{(x_{i}, x_{j\pm1}), (x_{i\pm1}, x_{j}), (x_{i\pm1}, x_{j-1}), (x_{i\pm1}, x_{j+1})\}\cap \Omega(t_{0}). 
\eequ
Thus, denoting by $\Ncal$ the set of indices corresponding to these active locations, we have that the corrector flux is calculated as
\bequ
\F^{*p+\frac{1}{2}}_{i,j}=\frac{1}{card(\Ncal)}\sum\limits_{(\sigma,\zeta)\in \Ncal}\tilde{\F}^{p+\frac{1}{2}}_{\sigma,\zeta},
\eequ
ultimately enabling us to use the trapezoidal approximation to obtain the corrected value for $c$  at $t_{p+\frac{1}{2}}$ as
\bequ\label{correctorPartOne}
c^{p+\frac{1}{2}}_{i,j}=c^{p}_{i,j}+\frac{\delta t}{4}\big[H(\F^{p}_{i,j},c^{p}_{i,j},\textbf{u}^{p}_{i,j})+H(\F^{*p+\frac{1}{2}}_{i,j},\tilde{c}^{p+\frac{1}{2}}_{i,j},\tilde{\textbf{u}}^{p+\frac{1}{2}}_{i,j})\big]
\eequ
where  $\tilde{\textbf{u}}^{p+\frac{1}{2}}_{i,j}:=[\tilde{c}^{p+\frac{1}{2}}_{i,j}, F^{p}_{i,j},l^{p}_{i,j}]$, $\forall i,j=1...M$. Finally, we use the average
\bequ
\bar{c}^{p+\frac{1}{2}}_{i,j}:=\frac{c^{p}_{i,j}+c^{p+\frac{1}{2}}_{i,j}}{2}
\eequ
to re-evaluate the flux at $t_{p+\frac{1}{2}}$,
namely $\F^{p+\frac{1}{2}}$ (corresponding to the average values $\bar{c}^{p+\frac{1}{2}}$) and then to initiate the predictor-corrector steps described above on this new time interval $[t_{p+\frac{1}{2}}, t_{p+1}]$. Thus, following the predictor step, we first obtain the predicted values at at $t_{p+1}$, namely
\bequ\label{predictoPartTwo}
\tilde{c}^{p+1}_{i,j}=\bar{c}^{p+\frac{1}{2}}_{i,j}+\frac{\delta t}{2}H(\F^{p+\frac{1}{2}}_{i,j},\bar{c}^{p+\frac{1}{2}}_{i,j},\bar{\textbf{u}}^{p+\frac{1}{2}}_{i,j}) 
\eequ
where $\bar{\textbf{u}}^{p+\frac{1}{2}}_{i,j}:=[\bar{c}^{p+\frac{1}{2}}_{i,j}, F^{p}_{i,j},l^{p}_{i,j}]$, $\forall i,j=1...M$. Finally, we correct these values at $t_{p+1}$ with the same non-local trapezoidal-type corrector as described in \eqref{correctorPartOne}, here involving the corrector flux calculated as average of the predicted flux values $\tilde{\F}^{p+1}$ (corresponding to the predicted values $\tilde{c}^{p+1}$) at the active neighbouring locations given in \eqref{activeLocations}, namely
\bequ
\F^{*p+1}_{i,j}=\frac{1}{card(\Ncal)}\sum\limits_{(\sigma,\zeta)\in \Ncal}\tilde{\F}^{p+1}_{\sigma,\zeta}.
\eequ
Thus, this last corrector step gives us ultimately the values that we accept at $t_{p+1}$, namely
\bequ
c^{p+1}_{i,j}=\bar{c}^{p+1}_{i,j}+\frac{\delta t}{4}\big[H(\F^{p+\frac{1}{2}}_{i,j},\bar{c}^{p+\frac{1}{2}}_{i,j},\bar{\textbf{u}}^{p+\frac{1}{2}}_{i,j}) +H(\F^{*p+1}_{i,j},\tilde{c}^{p+1}_{i,j},\tilde{\textbf{u}}^{p+1}_{i,j})\big]
\eequ
where $\tilde{\textbf{u}}^{p+1}_{i,j}:=[\tilde{c}^{p+1}_{i,j}, F^{p}_{i,j},l^{p}_{i,j}]$, $\forall i,j=1...M$.}

Lastly, for the discretisation of \dt{\eqref{globalMacroSystem-F} and \eqref{globalMacroSystem-l}, we follow the same predictor-corrector method as the one used in \cite{Dumitru_et_al_2013}, where we the corrector part uses simply a second-order trapezoidal scheme on $[t_{p}, t_{p+1}]$.}

\section{Computational Simulations and Results}
To illustrate our model, we consider the region $Y:=[0,4] \times [0,4]$ \dt{and we start our dynamics by adopting here the same initial condition for $c$} as in \cite{Dumitru_et_al_2013}, namely
\bequ
c(x,0)=0.5\left(\text{exp}\left(-\frac{||x-(2,2)||^2_2}{0.03}\right)-\text{exp}(-28.125)\right)\left(\chi_{_{\Bila((2,2),0.5-\gamma)}} \ast \psi_{\gamma}\right),
\label{eq:canceric}
\eequ
where $\psi_{\gamma}$ is \dt{the} standard mollifier \dt{detailed in Appendix \ref{mollifierUsed}} that acts within a radius $\gamma <<\frac{\Delta x}{3}$ from $\partial \Bila((2,2),0.5-\gamma)$ to smooth out the characteristic function $\chi_{_{\Bila((2,2),0.5-\gamma)}}$. Thus, initially, the cancer cell population occupies the region $\Omega(0)=\Bila((2,2),0.5)$ positioned in the centre of $Y$. 

\begin{figure}[ht!]
    \centering
    \begin{subfigure}{0.5\linewidth}
        \includegraphics[width=\linewidth]{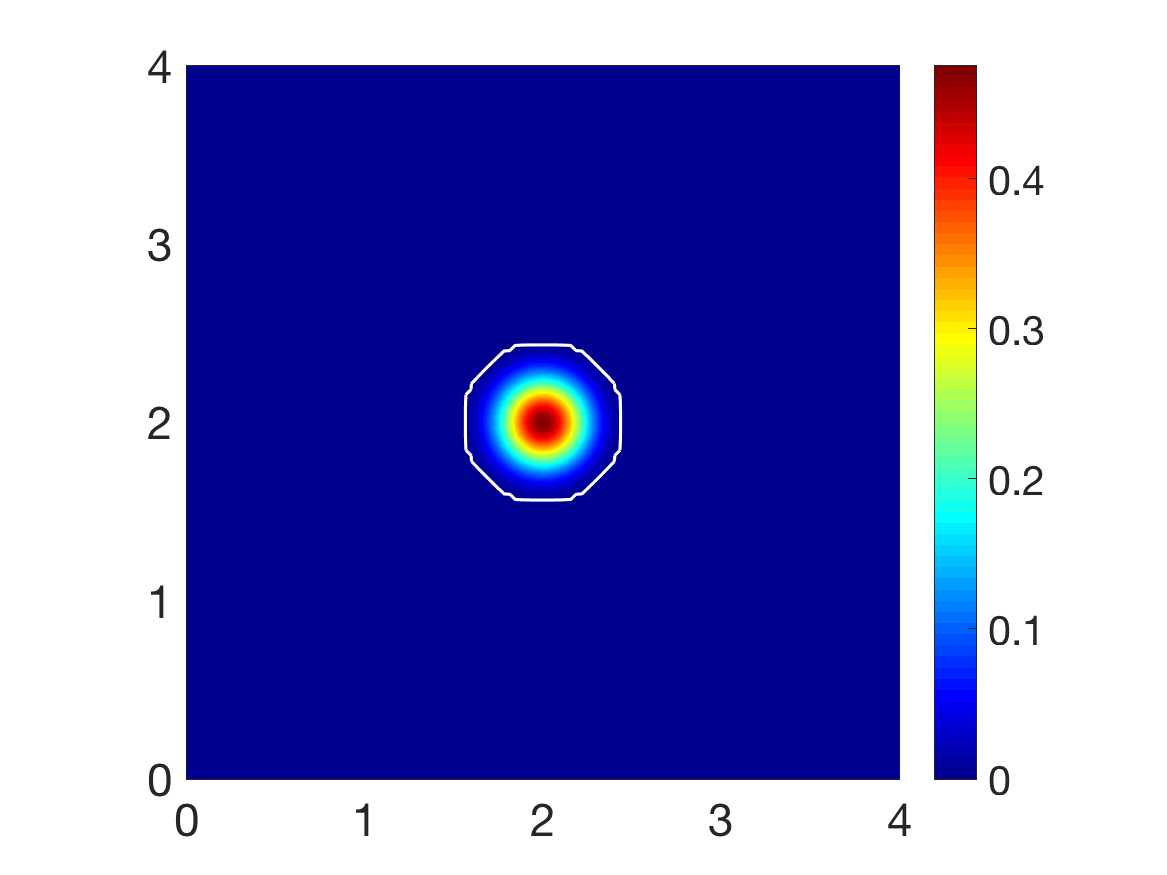}
        \caption{\emph{initial cancer distribution}}
\end{subfigure}\hfil
    \begin{subfigure}{0.5\linewidth}
        \includegraphics[width=\linewidth]{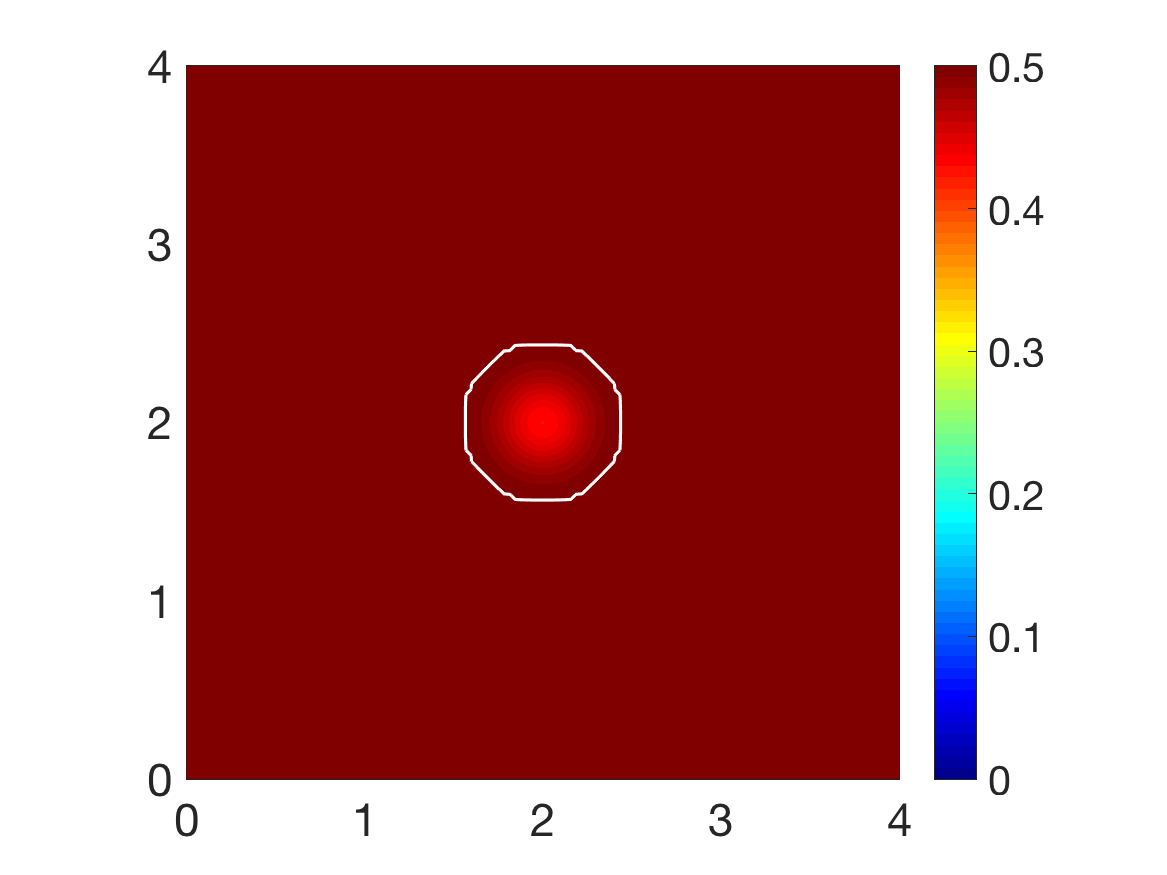}
        \caption{\emph{initial ECM density: $F(\cdot,0)+l(\cdot,0)$}}
    \end{subfigure}\hfil
    
    \medskip
      \begin{subfigure}{0.5\linewidth}
        \includegraphics[width=\linewidth]{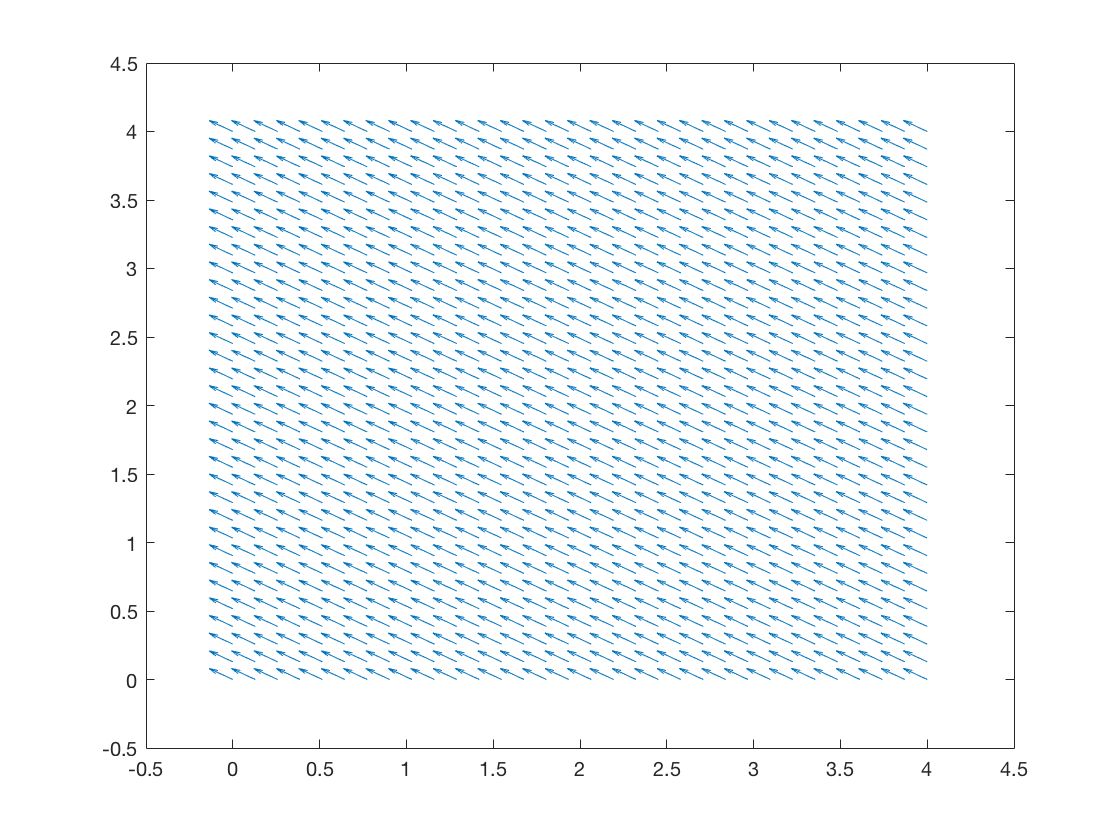}
        \caption{\emph{initial macro-scale fibres vector field}}
\end{subfigure}
    \caption{\emph{Initial conditions showing the distribution of cancer cells (a), the homogeneous density of ECM (b) with the invasive boundary of the tumour represented by the white contour, and the initial macroscopic fibre orientations per each micro-domain represented by a vector field (c). These vectors have been magnified from the usual size of the domain for better representation.}}
    \label{fig:homo}
\end{figure}

\paragraph{Initial condition for the ECM fibre component.}
For the initial distribution of the ECM \dt{fibre phase,} we \dt{consider first a generic micro-domain centred a $0$ of cell-scale size $\delta=h$, namely $\delta Y$,} and using the microscopic patterns of fibres defined in \eqref{eq:fib} and illustrated in Figure \ref{fig:fibre}, we replicate \dt{and centre this micro-fibre distribution in the cell-scale neighbourhood of any spatial location $(x_{i},x_{j})$ in the discretisation of  $Y$ on the corresponding micro-domain $\delta Y(x_{i},x_{j}):=\delta Y+(x_{i},x_{j})$. The maximal height of the micro-fibres considered here is appropriately calibrated uniformly across all micro-domains is so that the resulting macroscopic distribution of fibres $F(x,\cdot)$ represents a percentage $p=0.2$, of the mean density of the non-fibrous ECM phase. Therefore at initial time $t_0=0$, all fibre micro-domains $\delta Y(x_{i},x_{j})$ support identical distributions of micro-fibres, $\forall i, j= 1... M$} and as a consequence, every fibre orientation $\theta_{_{f}}((x_{i},x_{j}),0)$ exhibits the same initial orientation and magnitude, as shown in Figure \ref{fig:homo}(c). 

\dt{Finally, f}or the non-fibre ECM component, we consider both a homogeneous and a heterogeneous scenario, which will be detailed below.

\subsection{Homogeneous non-fibre ECM component}
The \dt{initial distribution of the} non-fibre ECM component, $l(x,0)$, will be in the first instance taken as the homogenous distribution, \dt{namely as} $l(x,0)=\text{min}\{0.5,1-c(x,0)\}$. The initial conditions of the cell population $c(x,0)$ given in \eqref{eq:canceric}, the full ECM density $v(x,0)=l(x,0)+F(x,0)$, and the resulting initial fibre orientations $\theta_{_{f}}(x,0)$ can be seen in Figure \ref{fig:homo}. The adhesive strength coefficients for cell-cell adhesion, cell-fibre adhesion and cell-non-fibre ECM adhesion, are taken here to be
\vspace{-0.1cm}
\bequ\label{norm_matrices}
\begin{array}{lll}
\textbf{S}_{max}=0.5,& \quad \textbf{S}_{cF}= 0.1& \quad \text{and} \quad \textbf{S}_{cl}=0.01, \\[-0.1cm]
\end{array}
\eequ
respectively. 

\begin{figure}[h]
    \centering 
\begin{subfigure}{0.5\textwidth}
  \includegraphics[width=\linewidth]{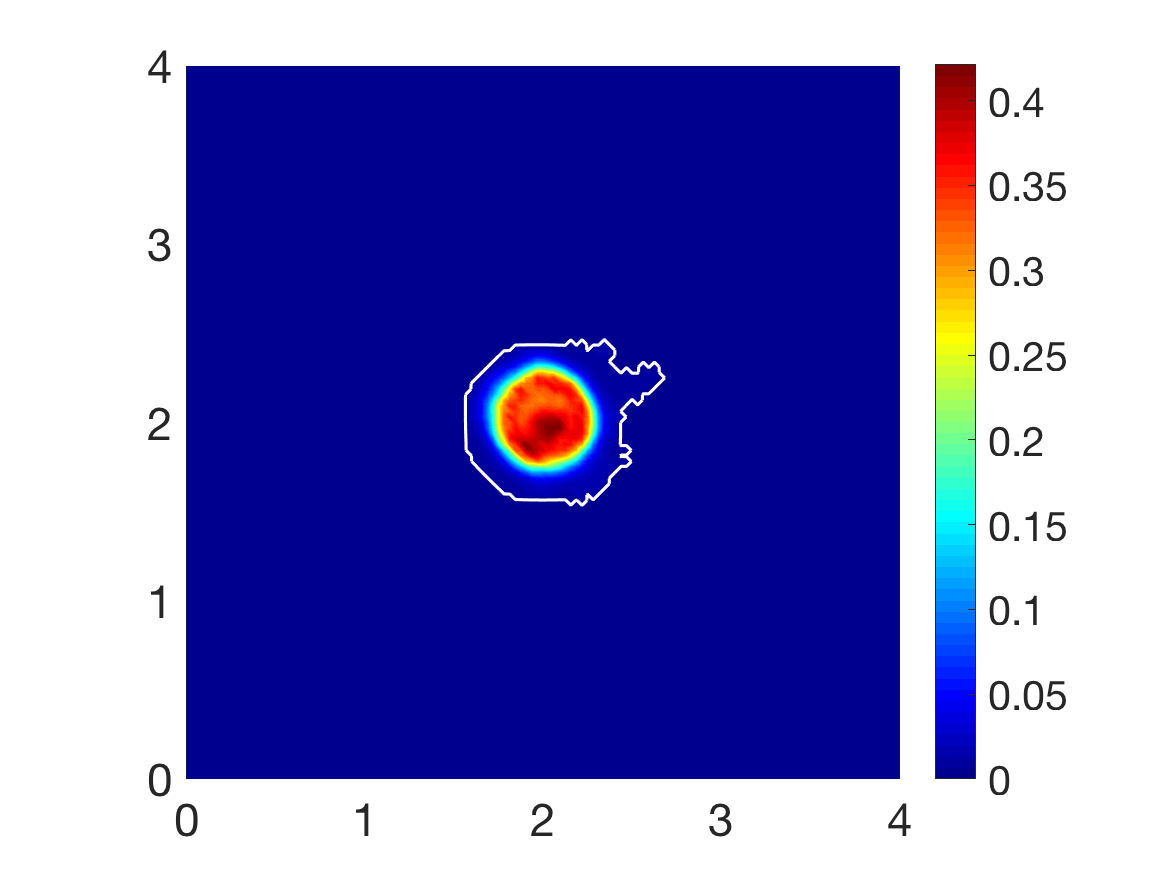}
  \caption{\emph{Cancer cell population}}
  \label{fig:1}
\end{subfigure}\hfil 
\begin{subfigure}{0.5\textwidth}
  \includegraphics[width=\linewidth]{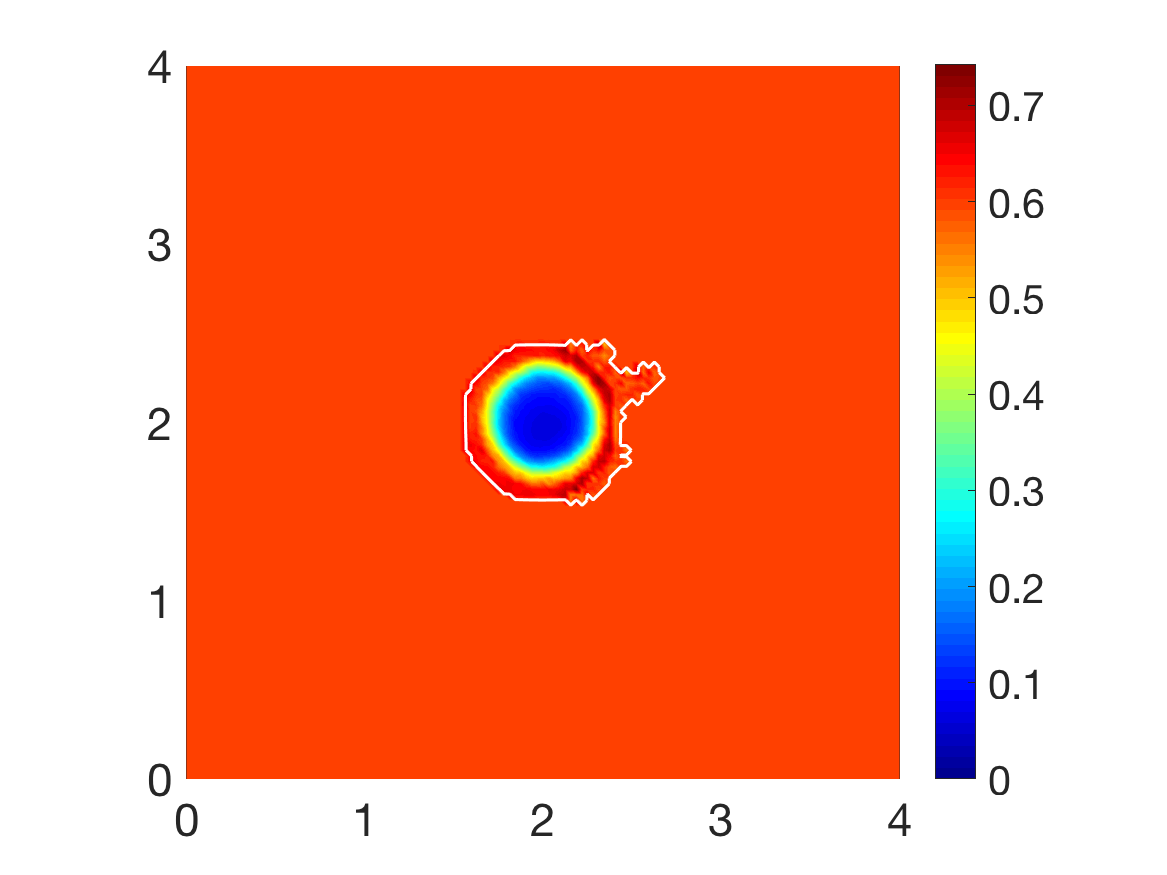}
  \caption{\emph{Matrix distribution}}
  \label{fig:2}
\end{subfigure}\hfil 

\medskip
\begin{subfigure}{0.5\textwidth}
  \includegraphics[width=\linewidth]{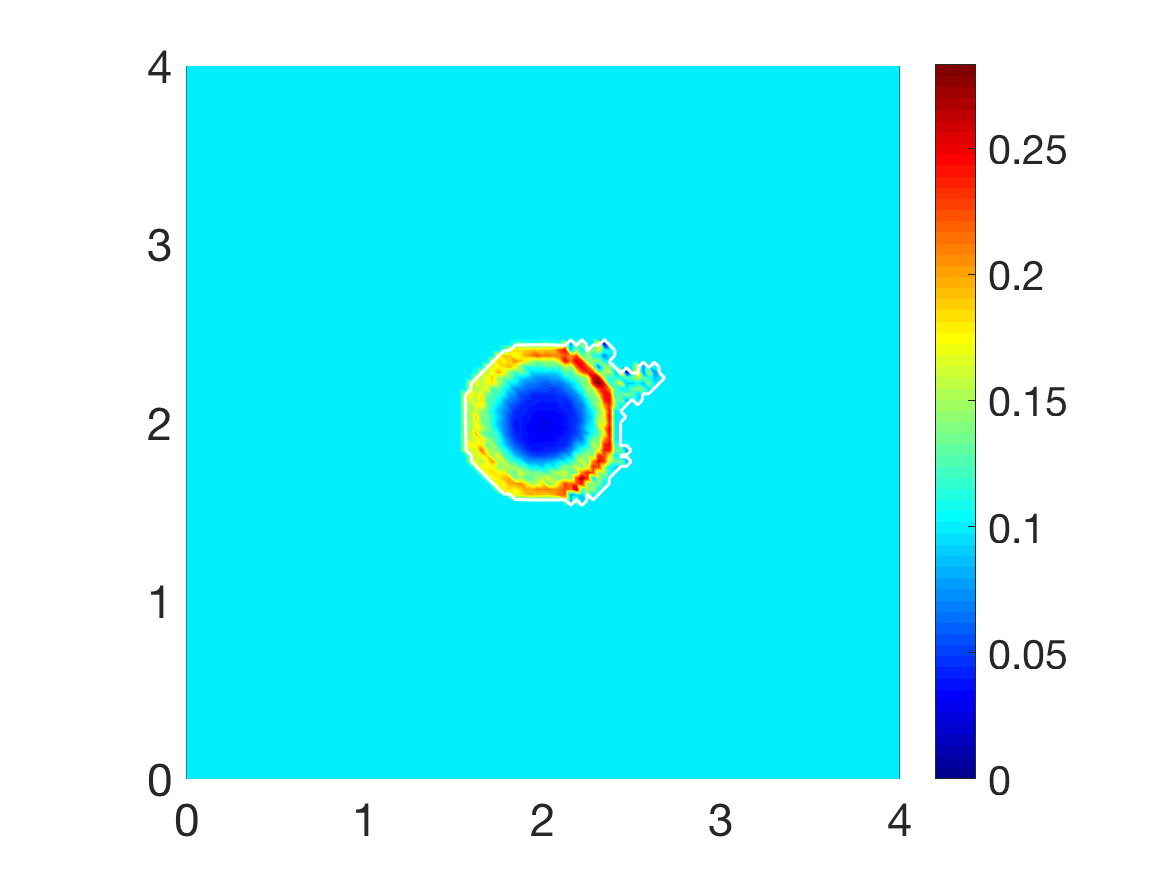}
  \caption{\emph{Macroscopic fibre density}}
  \label{fig:3}
  \end{subfigure}\hfil 
\begin{subfigure}{0.5\textwidth}
  \includegraphics[width=\linewidth]{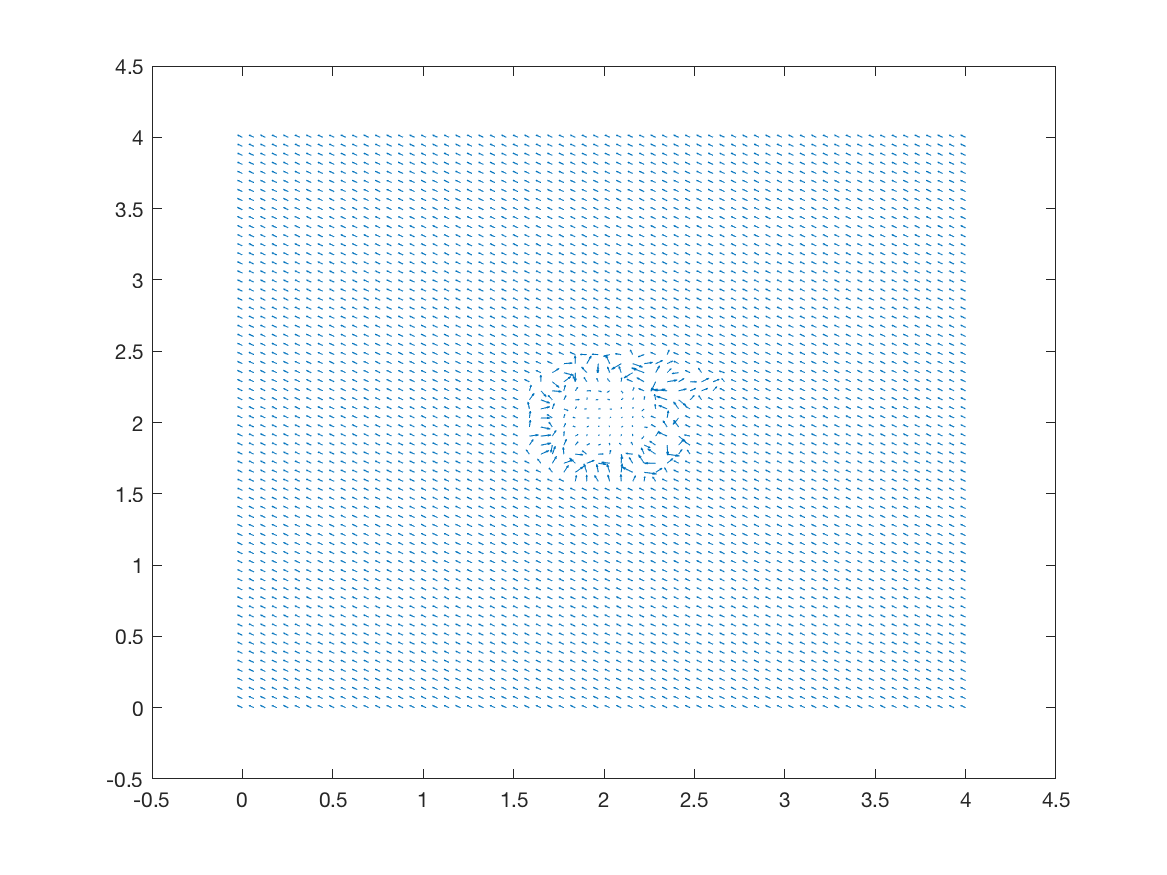}
  \caption{\emph{Fibre orientation - coarsened 2 fold}}
  \label{fig:4}
\end{subfigure}\hfil 

\medskip
\begin{subfigure}{0.5\textwidth}
  \includegraphics[width=\linewidth]{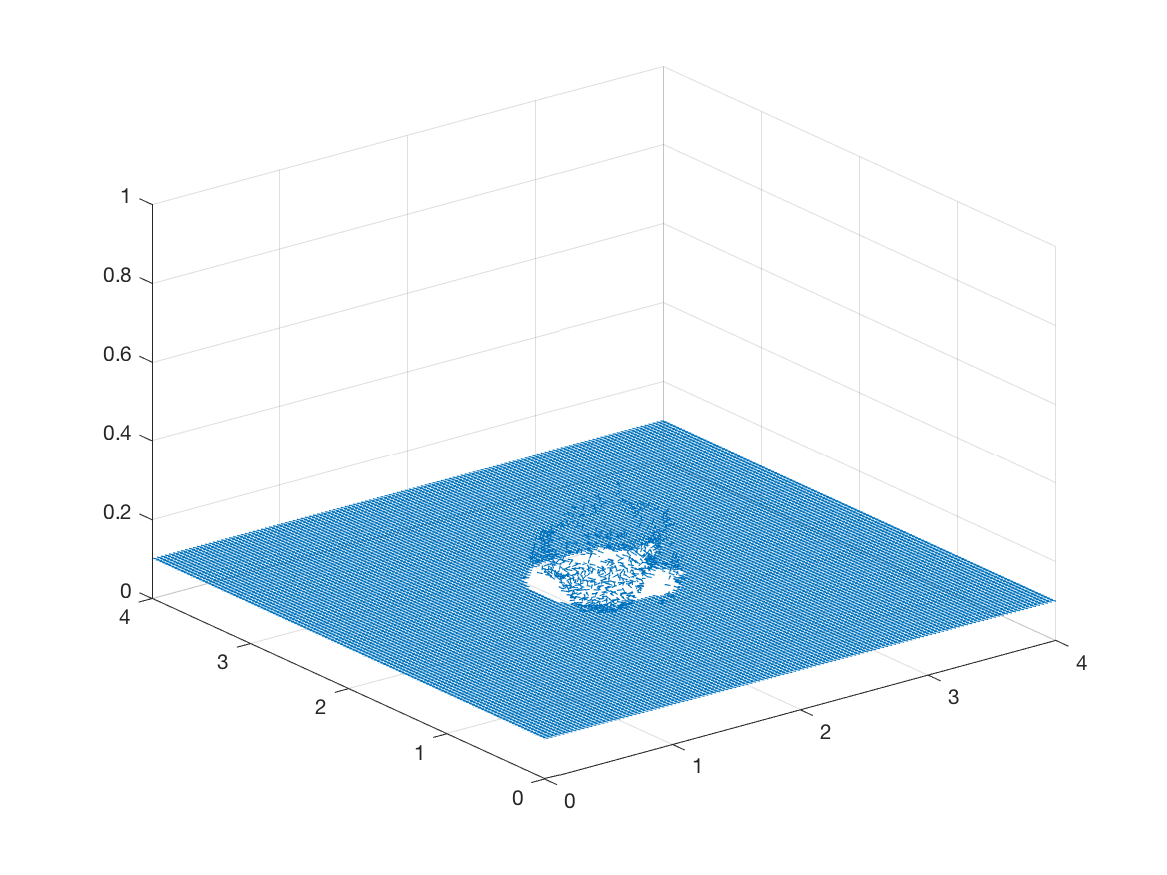}
  \caption{\emph{3D fibre vector field}}
  \label{fig:3}
  \end{subfigure}\hfil 
\begin{subfigure}{0.5\textwidth}
  \includegraphics[width=\linewidth]{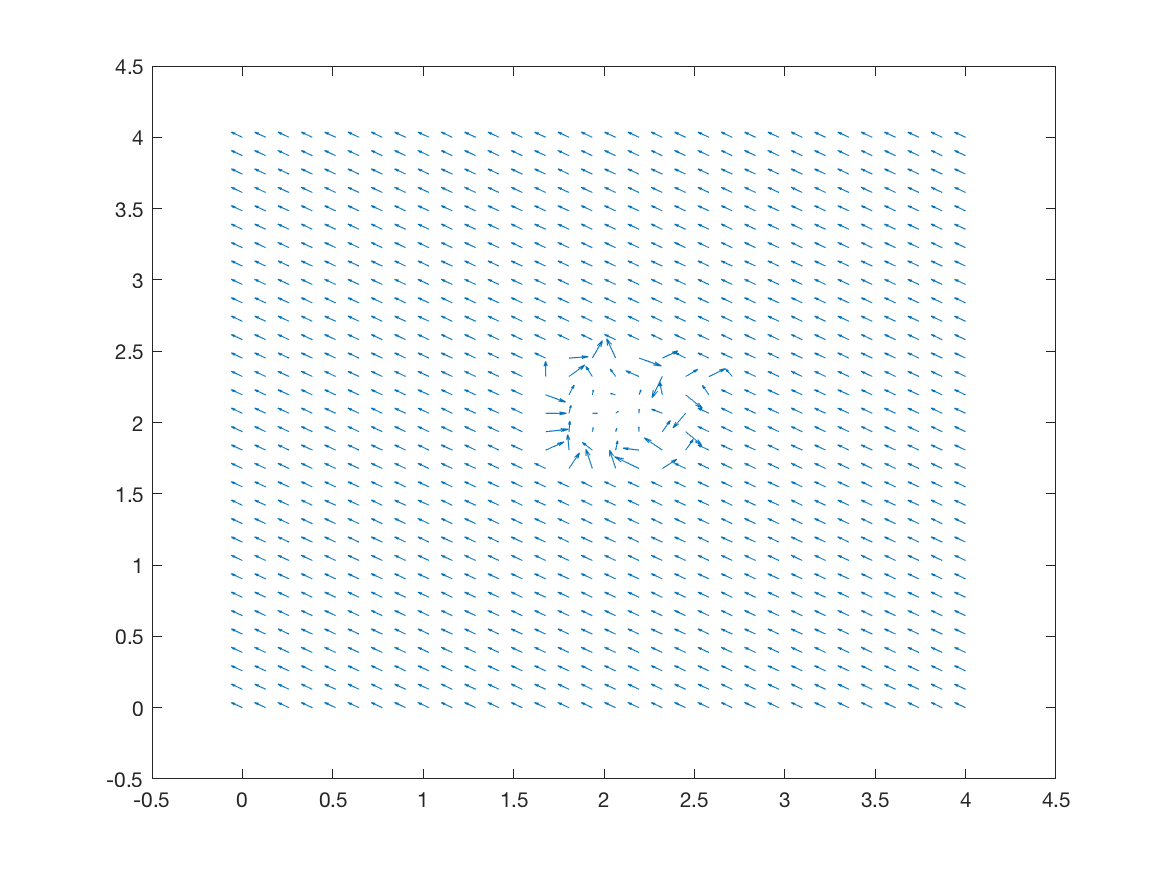}
  \caption{\emph{Fibre orientation - coarsened 4 fold}}
  \label{fig:4}
\end{subfigure}\hfil 

\caption{Simulations at stage $20\Delta t$ with a homogenous distribution of the non-fibre ECM component.}
\label{fig:homo20}
\end{figure}

Using the parameter set $\Sigma$ from Appendix \ref{paramSection} and adhesion coefficients (\ref{norm_matrices}), in Figure \ref{fig:homo20} we show the computational results at macro-micro stage $20 \Delta t$ for the evolution of: the cancer cell population in subfigure \ref{fig:homo20}(a); the full ECM density in subfigure \ref{fig:homo20}(b); the macro-scale fibre magnitude in subfigure \ref{fig:homo20}(c); the vector field \dt{of oriented fibres} at two different resolutions, namely, coarsened twice and coarsened fourfold in subfigures \ref{fig:homo20}(d) and \ref{fig:homo20}(f), respectively; and a 3D plot of the macroscopic oriented fibres in \ref{fig:homo20}(e). 

\begin{figure}[h]
    \centering 
\begin{subfigure}{0.5\textwidth}
  \includegraphics[width=\linewidth]{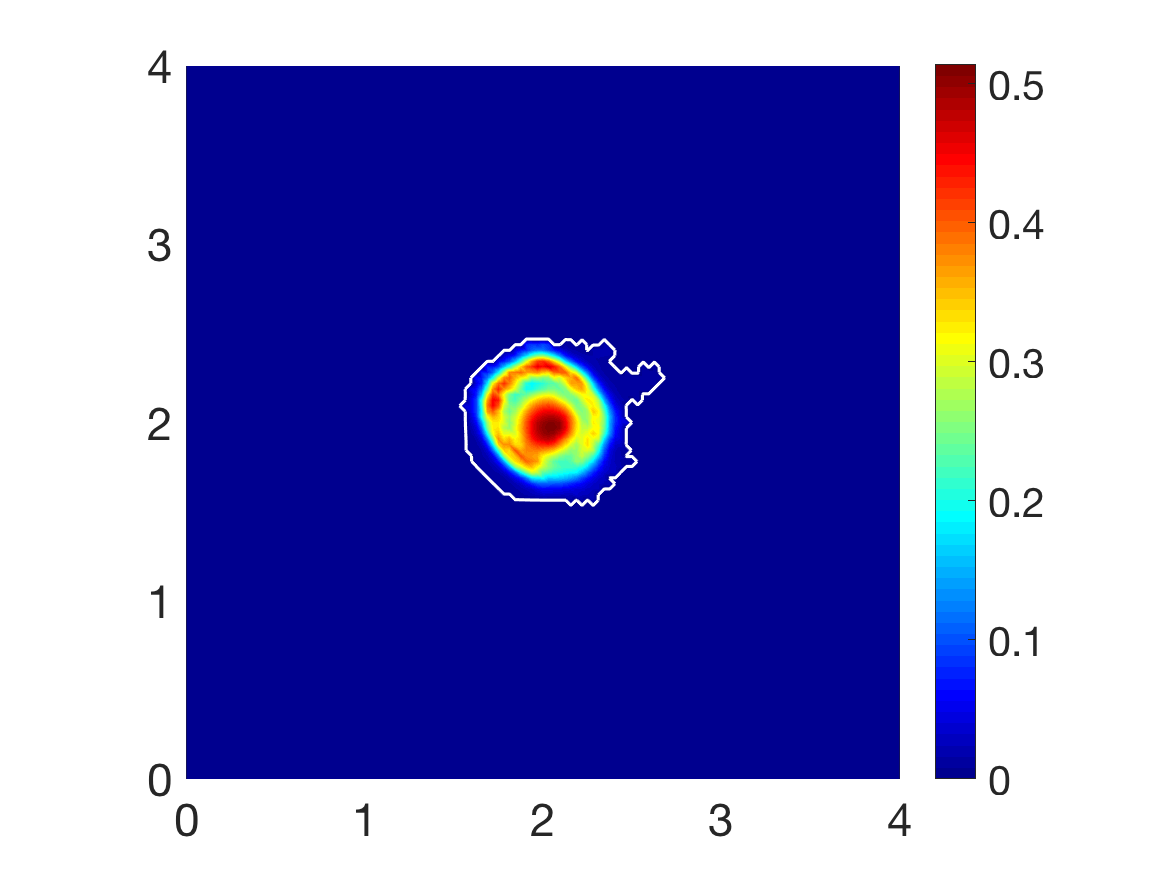}
  \caption{\emph{Cancer cell population}}
  \label{fig:1}
\end{subfigure}\hfil 
\begin{subfigure}{0.5\textwidth}
  \includegraphics[width=\linewidth]{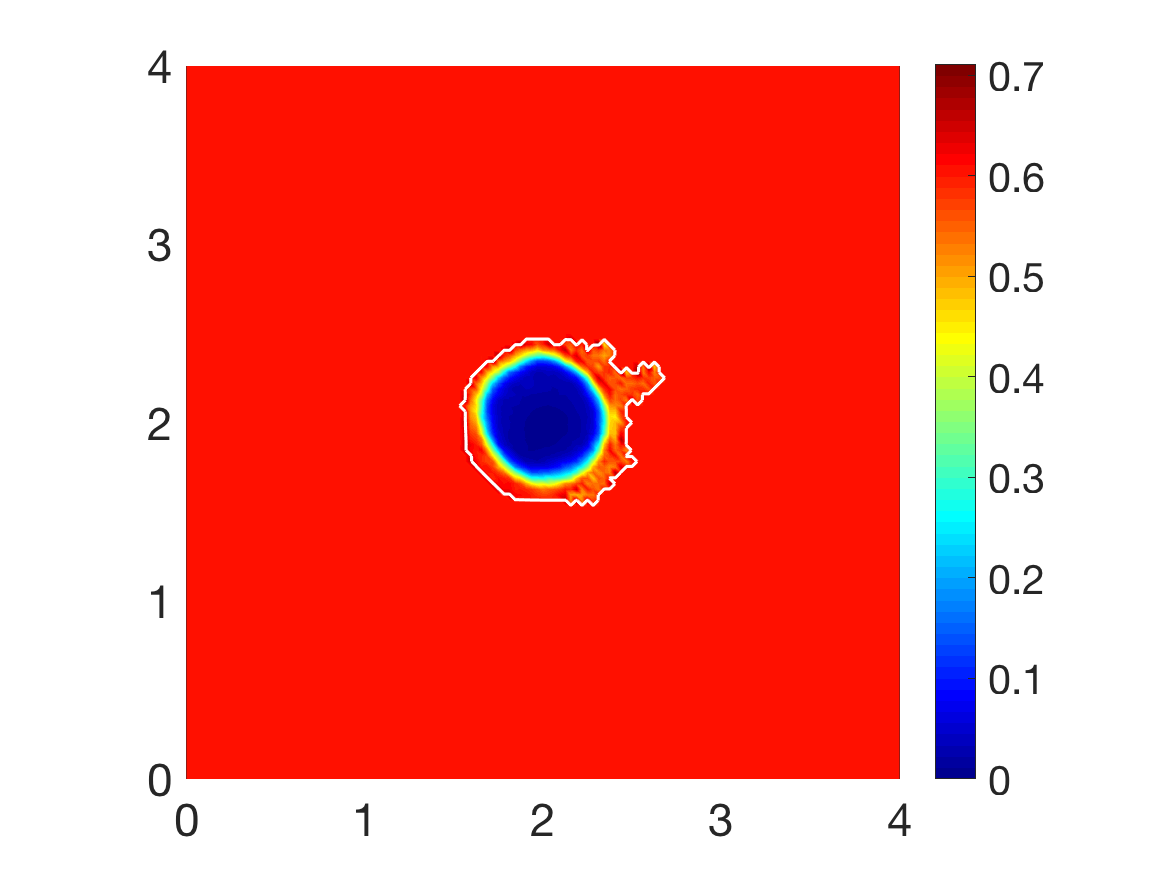}
  \caption{\emph{Matrix distribution}}
  \label{fig:2}
\end{subfigure}\hfil 

\medskip
\begin{subfigure}{0.5\textwidth}
  \includegraphics[width=\linewidth]{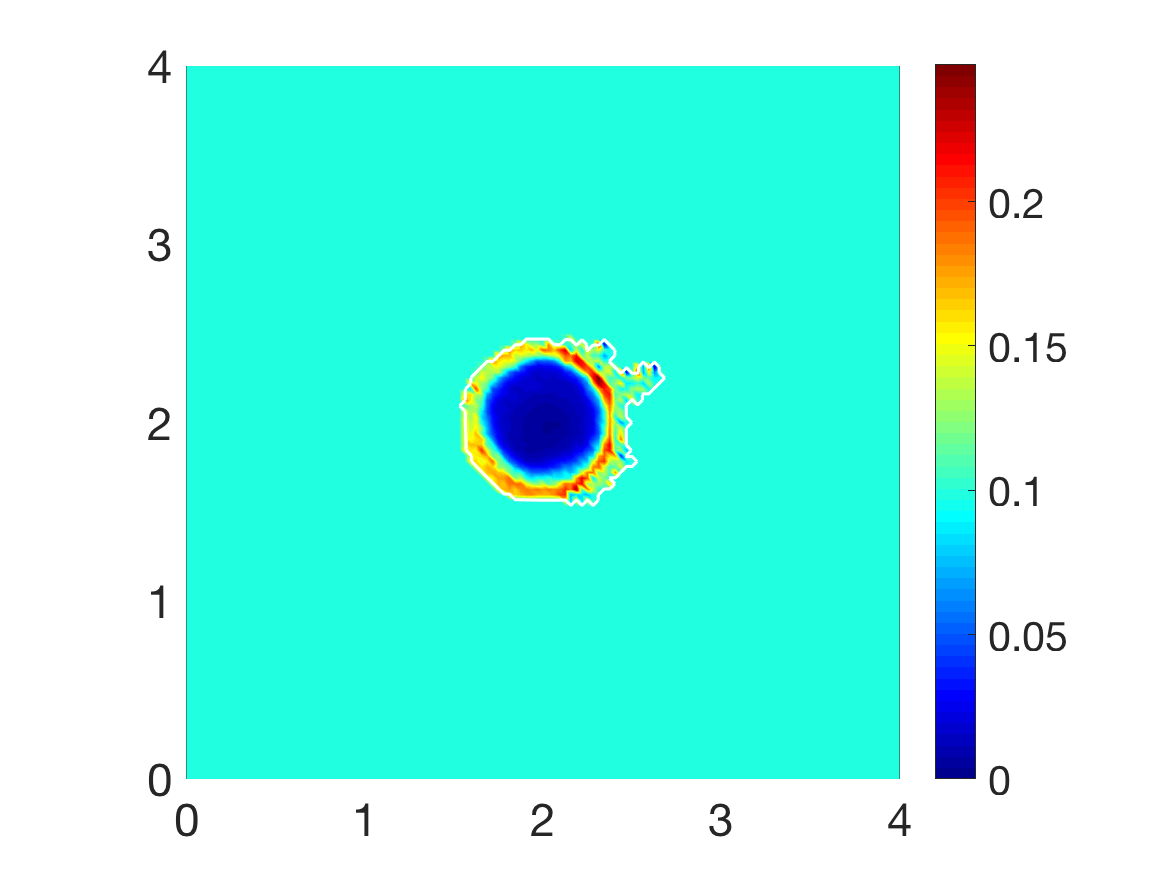}
  \caption{\emph{Macroscopic fibre density}}
  \label{fig:3}
  \end{subfigure}\hfil 
\begin{subfigure}{0.5\textwidth}
  \includegraphics[width=\linewidth]{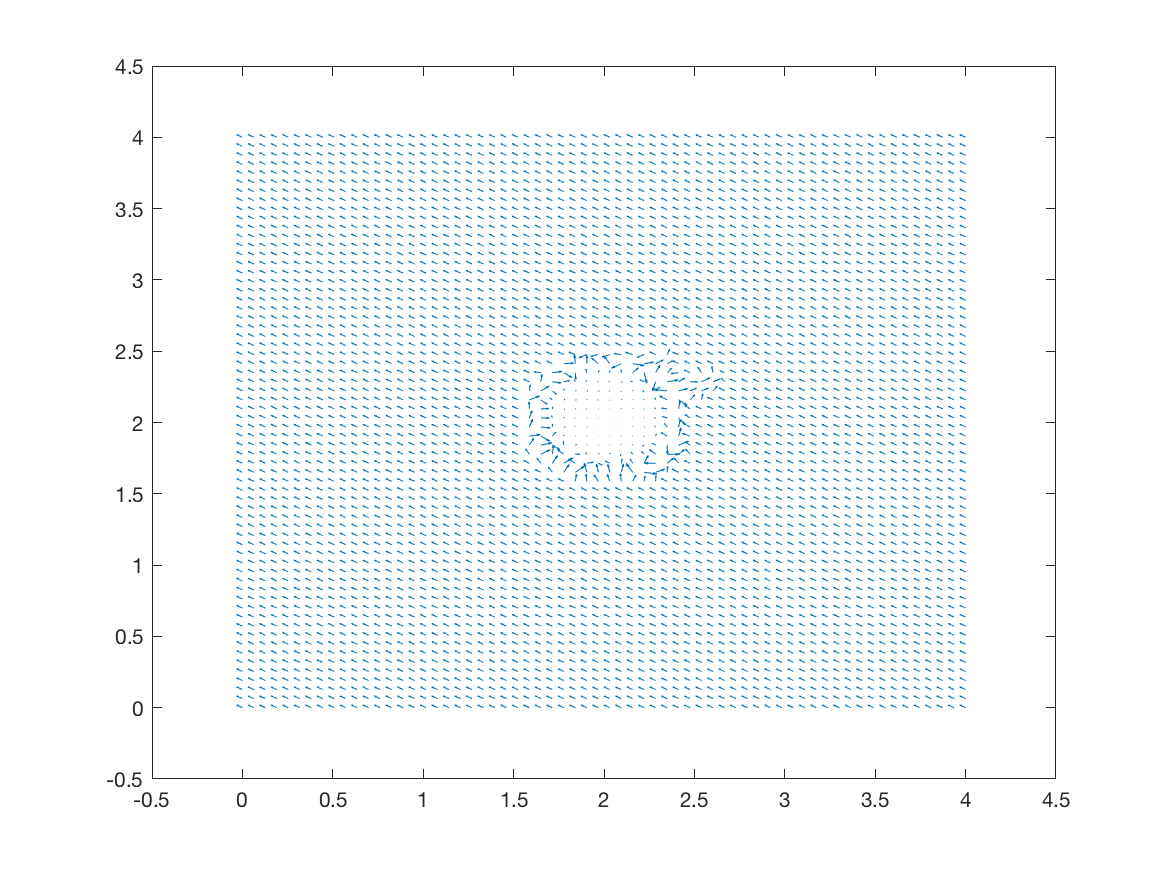}
  \caption{\emph{Fibre orientation - coarsened 2 fold}}
  \label{fig:4}
\end{subfigure}\hfil 

\medskip
\begin{subfigure}{0.5\textwidth}
  \includegraphics[width=\linewidth]{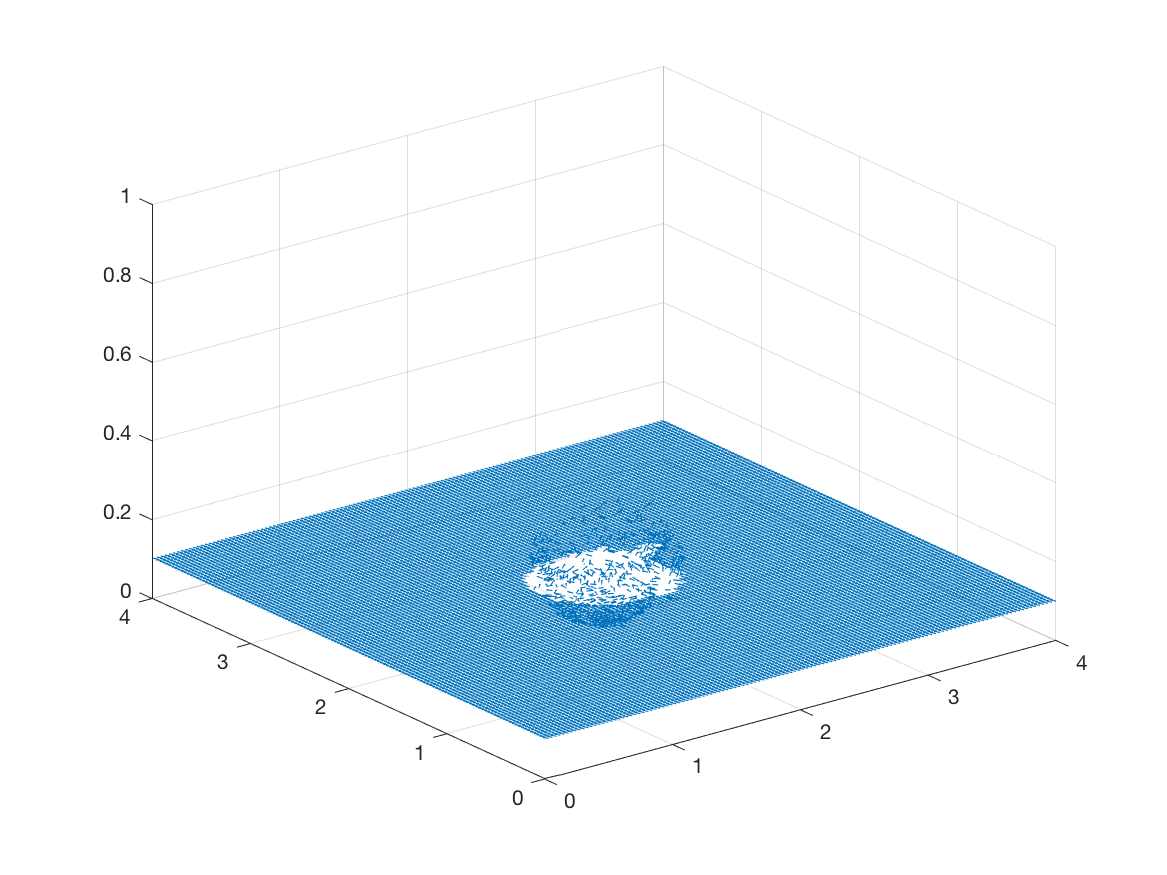}
  \caption{\emph{3D fibre vector field}}
  \label{fig:3}
  \end{subfigure}\hfil 
\begin{subfigure}{0.5\textwidth}
  \includegraphics[width=\linewidth]{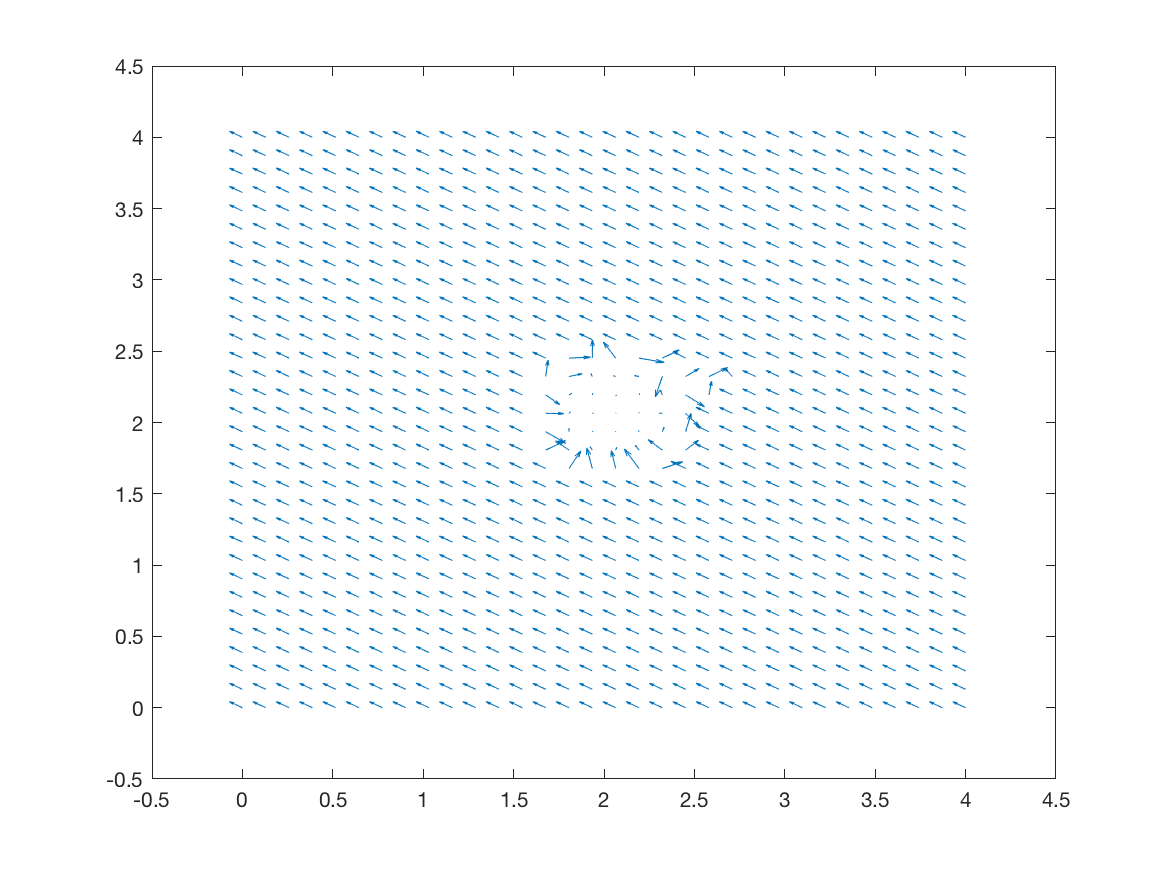}
  \caption{\emph{Fibre orientation - coarsened 4 fold}}
  \label{fig:4}
\end{subfigure}\hfil 

\caption{Simulations at stage $40\Delta t$ with a homogenous distribution of the non-fibrous part of the matrix}
\label{fig:homo40}
\end{figure}

Comparing with the initial distributions of cancer cells and ECM displayed in Figure \ref{fig:homo}, the main body of the tumour is increasing in size, whilst decreasing in overall density, spreading the initial distribution outwards \dt{and} creating a plateau of cancer cells, as shown in subfigure \ref{fig:homo20}(a). \dt{While in the absence of fibres the boundary of the tumour was expanding isotropically in the case of homogenous ECM, as showed in \cite{Dumitru_et_al_2013,shutt_chapter}, a different situation we witness here in the case of homogeneous non-fibre ECM as the presence of the oriented fibres phase of ECM is now taken into consideration}. \dt{Specifically, t}he cancer cell invasion becomes anisotropic, leading to lobular patterns and having the fibres reaching outwards in the boundary regions of faster tumour progression. This behaviour is clarified by the fibre vector plot \ref{fig:homo20}(d) where the orientations of the redistributed fibres can be seen to point in the direction of this lobule on the invasive edge. The orientation of the fibres is strongly affected during their rearrangement, with their behaviour dependent on the initial macroscopic density of fibres and the spatial flux of the cancer cells. This flux carries a higher weight than the distribution of fibres and thus the cells ultimately have governance over the direction of realignment. \dt{Finally, Alongside the fibre realignment, the cancer cells also degrade the fibres, this leading to} a low density central region of fibres \ref{fig:homo20}(c). 

\dt{As} the simulation continue to stage $40 \Delta t$, the \dt{initial} main body of the tumour \dt{(consisting of a high density region of cells in the centre of $Y$)} is spreading out, following the initial orientation of the fibres, \dt{giving rise to lobular progression pattern for the cell population} in this direction, \dt{as shown in} Figure \ref{fig:homo40}(a). The boundary of the tumour has undergone minor changes \dt{with respect to stage $20 \Delta t$ shown in Figure \ref{fig:homo20} }, the main tumour dynamics occurring mainly  on the central cluster of cells. The non-fibrous part of the ECM is further degraded under the presence of cancer cells \ref{fig:homo40}(b), and the fibres are being pushed to the boundary of the tumour \ref{fig:homo40}(c), creating a larger \dt{region of} low density \dt{ECM}.

\begin{figure}[ht!]
    \centering
    \begin{subfigure}{0.5\linewidth}
        \includegraphics[width=\linewidth]{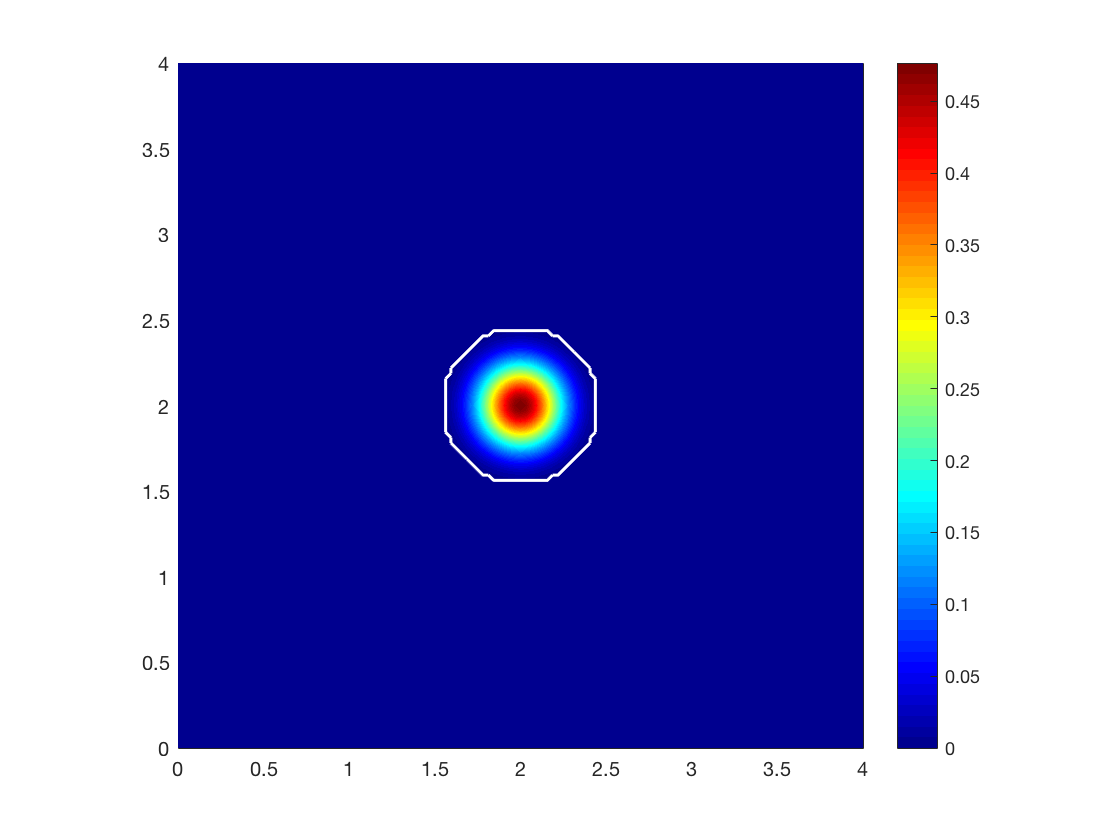}
        \caption{\emph{initial cancer distribution}}
\end{subfigure}\hfil
    \begin{subfigure}{0.5\linewidth}
        \includegraphics[width=\linewidth]{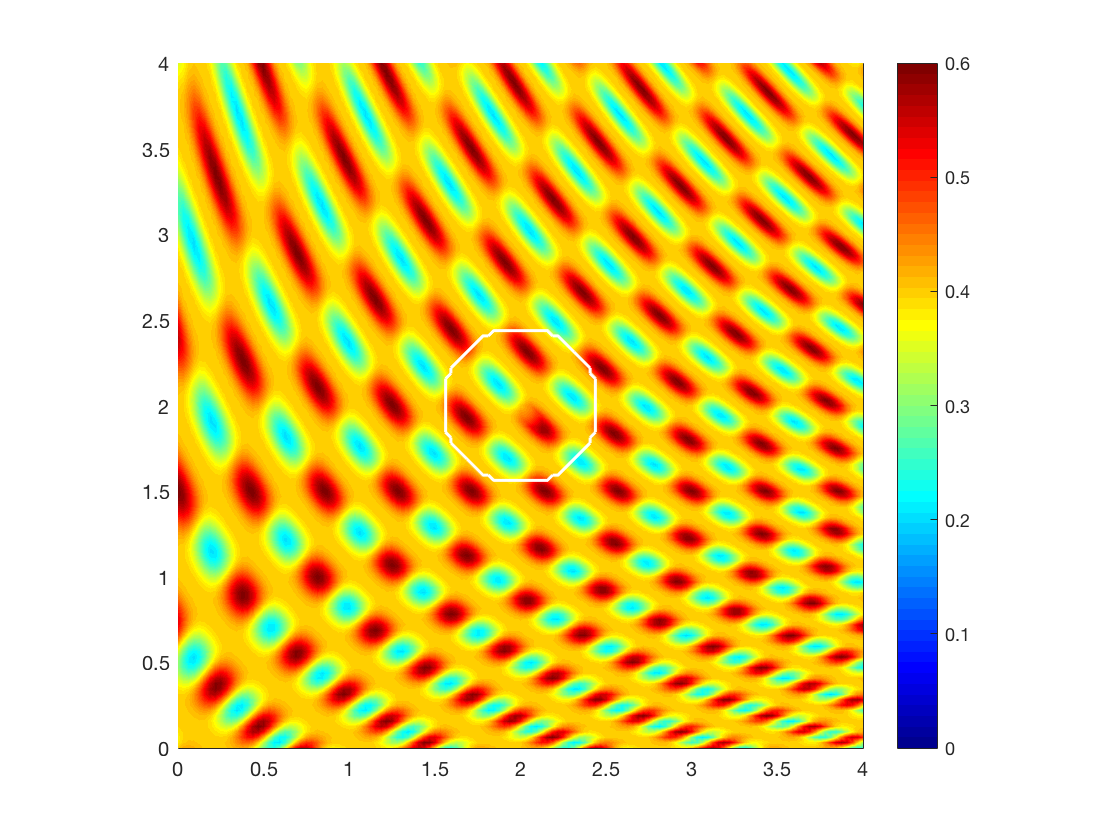}
        \caption{\emph{Initial ECM density}}
    \end{subfigure}\hfil
    
    \medskip
      \begin{subfigure}{0.5\linewidth}
        \includegraphics[width=\linewidth]{ic_fibre_small.png}
        \caption{\emph{Initial vector field of fibre directions}}
\end{subfigure}
    \caption{\emph{Initial conditions showing the distribution of cancer cells (a), the heterogeneous density of ECM (b) with the invasive boundary of the tumour represented by the white contour, and the initial macroscopic fibre orientations per each micro-domain represented by a vector field (c). These vectors have been magnified from the usual size of the domain for better representation.}}
    \label{fig:smallhetero}
\end{figure}

\subsection{Heterogeneous non-fibrous ECM component}
We now introduce an initially heterogeneous non-\dt{fibre} ECM component. Whilst maintaining \dt{the same} initial conditions for $c(x,0)$ specified in \eqref{eq:canceric} as well as for the initial distributions of ECM micro-fibres \dt{(illustrated in Figure \ref{fig:fibre})}, the heterogeneity of the non-fibre ECM phase will be structured in a similar manner to \cite{Domschke_et_al_2014,shutt_chapter} using the initial condition 
\begin{equation}
l(x,0)=\text{min}\left\{ h(x_1,x_2), 1- c(x,0)\right\},
\label{eq:matrix_IC}
\end{equation}
where 
\begin{align*}
h(x_1,x_2)&=\frac{1}{2}+\frac{1}{4}\text{sin}(\zeta x_1 x_2)^3 \cdot \text{sin}(\zeta \frac{x_2}{x_1}),  \\
(x_1,x_2)&= \frac{1}{3}(x+1.5) \ \in [0,1]^2 \ \text{for} \ x \in D, \quad \zeta = 7\pi.
\end{align*}
These initial conditions can be seen in Figure \ref{fig:smallhetero}. 

\begin{figure}[h]
    \centering 
\begin{subfigure}{0.5\textwidth}
  \includegraphics[width=\linewidth]{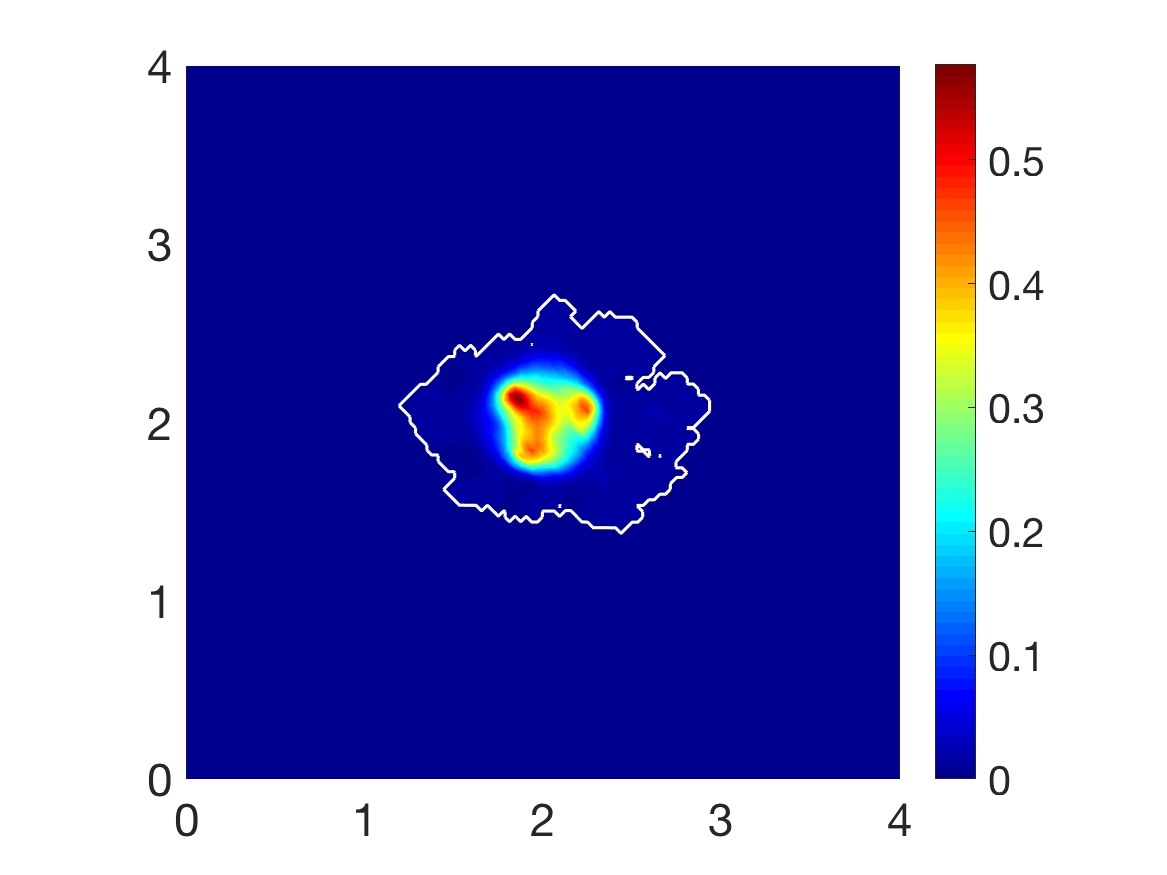}
  \caption{\emph{Cancer cell population}}
  \label{fig:1}
\end{subfigure}\hfil 
\begin{subfigure}{0.5\textwidth}
  \includegraphics[width=\linewidth]{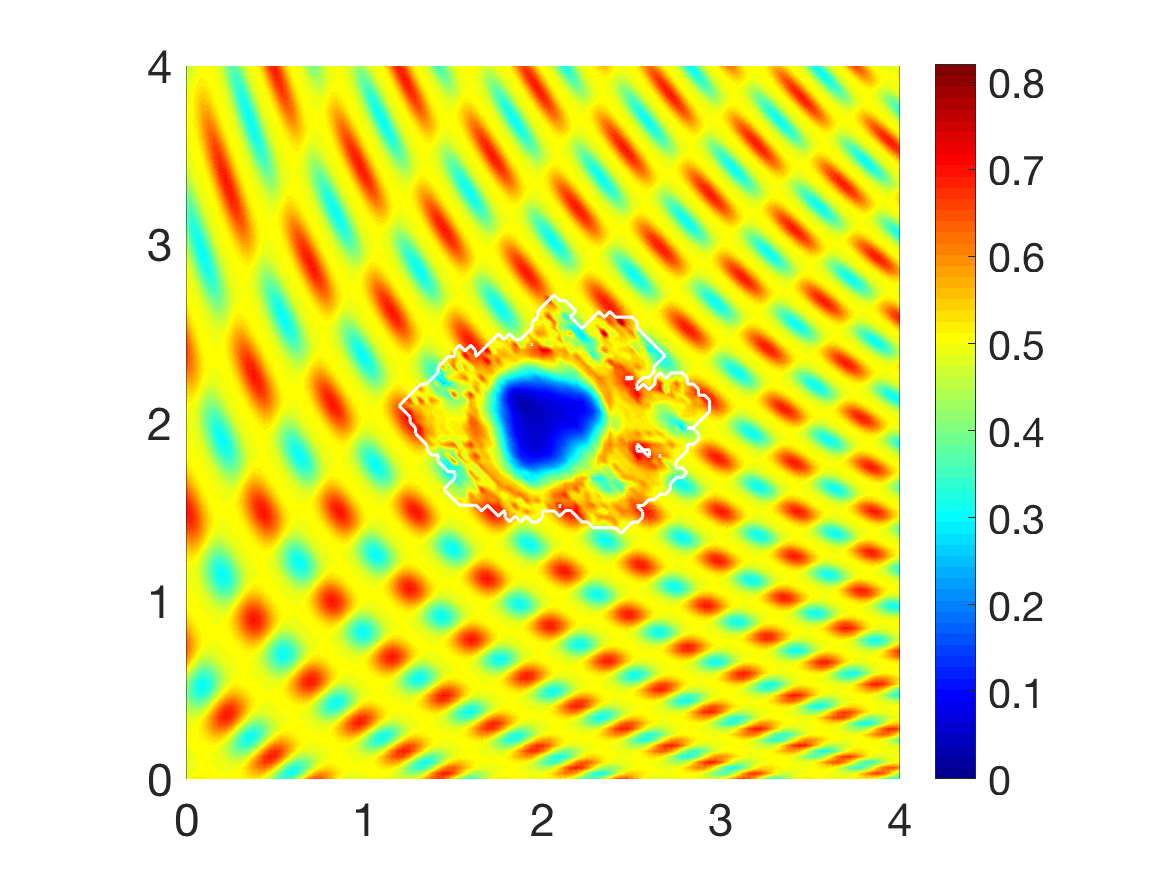}
  \caption{\emph{Matrix distribution}}
  \label{fig:2}
\end{subfigure}\hfil 

\medskip
\begin{subfigure}{0.5\textwidth}
  \includegraphics[width=\linewidth]{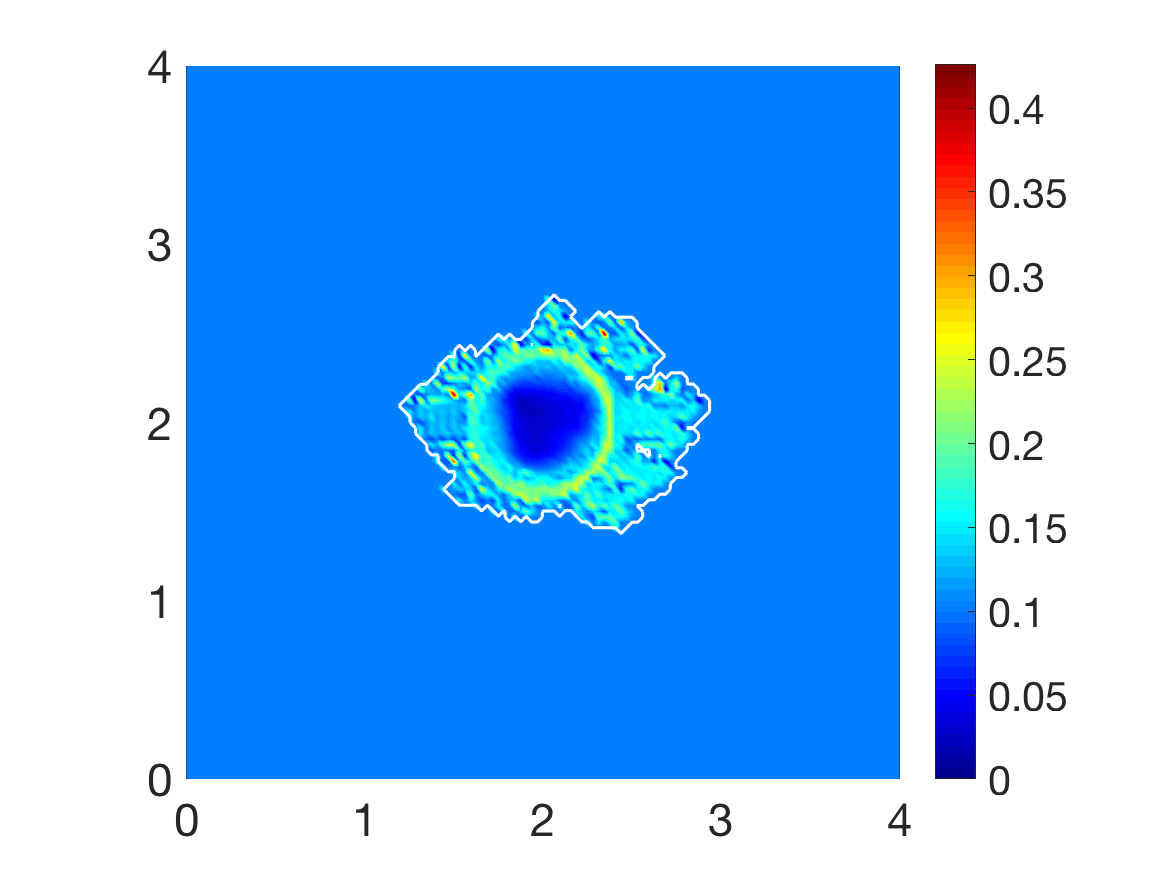}
  \caption{\emph{Fibre magnitude density}}
  \label{fig:3}
  \end{subfigure}\hfil 
\begin{subfigure}{0.5\textwidth}
  \includegraphics[width=\linewidth]{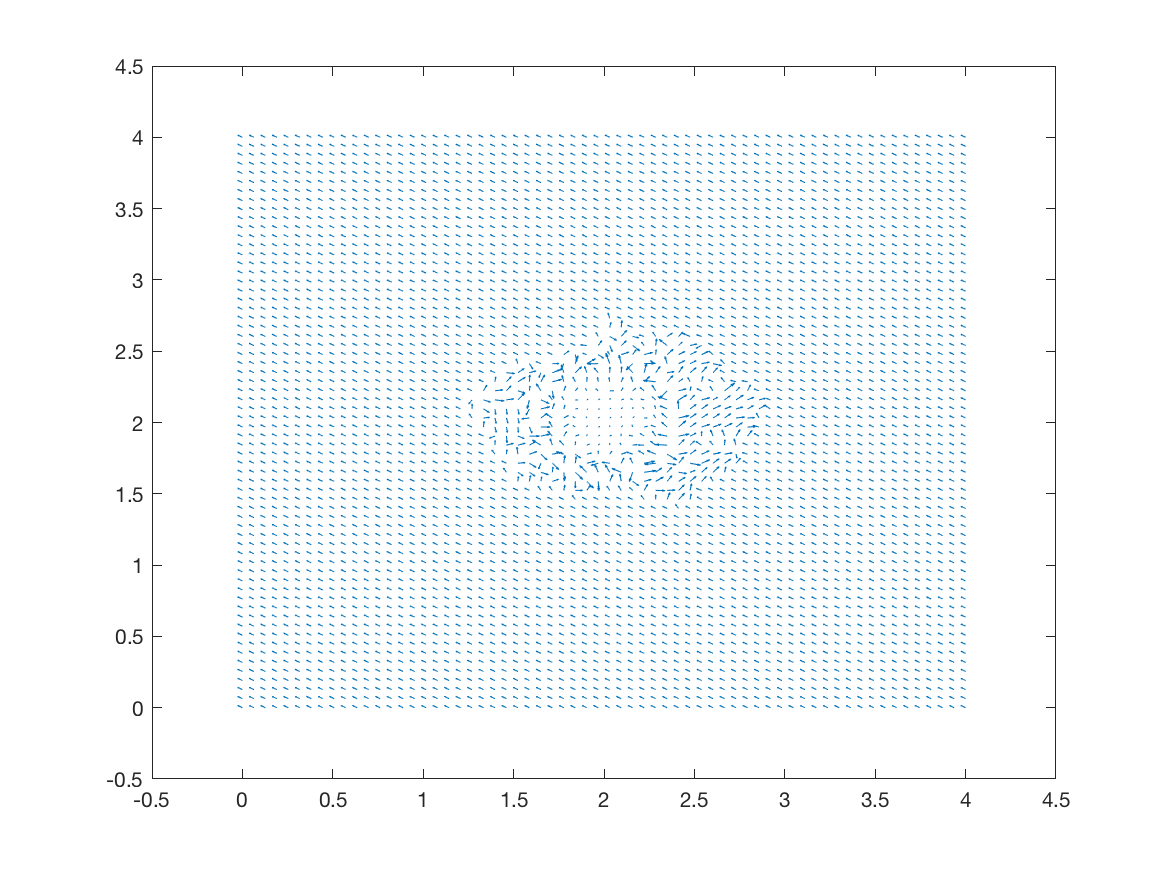}
  \caption{\emph{Fibre vector field - coarsened 2 fold}}
  \label{fig:4}
\end{subfigure}\hfil 

\medskip
\begin{subfigure}{0.5\textwidth}
  \includegraphics[width=\linewidth]{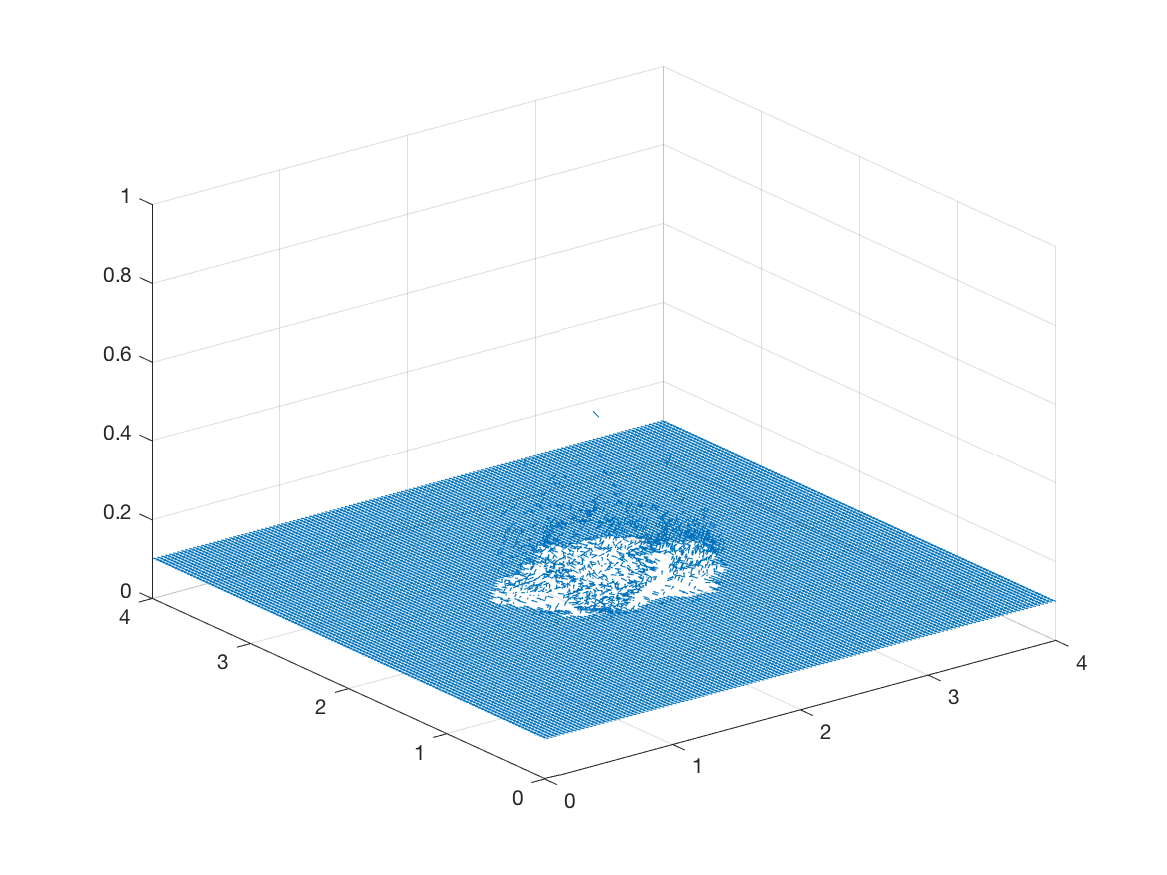}
  \caption{\emph{3D fibre vector field}}
  \label{fig:3}
  \end{subfigure}\hfil 
\begin{subfigure}{0.5\textwidth}
  \includegraphics[width=\linewidth]{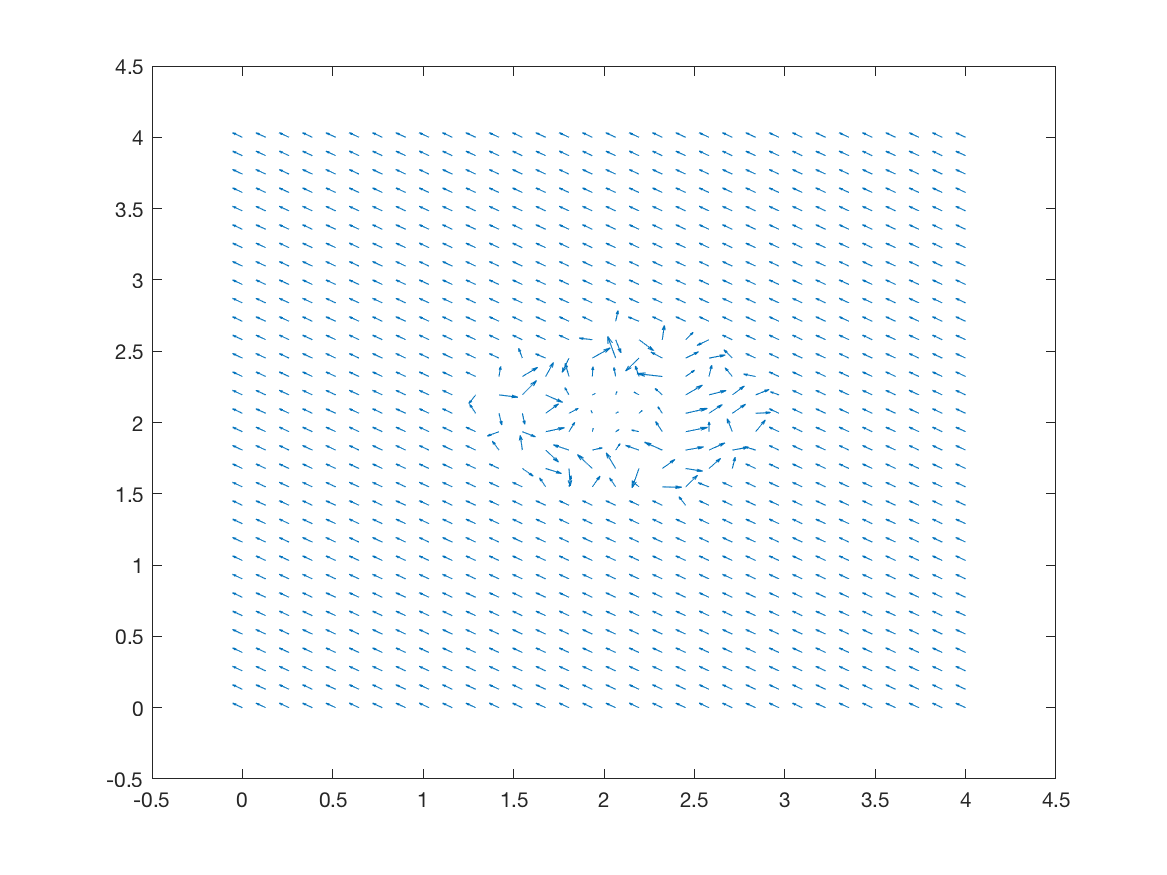}
  \caption{\emph{Fibre vector field - coarsened 4 fold}}
  \label{fig:4}
\end{subfigure}\hfil 

\caption{Simulations at stage $20\Delta t$ with a heterogeneous distribution of the non-fibrous part of the matrix.}
\label{fig:smallhetero20}
\end{figure}

\begin{figure}[h]
    \centering 
\begin{subfigure}{0.5\textwidth}
  \includegraphics[width=\linewidth]{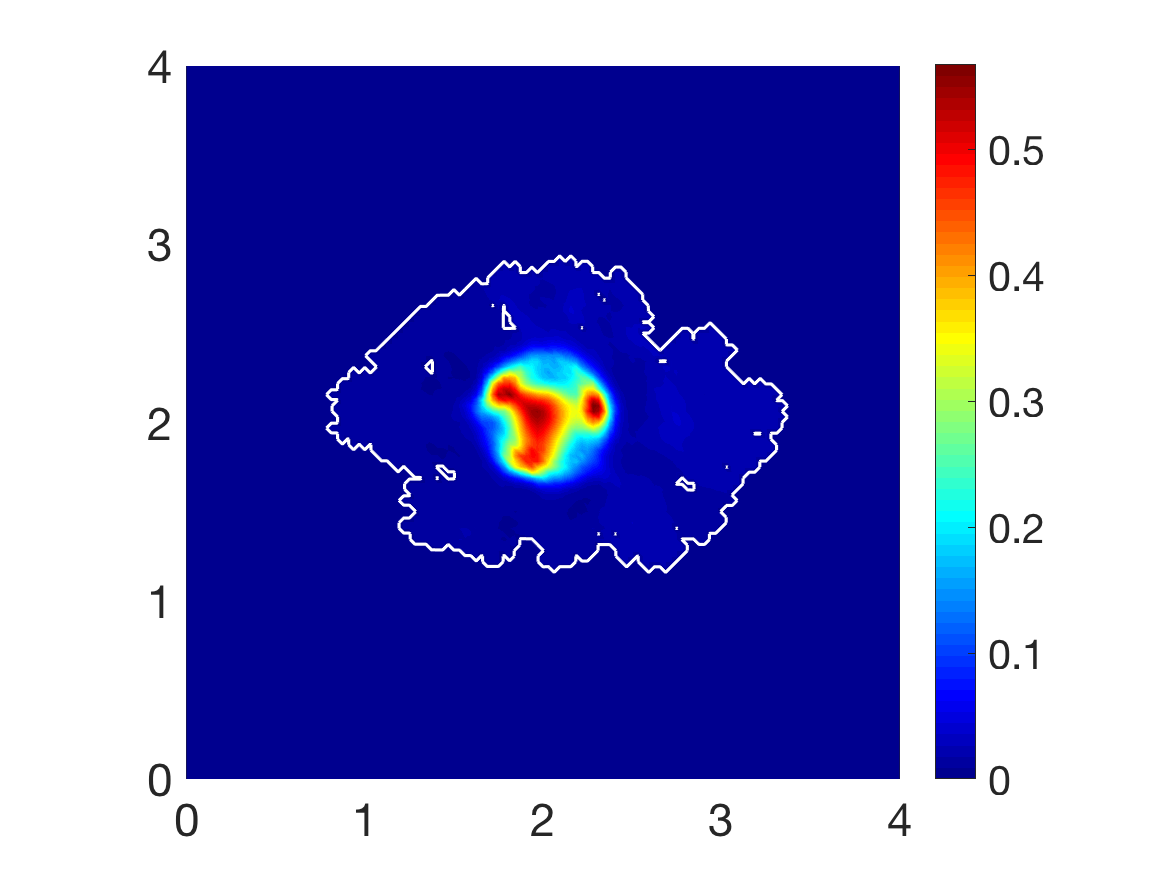}
  \caption{\emph{Cancer cell population}}
  \label{fig:1}
\end{subfigure}\hfil 
\begin{subfigure}{0.5\textwidth}
  \includegraphics[width=\linewidth]{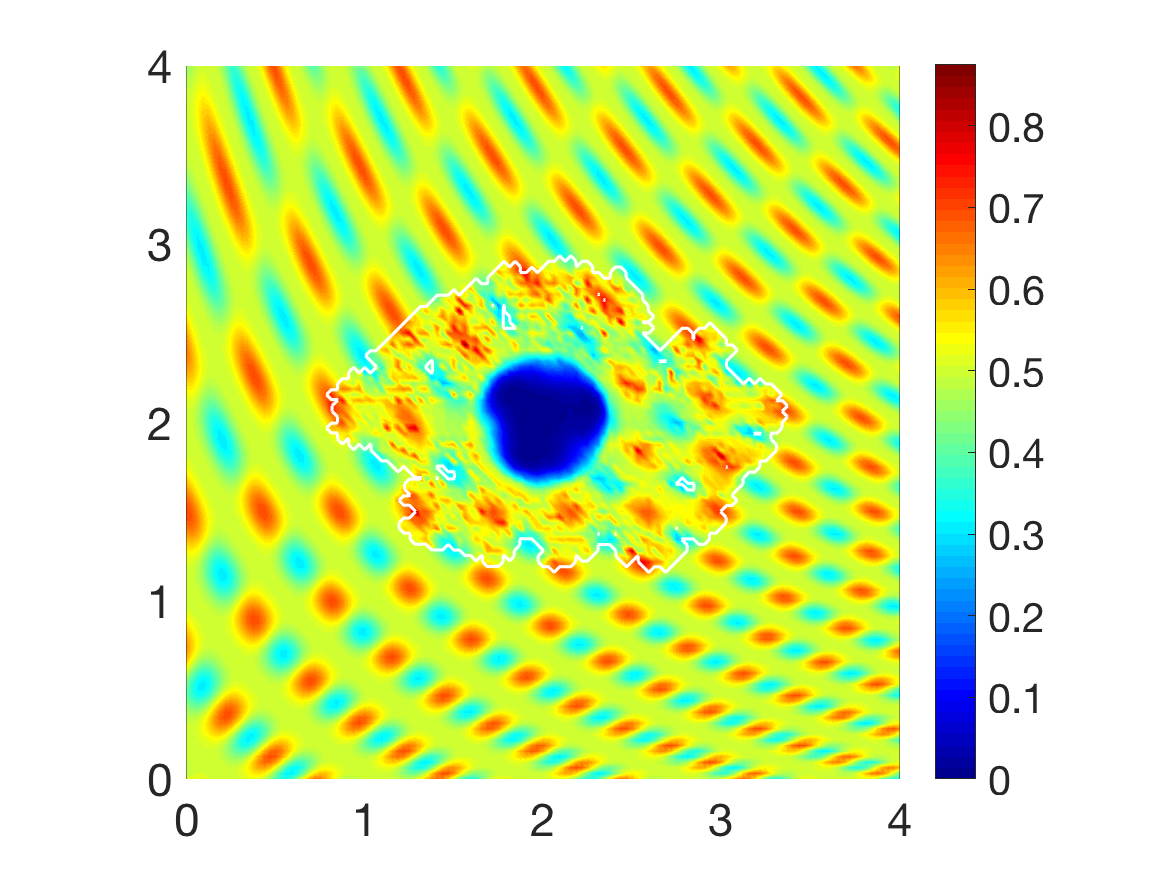}
  \caption{\emph{Matrix distribution}}
  \label{fig:2}
\end{subfigure}\hfil 

\medskip
\begin{subfigure}{0.5\textwidth}
  \includegraphics[width=\linewidth]{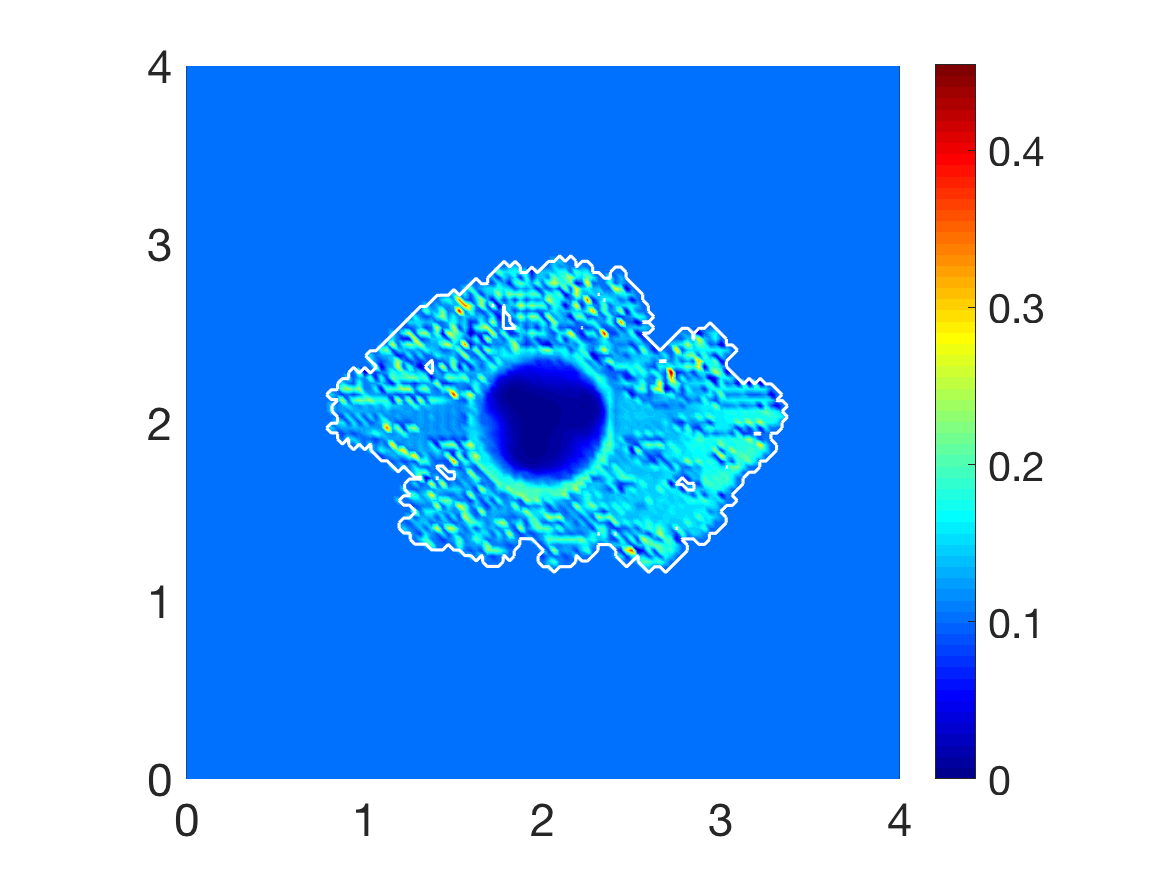}
  \caption{\emph{Macroscopic Fibre density}}
  \label{fig:3}
  \end{subfigure}\hfil 
\begin{subfigure}{0.5\textwidth}
  \includegraphics[width=\linewidth]{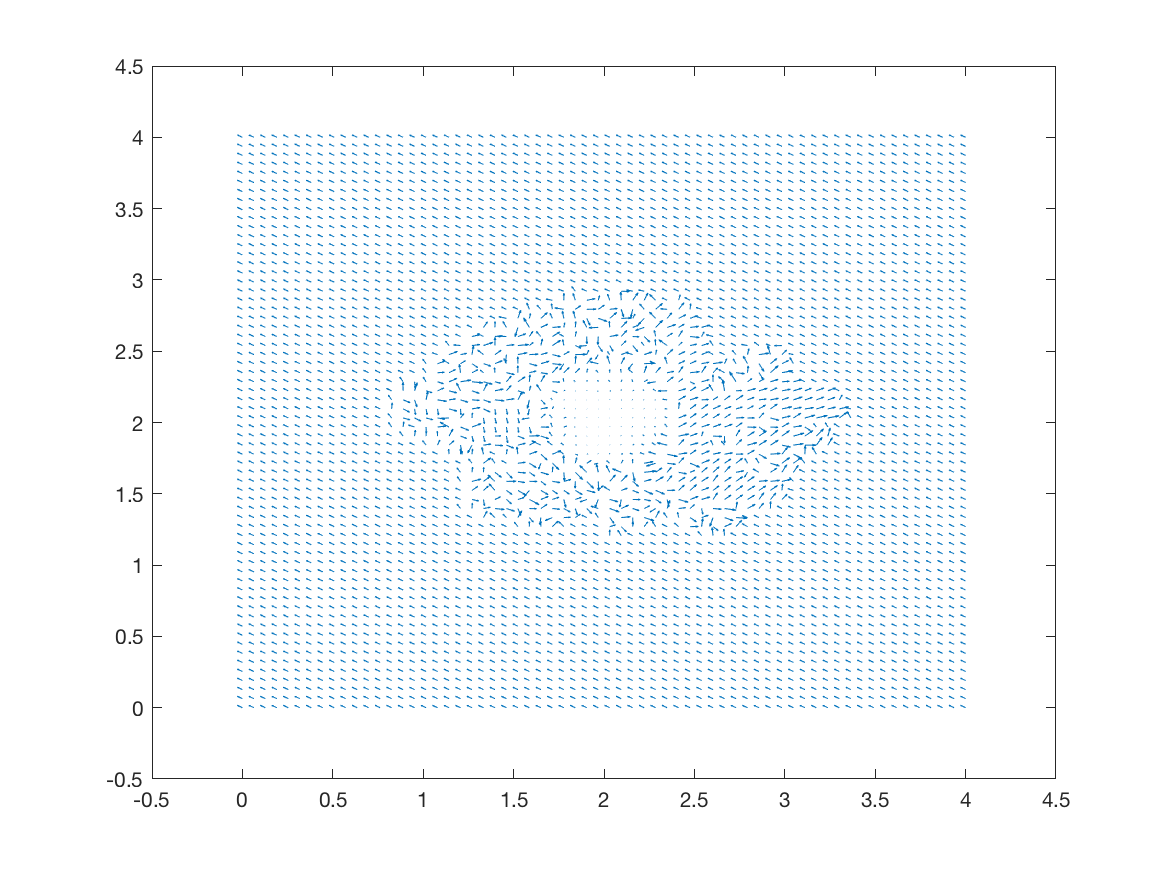}
  \caption{\emph{Fibre orientation - coarsened 2 fold}}
  \label{fig:4}
\end{subfigure}\hfil 

\medskip
\begin{subfigure}{0.5\textwidth}
  \includegraphics[width=\linewidth]{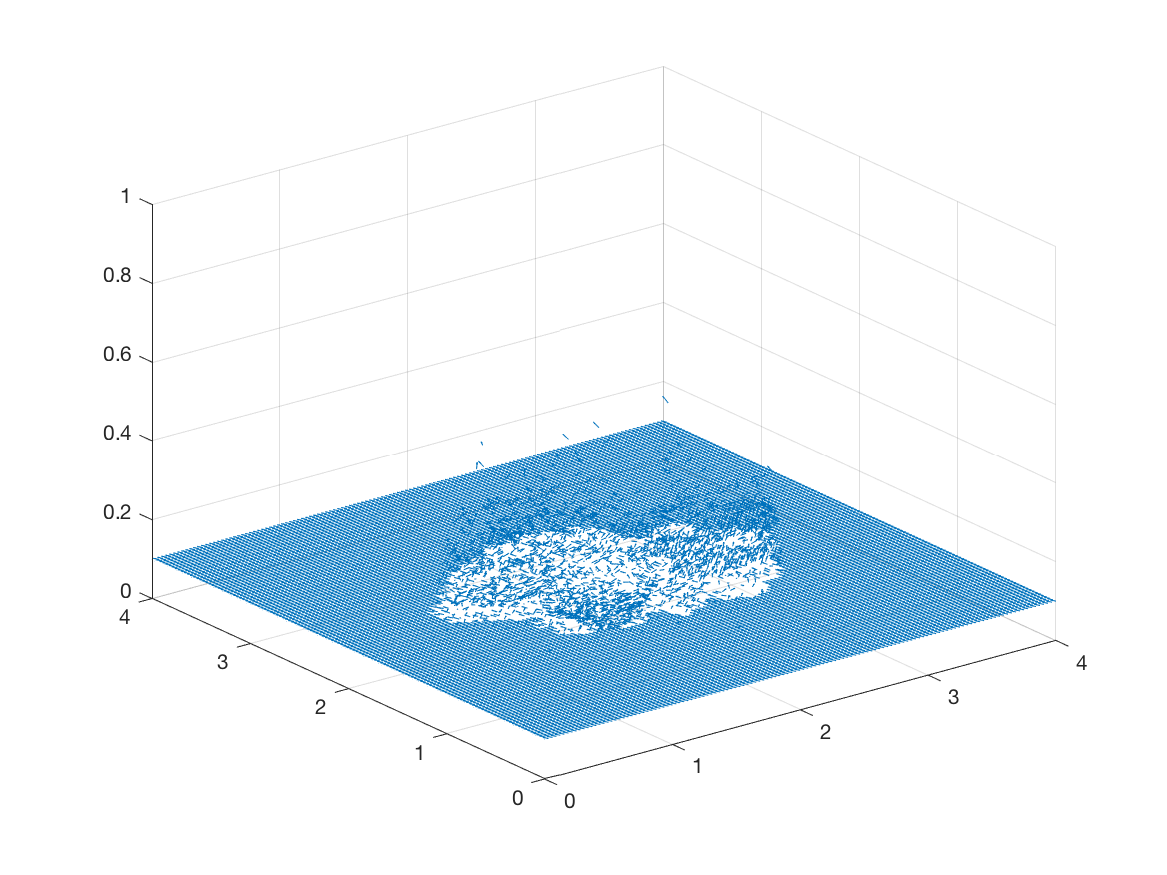}
  \caption{\emph{3D fibre vector field}}
  \label{fig:3}
  \end{subfigure}\hfil 
\begin{subfigure}{0.5\textwidth}
  \includegraphics[width=\linewidth]{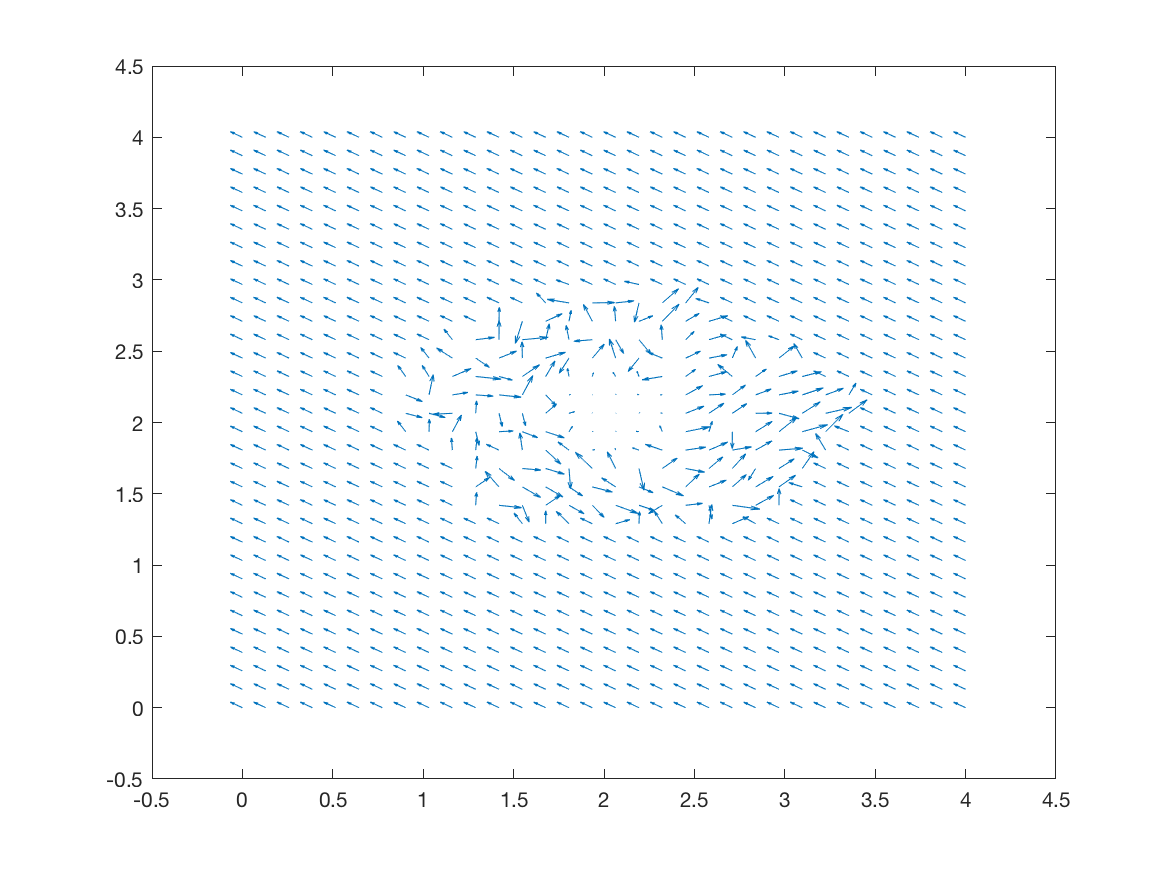}
  \caption{\emph{Fibre orientation - coarsened 4 fold}}
  \label{fig:4}
\end{subfigure}\hfil 

\caption{Simulations at stage $40\Delta t$ with a heterogeneous distribution of the non-fibrous part of the matrix.}
\label{fig:smallhetero40}
\end{figure}

Computational results at stage $20 \Delta t$ are shown in Figure \ref{fig:smallhetero20}, using the initial conditions shown in Figure \ref{fig:smallhetero} and the parameter set $\Sigma$ with the adhesive terms (\ref{norm_matrices}).  Due to the initial distribution of the non-fibrous component of the matrix, there are patches of high and low density areas, and \dt{regions of high} tumour \dt{density correspond to the areas of high} degradation of fibres and the surrounding \dt{non-fibre ECM} \ref{fig:smallhetero20}(b). The proliferating edge of the tumour is expanding in a lobular fashion, reaching out to the high density patches and encasing the low density regions in the process, \dt{as} the higher \dt{ECM} density equates to increased opportunity for cell-adhesion. This is the natural direction in which the tumour cells try to invade, pushing out from its centre and into the surrounding matrix, \dt{and causing the} tumour \dt{to} encircles itself with \dt{a region of higher magnitude} fibres, as shown in \ref{fig:smallhetero20}(c)-(f). The macroscopic orientation of the fibres is refashioned \ref{fig:smallhetero20}(d) \dt{as} the cancer cells have rearranged \dt{and degraded} the fibres, \dt{leading to significant changes in the fibre orientations and magnitude patterns near the boundary of the tumour with respect to their initial state, and causing them both to increase their magnitude and to point generally toward the the fast invading regions of the cancer boundary}. \dt{While} the fibre are being pushed and rearranged by the cancer cells outwards, away from the main body of the tumour, \dt{in regions of high cancer density, these are degraded, as evidenced by the low distribution of fibres in the centre of the tumour, presented in subfigures \ref{fig:smallhetero20}(c)-(f).}

Figure \ref{fig:smallhetero40} illustrates simulations plotted at stage $40 \Delta t$. The main body of the tumour is beginning to form new high distribution regions within the highly degraded patch of ECM, as shown in \ref{fig:smallhetero40}(a)-(b). This build up of cells is due to \dt{increasingly higher magnitudes for rearranged fibres with invasion favourable orientations, which result into a significantly higher effect of cell-fibre adhesion leading to increased transport of cells towards those areas.} Islands are forming within the boundary of the tumour away from the primary tumour mass due to \dt{low ECM density in those regions, which result in weak levels of both cell-non-fibre ECM and cell-fibre adhesion, and as a consequence the cells take longer time to advance upon these regions}. \dt{As shown in Figures \ref{fig:smallhetero40}(c)-(f)}, the fibres persevere in surrounding the tumour, with their oriented fibres \dt{on the central part of the tumour (corresponding to regions of very high cancer cell density)} continuing to \dt{be strongly degraded and dominated in their direction by the flow $\F$}. Again, \dt{high density regions of ECM fibres} equates to more opportunities for cell-fibre adhesion, thus creating a preferential direction of invasion. The cancer cells are rearranging the fibres to follow this direction, allowing them an easier route of invasion. \dt{As shown in Figure \ref{fig:smallhetero40}(c)}, by the gathering of fibre distributions away from the \dt{tumour} centre, it is evident that the cancer cells are pushing the fibres outwards to the boundary of the tumour and in the direction of the invasion \dt{front}, as found also in \cite{Pinner_2007} . 

\subsection{Increased cell-fibre adhesion \dt{within the heterogeneous non-fibre ECM phase scenario}}
\begin{figure}[ht!]
    \centering 
\begin{subfigure}{0.5\textwidth}
  \includegraphics[width=\linewidth]{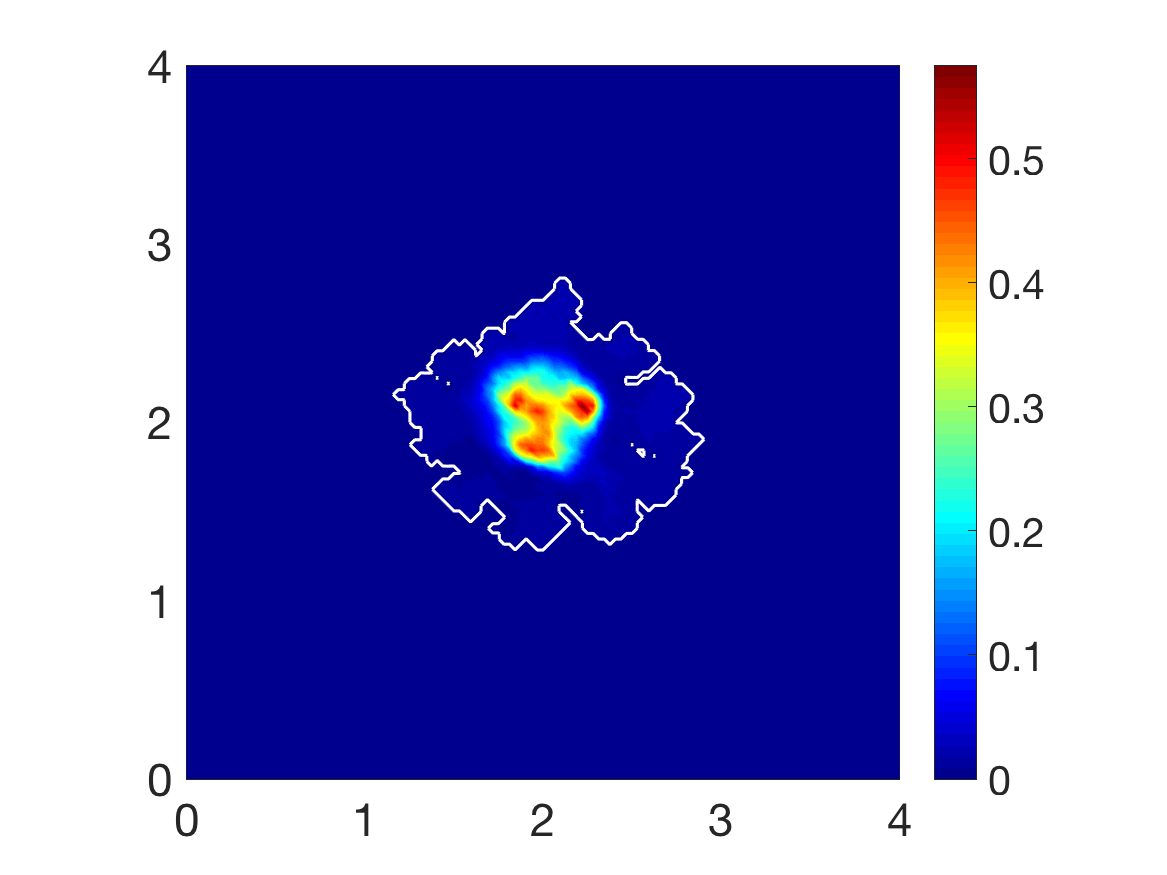}
  \caption{\emph{Cancer cell population}}
  \label{fig:1}
\end{subfigure}\hfil 
\begin{subfigure}{0.5\textwidth}
  \includegraphics[width=\linewidth]{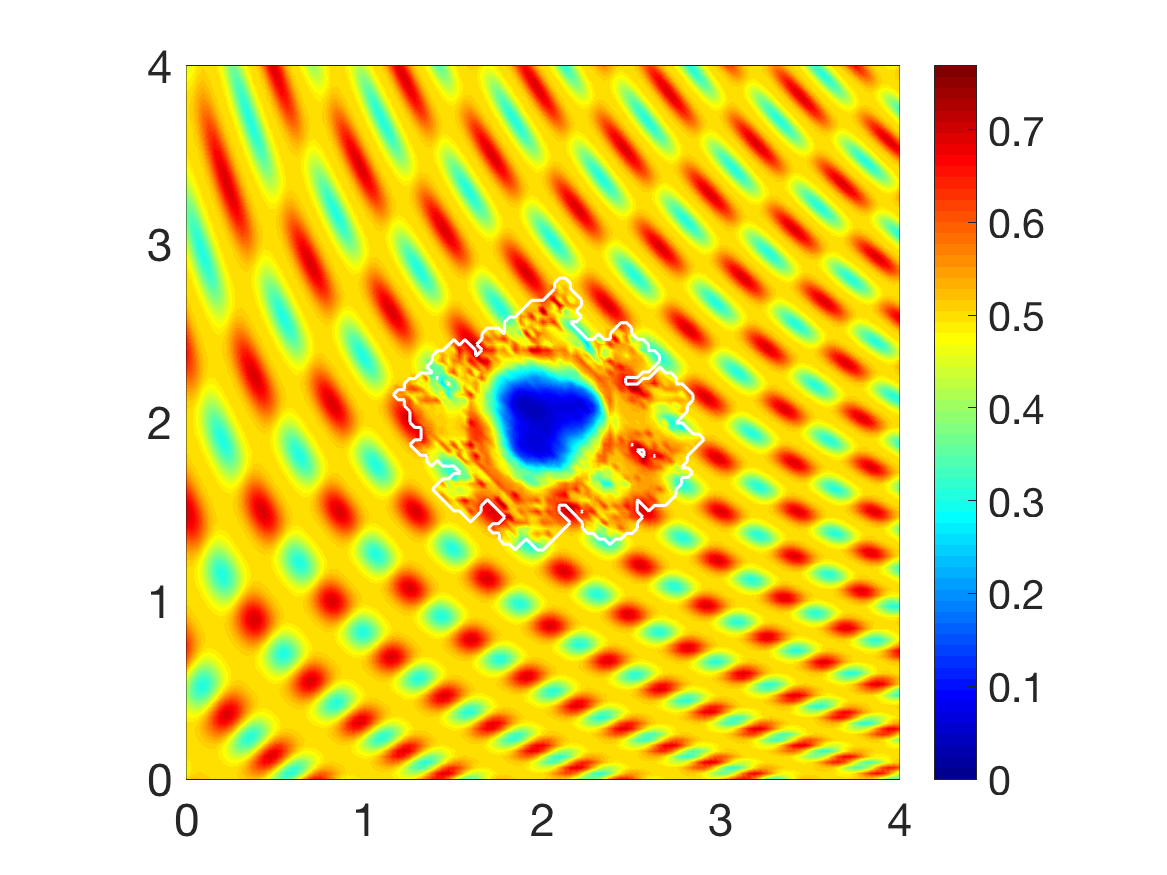}
  \caption{\emph{Matrix distribution}}
  \label{fig:2}
\end{subfigure}\hfil 

\medskip
\begin{subfigure}{0.5\textwidth}
  \includegraphics[width=\linewidth]{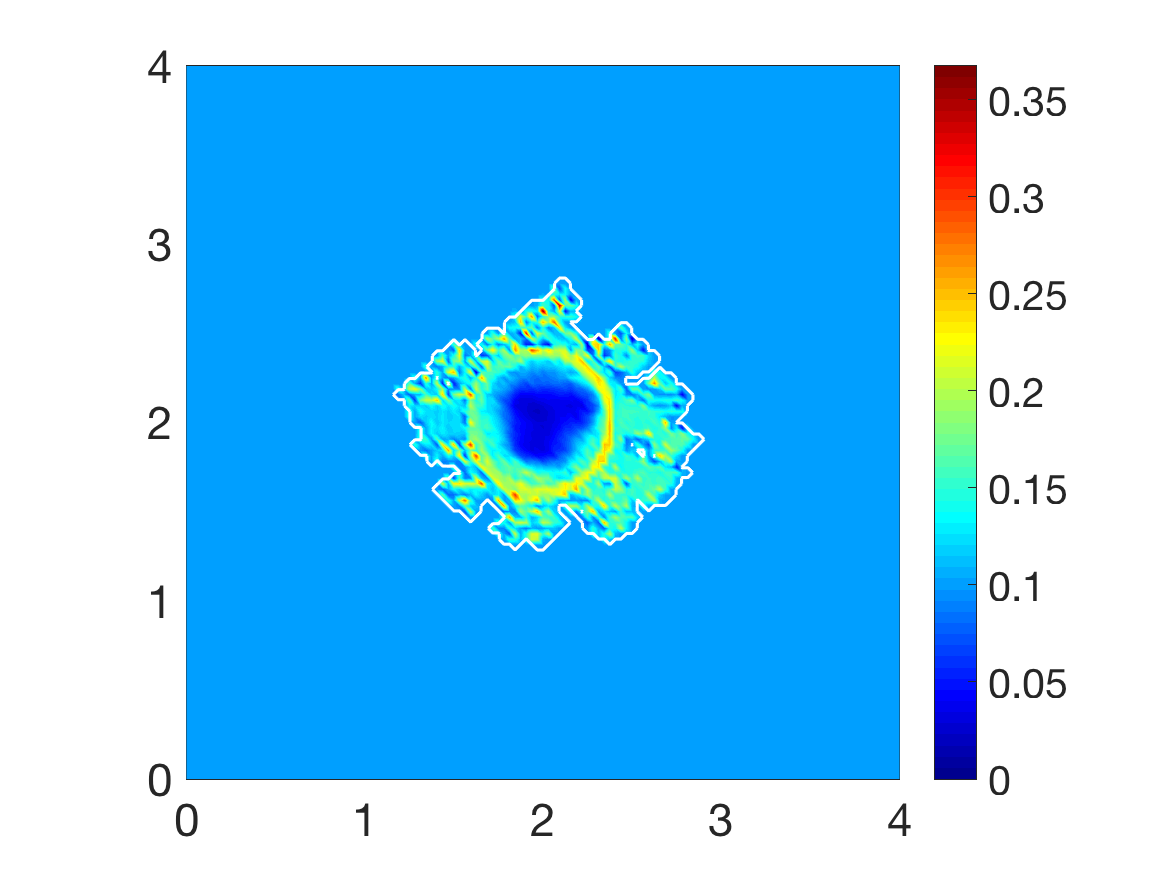}
  \caption{\emph{Macroscopic fibre density}}
  \label{fig:3}
  \end{subfigure}\hfil 
\begin{subfigure}{0.5\textwidth}
  \includegraphics[width=\linewidth]{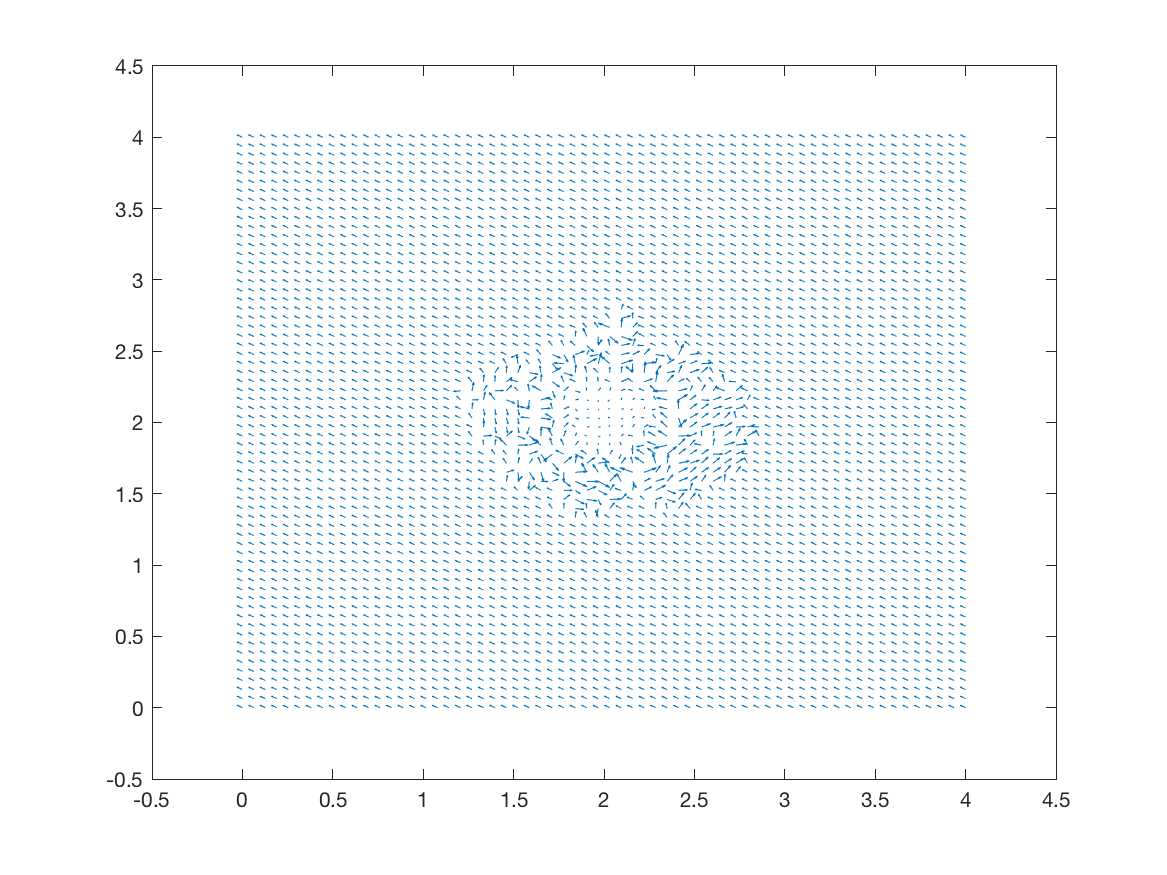}
  \caption{\emph{Fibre orientation - coarsened 2 fold}}
  \label{fig:4}
\end{subfigure}\hfil 

\medskip
\begin{subfigure}{0.5\textwidth}
  \includegraphics[width=\linewidth]{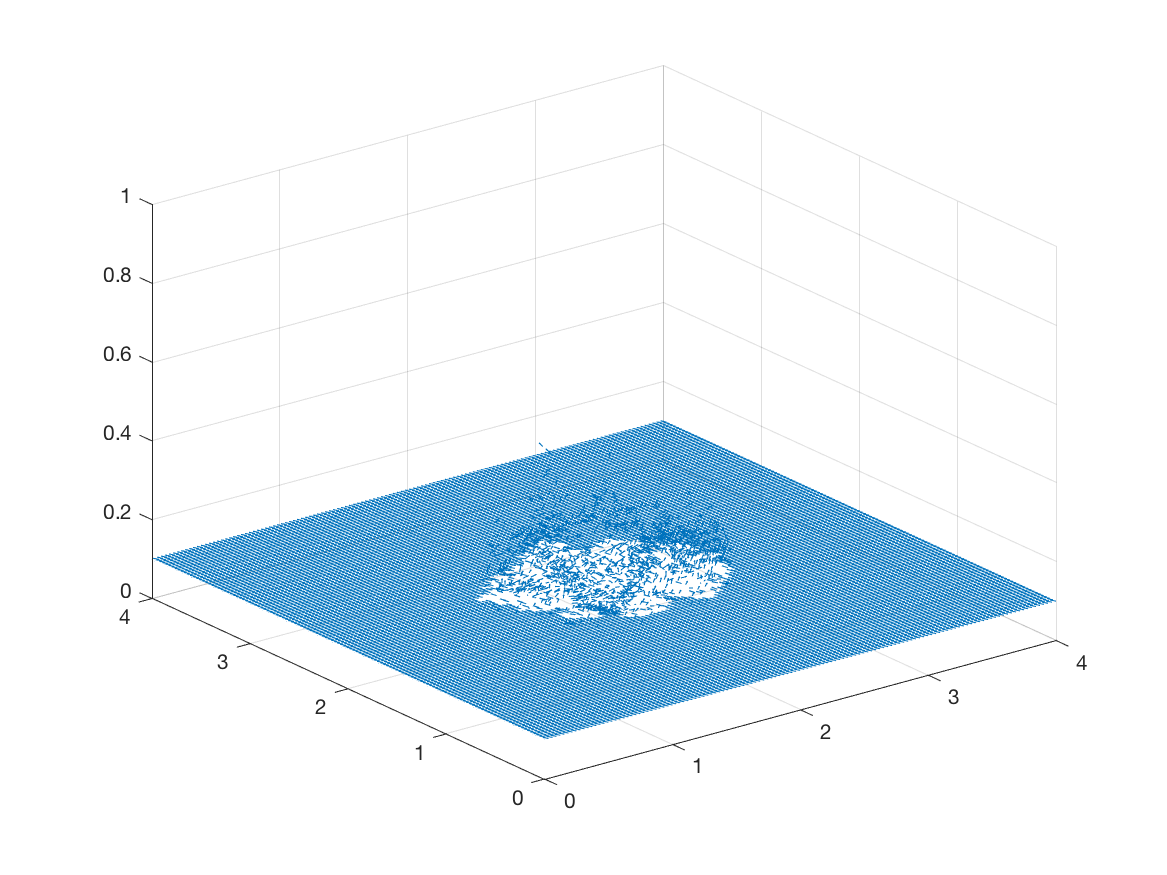}
  \caption{\emph{3D fibre vector field}}
  \label{fig:3}
  \end{subfigure}\hfil 
\begin{subfigure}{0.5\textwidth}
  \includegraphics[width=\linewidth]{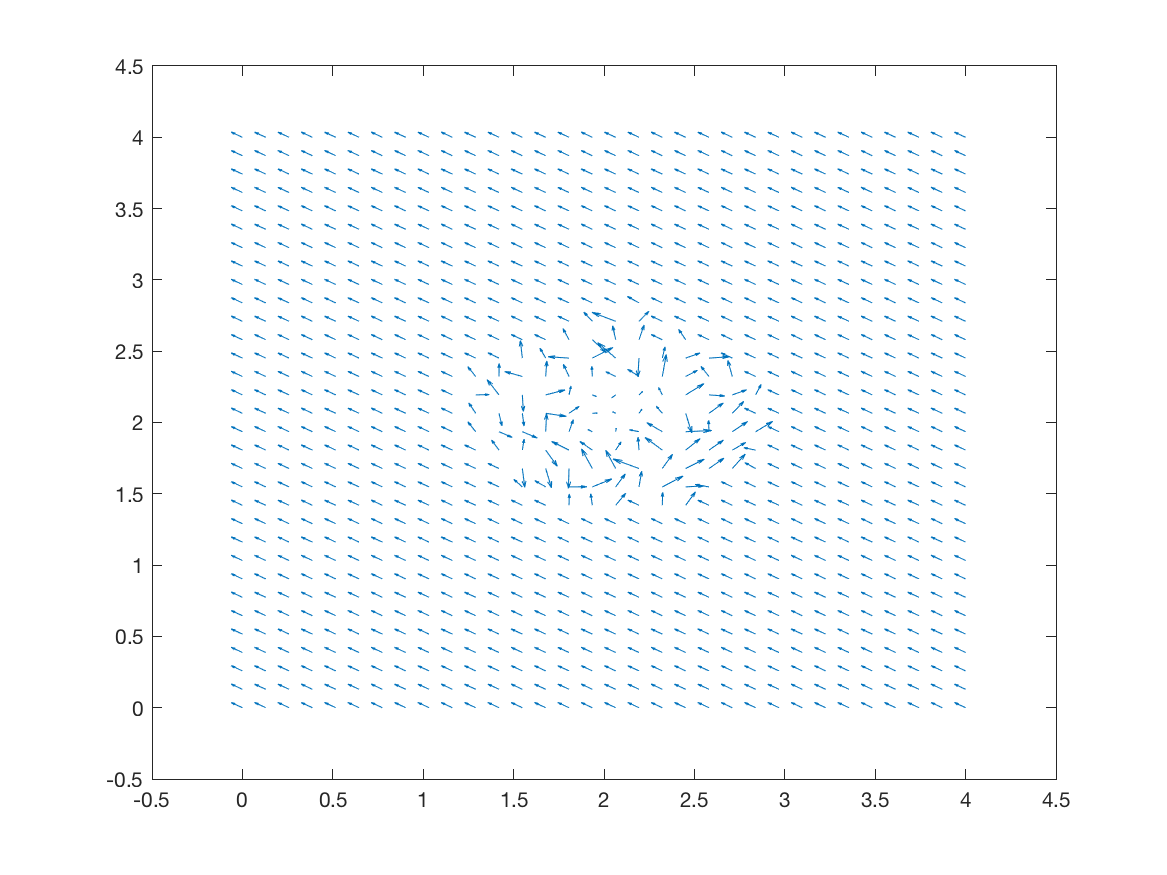}
  \caption{\emph{Fibre orientation - coarsened 4 fold}}
  \label{fig:4}
\end{subfigure}\hfil 

\caption{Simulations at stage $20\Delta t$ with a heterogeneous distribution of the non-fibrous part of the matrix and increased cell-fibre adhesion.}
\label{fig:30fib20}
\end{figure}

\begin{figure}[h]
    \centering 
\begin{subfigure}{0.5\textwidth}
  \includegraphics[width=\linewidth]{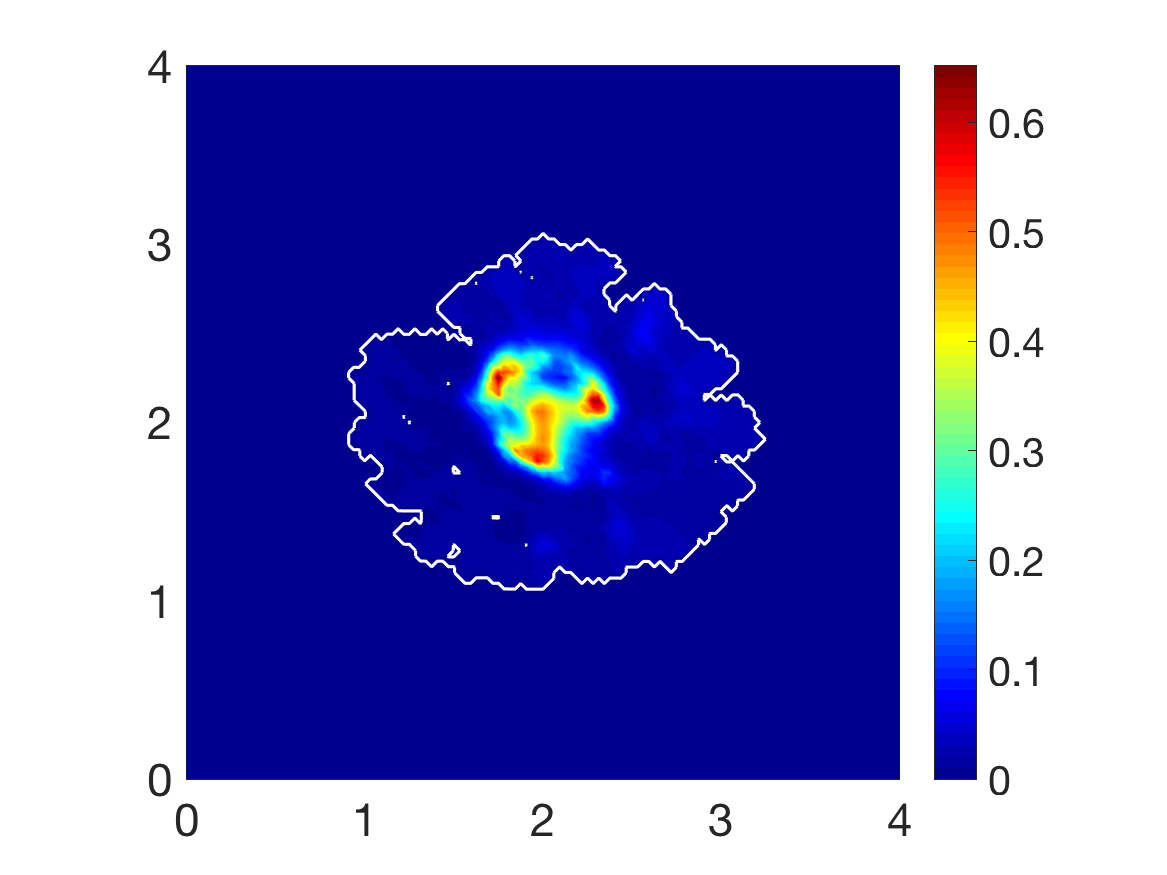}
  \caption{\emph{Cancer cell population}}
  \label{fig:1}
\end{subfigure}\hfil 
\begin{subfigure}{0.5\textwidth}
  \includegraphics[width=\linewidth]{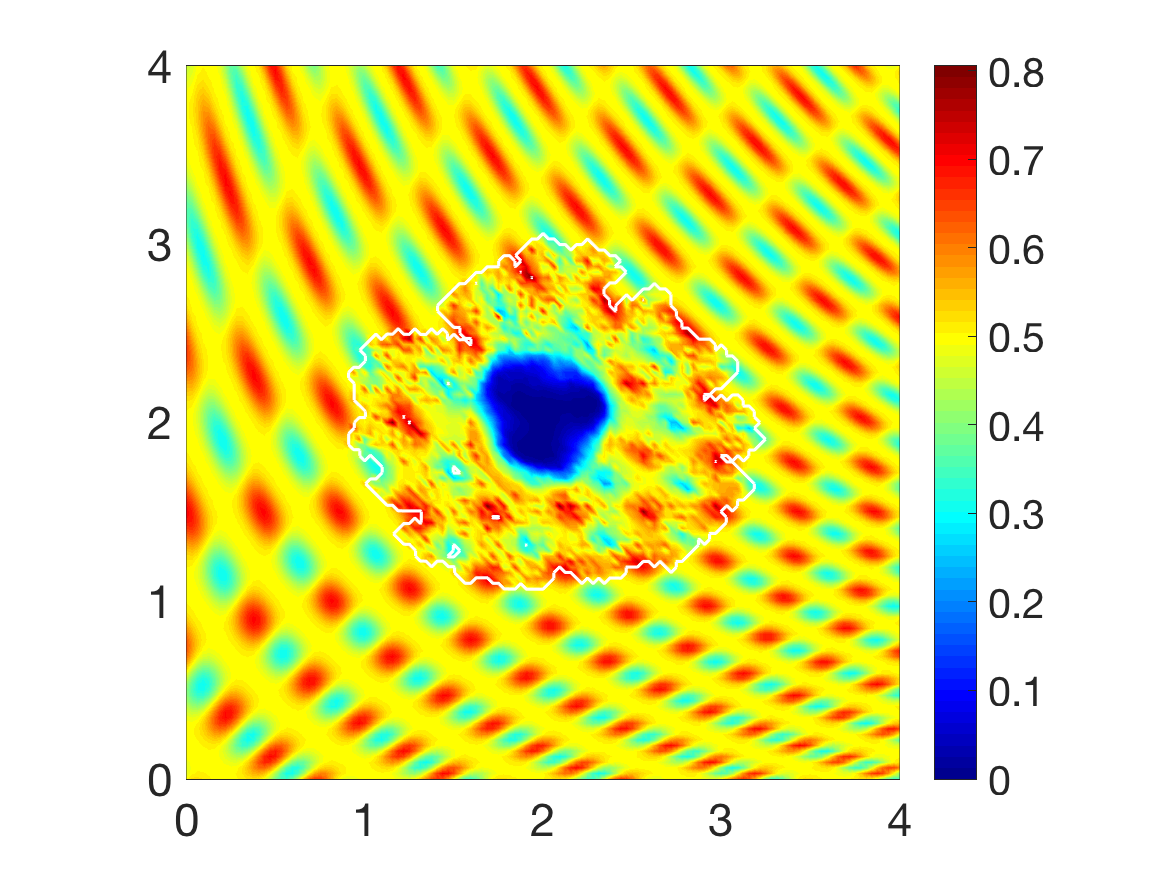}
  \caption{\emph{Matrix distribution}}
  \label{fig:2}
\end{subfigure}\hfil 

\medskip
\begin{subfigure}{0.5\textwidth}
  \includegraphics[width=\linewidth]{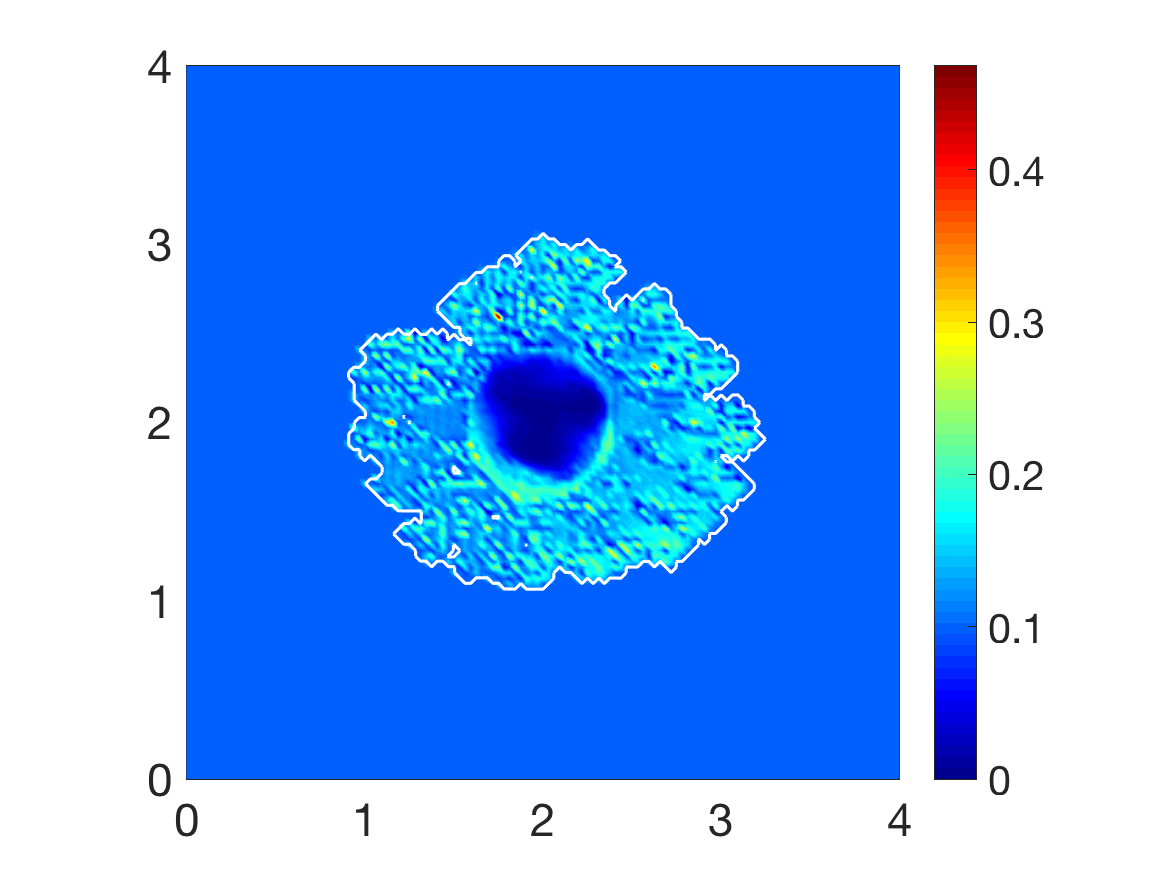}
  \caption{\emph{Macroscopic fibre density}}
  \label{fig:3}
  \end{subfigure}\hfil 
\begin{subfigure}{0.5\textwidth}
  \includegraphics[width=\linewidth]{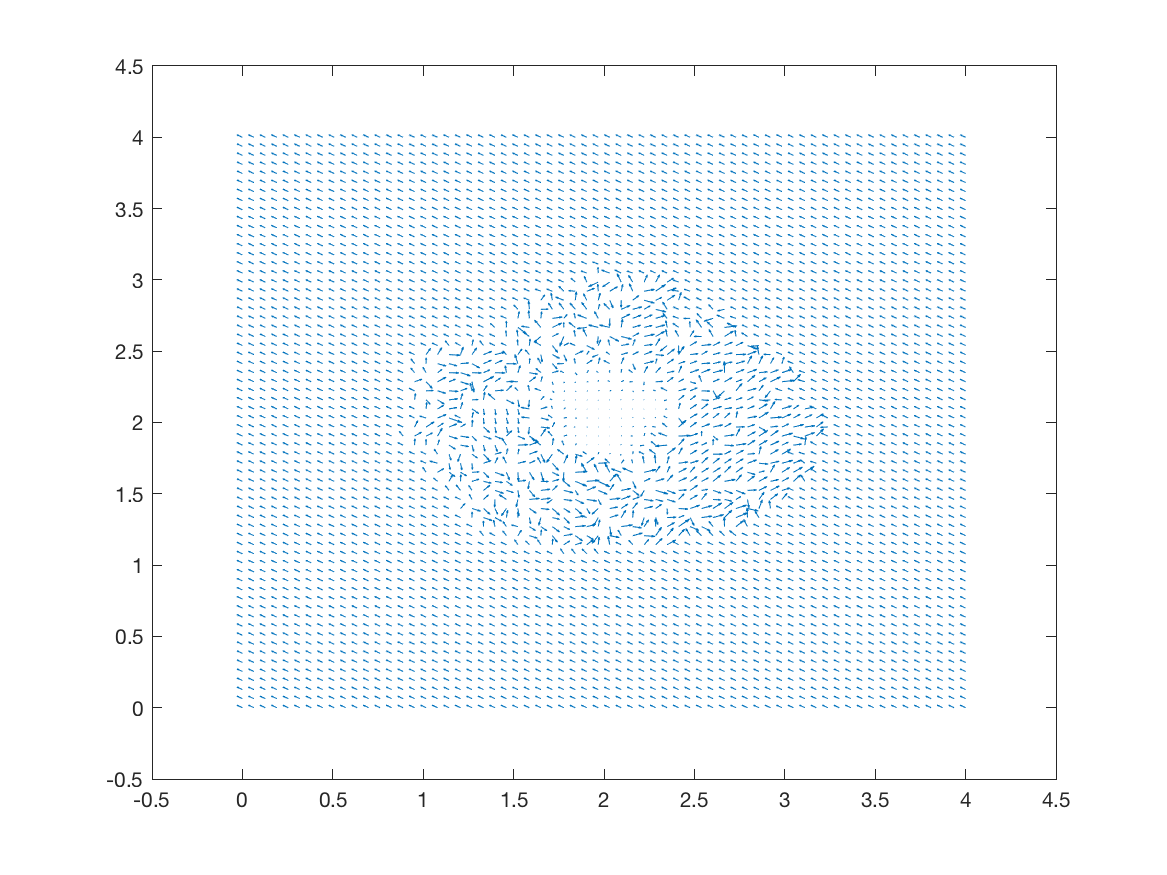}
  \caption{\emph{Fibre orientation - coarsened 2 fold}}
  \label{fig:4}
\end{subfigure}\hfil 

\medskip
\begin{subfigure}{0.5\textwidth}
  \includegraphics[width=\linewidth]{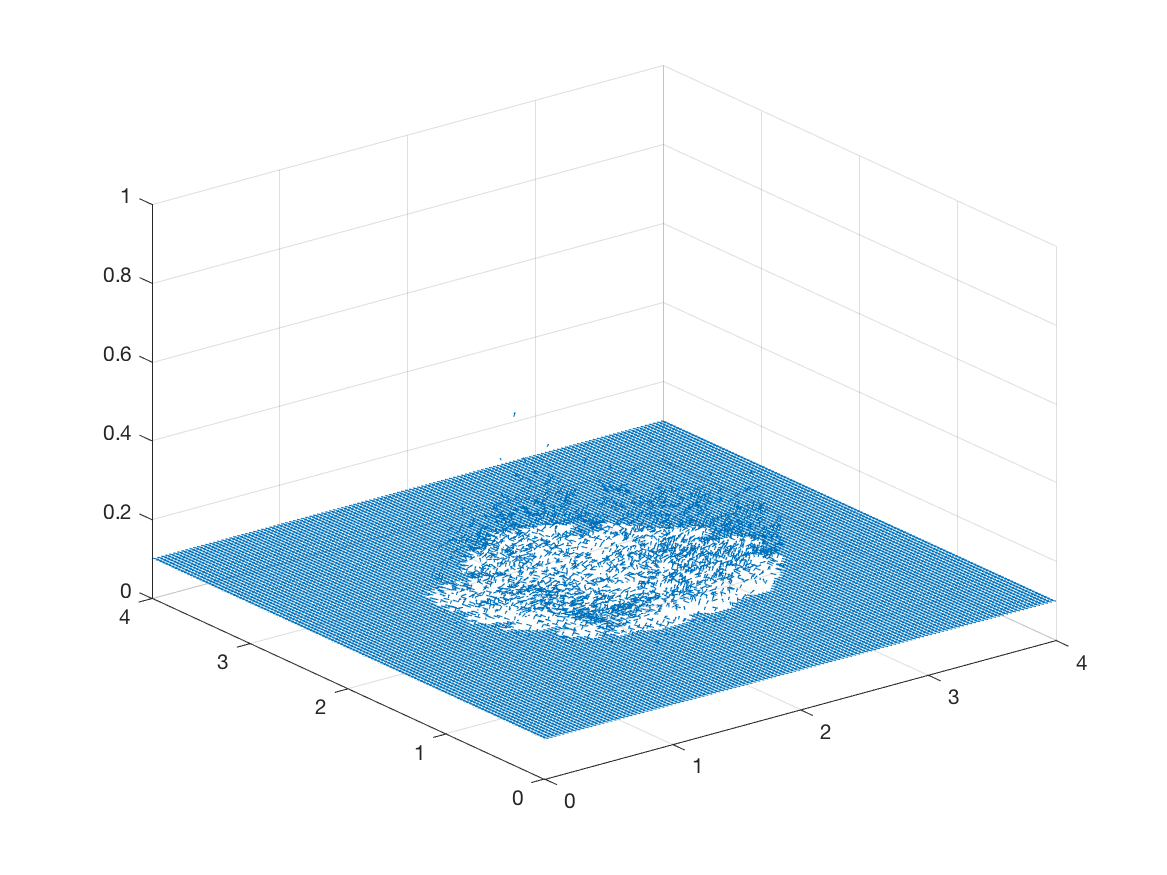}
  \caption{\emph{3D fibre vector field}}
  \label{fig:3}
  \end{subfigure}\hfil 
\begin{subfigure}{0.5\textwidth}
  \includegraphics[width=\linewidth]{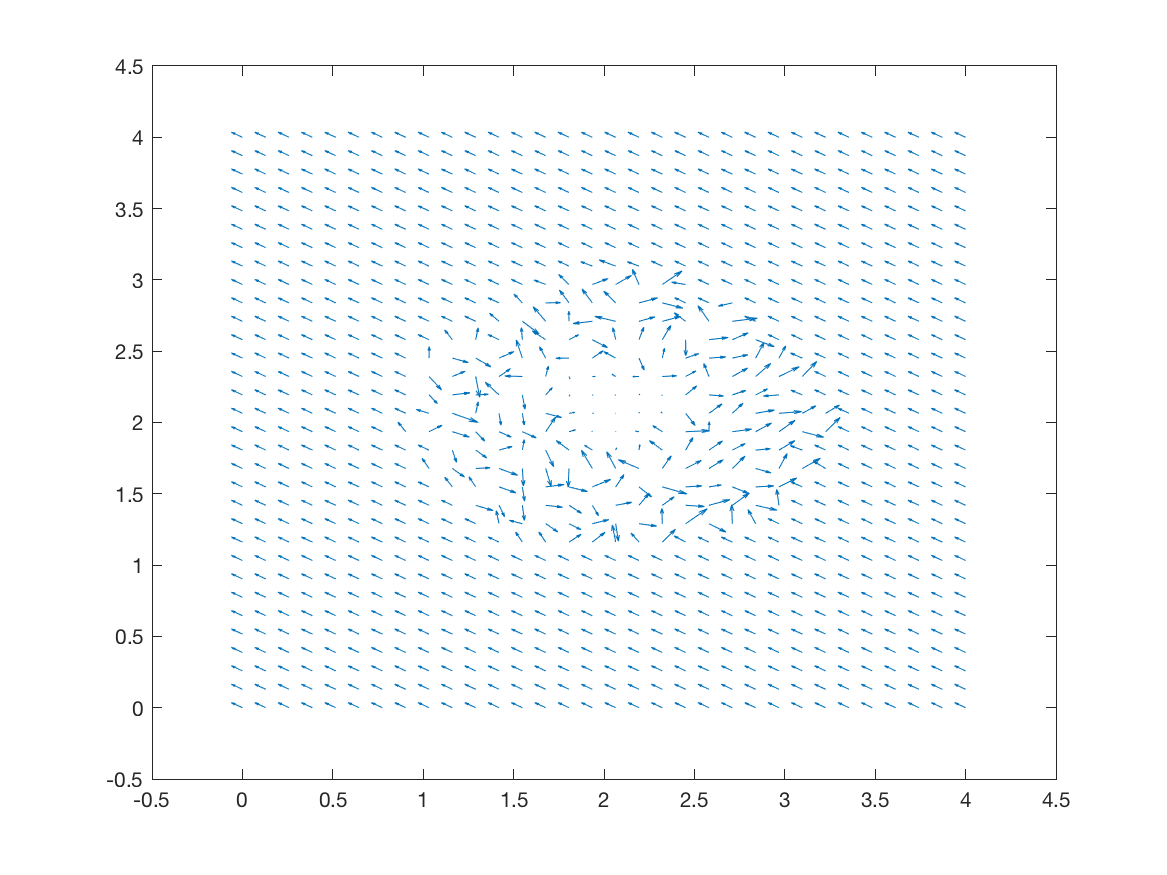}
  \caption{\emph{Fibre orientation - coarsened 4 fold}}
  \label{fig:4}
\end{subfigure}\hfil 

\caption{Simulations at stage $40\Delta t$ with a heterogeneous distribution of the non-fibrous part of the matrix and increased cell-fibre adhesion.}
\label{fig:30fib40}
\end{figure}

As we explore the effect of the heterogeneous \dt{two phase} ECM, \dt{it is important to consider the relation between the tumour progression and increased cell-fibre adhesion}. By increasing cell-fibre adhesion, we expect the cancer cells to advance further into their surrounding environment. \dt{For that, we \dt{double the} cell-fibre adhesion, \dt{taking} $\textbf{S}_{cF}=0.2$, while maintaining }the same initial conditions (i.e., for cancer cell population, those given in \eqref{eq:canceric}; for fibre ECM phase, those defined in \eqref{eq:fib} and illustrated in Figure \ref{fig:fibre}; and for heterogeneous non-fibrous ECM, those given in \eqref{eq:matrix_IC}) along with the parameters defined in the parameter set $\Sigma$ \dt{and} the cell-cell and cell-ECM-non-fibres adhesion as stated in (\ref{norm_matrices}). \dt{T}he computational results shown \dt{for this situation} in Figures \ref{fig:30fib20} and \ref{fig:30fib40}, \dt{show that} the fibres now have more influence over the route of invasion \dt{and its pattern}, presenting a leading edge with a highly lobular structure. There are small islands present in high density regions of the ECM, suggesting the density is simply too high and first requires more degradation before the cells can completely engulf this area. The overall degradation of the matrix has remained centralised to the \dt{central part} of the tumour as before. \dt{Similar to the previous cases considered here}, the cancer cells have rearranged the fibres \dt{also in this situation, strongly degrading the fibres in the regions with very high cancer density} \ref{fig:30fib20}(c), whilst the \dt{magnitude and} orientation of the fibres \dt{situated closer to the tumour periphery} have been altered and are positioned similar to the case where cell-fibre adhesion $\textbf{S}_{cF}=0.1$, \dt{pointing clearly outwards in boundary regions of increased invasive behaviour}. 

Figure \ref{fig:30fib40} displays computations at stage $40 \Delta t$. \dt{An important} difference between Figures \ref{fig:30fib20}(a) and \ref{fig:30fib40}(a) is \dt{observed} within the main body of the tumour. When the cell-fibre adhesion coefficient is increased, the \dt{central part} of the tumour has an overall higher distribution, and is being pulled in different directions, \dt{as} illustrated by the \dt{three} areas of increased cell distribution. Shown in Figure \ref{fig:30fib40}(d), the fibre orientations have been realigned, \dt{and in boundary regions of faster invasion the cumulative fibres direction tends to become almost perpendicular} to the fibres from the peritumoural region. The cancer cells attempt to align the fibres with their own directional preference, i.e., outwards from the centre and towards the higher density regions of ECM where they have \dt{increased} opportunity for adherence. The cells continue to pursue this goal, as evidenced in Figure \ref{fig:30fib40}(b) where we see the leading edge advancing on the higher density areas of ECM. From \dt{our} simulations, \dt{we} noted that an increase in cell-fibre adhesion causes a larger overall invasion of the tumour, suggesting that the fibres presence plays an important role in the invasion of cancer.

\section{Conclusion}
We have presented a \dt{novel} multi-scale moving boundary model which builds on previous framework first developed in \cite{Dumitru_et_al_2013}. This multiscale model is developed to explore the adhesive dynamics of a cancer cell population within a two-phase heterogeneous \dt{ECM} \dt{and its impact over the overall invasion pattern during cancer growth and spread within the} surrounding \dt{human body tissue}. \dt{The ECM is considered here as being a mixture of two constitutive phases, namely a fibre and non-fibre phase.} \dt{We pay a special attention to the fibre phase, whose multiscale dynamics is explored and modelled in an integrated two-scale spatio-temporal fashion, with the cell-scale micro-dynamics being connected to the tissue-scale tumour dynamic through an emerging double feedback loop. To that end, we developed a novel multiscale model that explores on one hand the way the fibre \dt{micro-dynamics translates into the macro-scale level fibre dynamics (by providing \emph{on-the-fly} at tissue-scale a spatially-distributed vector field of oriented macroscopic fibre that have direct influence within the tumour progression)} and, on the other hand, the way in which the tissue-scale cancer cell population dynamics causes not only fibres degradation at macro-scale but also fibres rearrangement at microscale. Finally, the new multiscale model is embedded within the multiscale moving boundary framework exploring the leading edge proteolytic activity of matrix degrading enzymes introduced in \cite{Dumitru_et_al_2013}. Thus, we ultimately obtain a novel multi-scale modelling framework that combines two multiscale sub-systems that contribute to and share the same macro-dynamics, but have separate micro-scale processes that are simultaneously connected to the macro-dynamics through two independent feedback loops, with one of them addressing the cell-scale activity involved in the rearrangement of micro-fibres within the bulk of the tumour, and the second one exploring the proteolytic activity within a cell-scale neighbourhood of the tumour  boundary. } 

\dt{At the tissue scale, in order to explore the influence of the ECM fibre phase within the tissue-scale dynamics, besides the usual adhesion terms considered in \cite{Domschke_et_al_2014,Gerisch_Chaplain_2008} concerning cell-cell and cell-ECM-non-fibre adhesion, we derived and introduce a new non-local term in the macroscopic equation \eqref{globalMacroSystem-c} for tumour cell population that accounts for the cell-fibres adhesion. This new term, explores the critical influence that the macroscopic fibre vector field has over the direction of cellular adhesion in the macro-dynamics. Moreover, as this vector field of oriented ECM fibres is induced from the micro-scale distribution of micro-fibres,} a novel \textit{bottom-up} feedback link between cell- and tissue- scale dynamics has \dt{this way} been identified and explored mathematically. 

\dt{Further, while the cancer cells degrade both the non-fibre ECM and the fibre ECM components at macro-scale, it was important to observe that the of flux cancer cell $\F$ given in \eqref{spatial_flux} causes the rearrangement of the micro-fibres at micro-scale. To understand this, at any given macro-scale position $x\in \Omega(t)$ we explored this macro-micro interacting link on appropriately small cubic micro-domains centred at $x$, namely $\delta Y(x)$, where the distribution of the micro-fibres $f(z,t)$ (with $z\in \delta Y(x)$) induces naturally the fibre magnitude $F(x,t)$ and orientation $\theta_{_{f}}(x,t)$, and whose rigorous derivation and well-posedness was ensured in Section \ref{mfomECM}. Futhermore, while getting balanced by the initial macro-scale orientation of the existing fibres (induced from the distribution of the microfibres on $\delta Y(x)$, the macro-scale spatial flux $\F$ acts uniformly on the existing distribution of micro-fibres on any micro-domain $\delta Y(x)$, causing the micro-fibres initially distributed on $\delta Y(x)$ to be redistributed and rearranged in the resulting fibres relocation direction given in \eqref{eq:fibnu}. This fibres relocation direction was obtained as the contribution of the flux $\F(x,t)$ and the fibre vector field $\theta_{f}(x,t)$ that are weighted in accordance to the amount of cancer cells transported at $(x,t)$ and the magnitude of fibre that they meet at $(x,t)$, respectively. Finally, this relocation is accomplished to the extent in which the local microscopic conditions permit, these being explored here through an appropriately defined movement probability. This way, a top down link was established between the macro-dynamics and the dynamics fibres rearrangement at microscale.}

\dt{To address this new multiscale modelling platform computationally, we extended significantly the computational framework introduced in \cite{Dumitru_et_al_2013} by bringing in the implementation  of the interlinked two-scale fibre dynamics. To that end, besides the computational approach based on barycentric interpolation that the micro-scale fibres relocation process has required, the macro-solver needed several extension to accommodate the new modelling. To that end, alongside the formulation of a new approach to computing on the moving tumour domain, we proposed a new off-grid barycentric interpolation approach to calculate the new adhesion term, and finally we developed a novel non-local predictor-corrector numerical scheme to address the challenging macro-scale computational conditions created through the presence of a multiphase ECM that crucially includes the multiscale dynamics of the oriented fibres.} 
 
Using this \dt{multiscale computational approach for the proposed} model, we were able to simulate the multiscale nature of cancer invasion by exploring the link between the macroscopic spatial distribution and orientation of cancer cells and the matrix, and the microscopic rearrangement of fibres and micro-dynamics of MDEs that occur on the proliferating edge of the tumour. Overall, we considered the invasion of a cancer cell population within \dt{both} homogenous \dt{and heterogeneous} non-fibrous ECM phase, investigating the macro-scale dynamics of the cancer population and macroscopic densities of the ECM components, whilst considering their influence on both the micro-scale MDEs molecular dynamics occurring at the cell-scale along the invasive edge and also the microscopic fibre movement occurring within the boundary of the tumour. Finally, it is worth remarking at this stage that even in the homogeneous non-fibre ECM, the ECM as a whole will not be homogeneous, due to the presence of the oriented fibres that already lead to a constitutive heterogeneous ECM. 

The simulations presented in this paper have some similarities with previous work. We note a general lobular, fingering pattern for the progression of tumour boundary, aspect that \dt{was} observed also in \cite{Peng2016,Dumitru_et_al_2013} in the case of heterogeneous ECM. There is a noticeable increase in this behaviour when the coefficient of cell-fibre adhesion is increased, suggesting the microscopic fibres play a key role in the invasion process, aiding in the local progression of the tumour. It is shown throughout all simulations that, while being degraded by the cancer cells, the fibres are being pushed outwards from the centre of the domain  towards the boundary \cite{Pinner_2007}. This behaviour is known for amoeboid cell types, and particularly occurs in a loose/soft matrix \cite{Krakhmal_2015}, which is reminiscent of our model as the cancer cells flux rearranges the fibres continuously at micro-scale. We can conclude from our simulations that a heterogeneous ECM non-fibrous phase permits for an increase in tumour progression compared to an initial homogeneous distribution. It is clear that the \dt{ECM} fibres play \dt{an important} role during invasion, with an increase in cell-fibre adhesion displaying a larger overall region of invasion. This is in line with recent biological experiments that suggest the organisation of fibronectin fibrils promotes directional cancer migration \cite{Erdogan3799}.

\dt{Looking forward, this modelling framework enables the opportunity for addressing questions in a range of directions, such as:} accounting for cellular reactions within the proteolytic micro-scale dynamics involving the fibres, i.e., the chopping/degradation of the fibres by matrix metallo-proteinases (MMPs); \dt{exploring the process of }anchoring of collagen to cells and other components in the \dt{ECM}; as well as exploring the presence of a second cancer cell subpopulation. Further work will focus on both fibronectin and collagen, and their roles and functions,\dt{ ultimately aiming to gain a better understanding of} the full impact \dt{that these have within} tumour invasion.

\appendix
\section{Paths for fibres}\label{microFibresAPP_01sept2018}
As described in Section 2, \dt{on the} microscopic domains $\delta Y(x)$, \dt{we consider a fibre distribution given by a combination of five distinctive micro-fibres patterns that are defined along the following smooth paths $\{h_{j}\}_{j\in J}$:}
\bequd
h_{1}: z_{1}=z_{2};  \quad
h_{2}: z_{1}=\frac{1}{2}; \quad
h_{3}: z_{1}=\frac{1}{5}; \quad
h_{4}: z_{2}=\frac{2}{5};\quad \textrm{and}\quad
h_{5}: z_{2}=\frac{4}{5}\,.
\eequd
\section{The mollifier $\psi_{\gamma}$}\label{mollifierUsed}
The standard mollifier $\psi_{\gamma}:\R^{N}\to\R_{+}$ (which was used also in \cite{Dumitru_et_al_2013}) is defined as usual, namely
\[
\psi_{\gamma}(x):=\frac{1}{\gamma^{N}}\psi\big(\frac{x}{\gamma}\big),
\]
where $\psi$ is the smooth compact support function given by
\bequd
\psi(x):=
\left\{
\begin{array}{cll}
\frac{exp\frac{1}{\nor{x}^{2}_{_{2}}-1}}{\int\limits_{\Bila(0,1)}exp\frac{1}{\nor{z}^{2}_{_{2}}-1}dz},& \quad if & x\in \Bila(0,1),\\[0.3cm]
0, & \quad if & x\not\in \Bila(0,1)
\end{array}
\right.
\eequd
\section{The radial kernel $\K(\cdot)$}\label{kernelAppendix}
\dt{To explore the influence on adhesion-driven migration decreases as the distance from $x+y$ to $x$ within the sensing region $B(x,r)$ increases, the expression of the radial dependent spatial kernel $\mathcal{K}(\cdot)$ appearing in  \eqref{adhesiveTermExpression} is taken here to be:} 
\vspace{-0.1cm}
\bequ
\mathcal{K}(r):=\frac{2\pi R^2}{3}\left(1-\frac{r}{2R}\right).
\eequ
\vspace{-0.5cm}
\begin{figure}[ht!]
    \centering
    \begin{subfigure}{0.4\linewidth}
        \includegraphics[width=\linewidth]{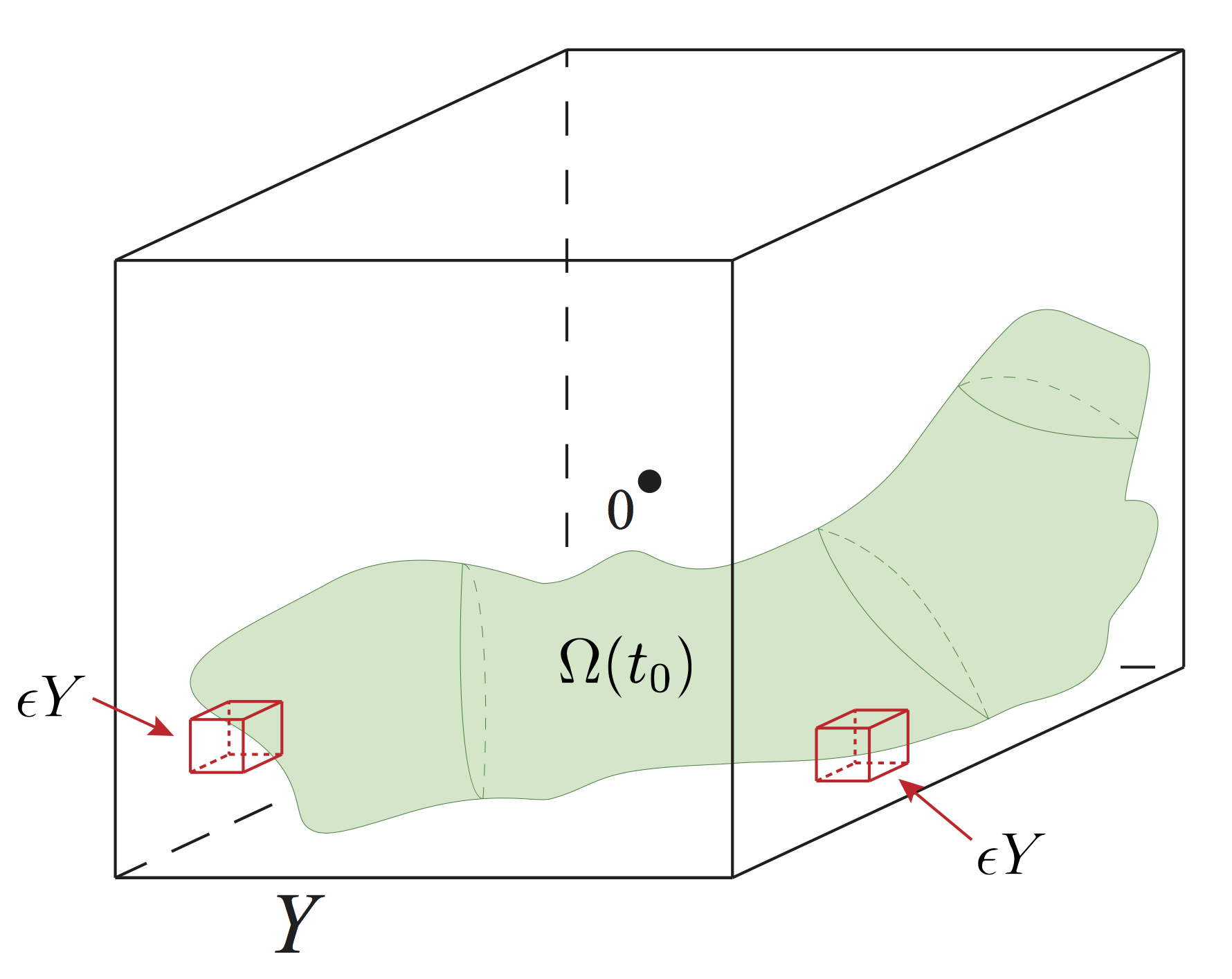}
        \caption{\emph{Spatial domain of tumour}}
\end{subfigure}\hfil
    \begin{subfigure}{0.6\linewidth}
        \includegraphics[width=\linewidth]{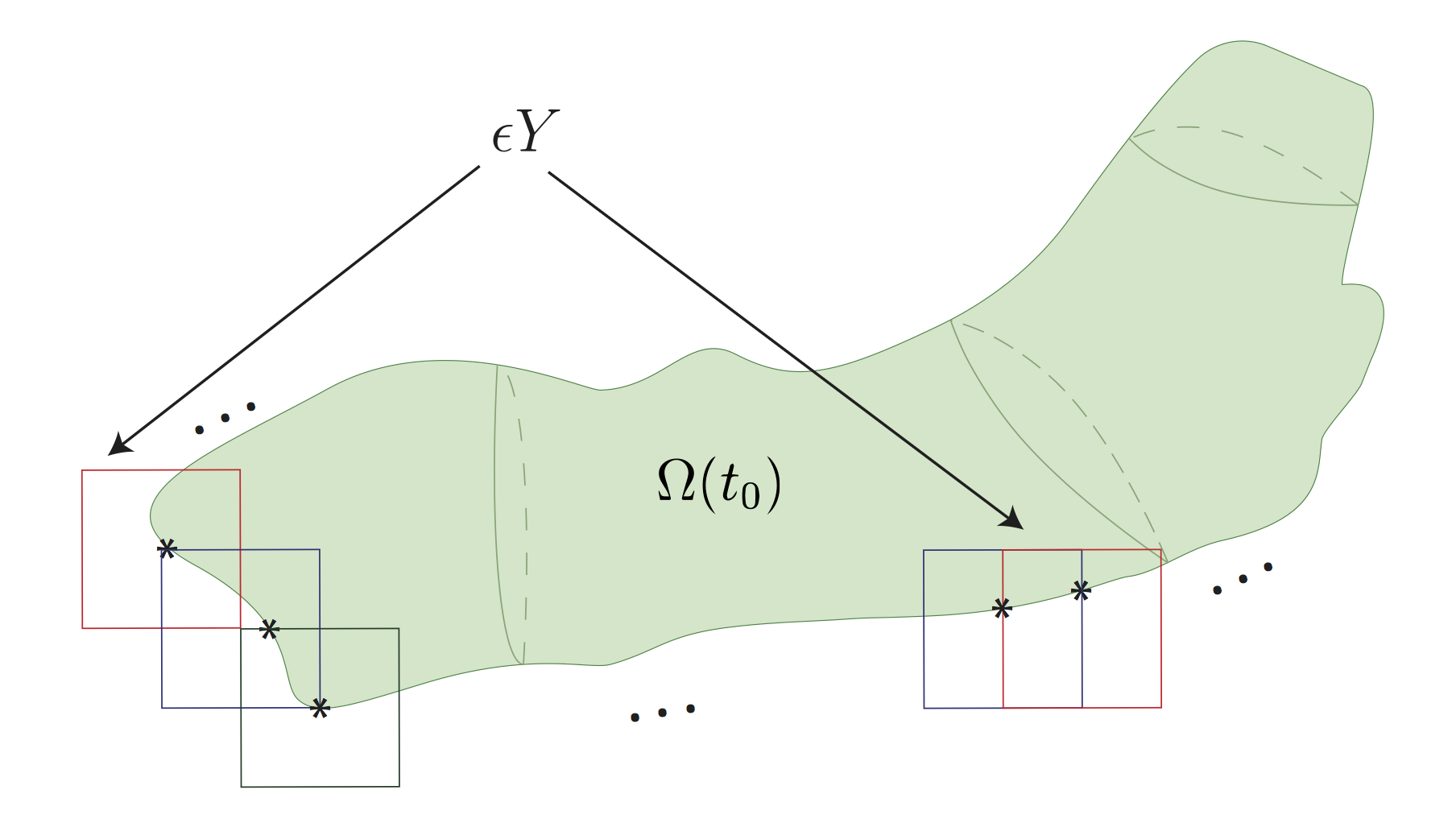}
        \caption{\emph{Boundary tracking with overlapping micro-domains}}
    \end{subfigure}\hfil
    \caption{\emph{Schematic diagrams showing in (a) the spatial cubic region $Y$ centred at the origin in $\mathbb{R}^3$. The solid red lines represent the family of macroscopic $\epsilon Y$ cubes placed on the boundary of the tumour $\partial \Omega (t_0)$, and the pale green region represents the mass of cancer cells $\Omega(t_0)$. The \dt{appropriately chosen bundle $\{\epsilon Y\}_{_{\epsilon Y}\in \P}$ of micro-domains introduced in in \cite{Dumitru_et_al_2013}} that covers the boundary $\partial\Omega(t_0)$ are shown in red and the corresponding half-way shifted cubes are illustrated by blue dashed lines.}}
    \label{fig:fullmodelintro}
\end{figure}
\vspace{-0.5cm}
\section{Key aspects within the multiscale \dt{moving-boundary} framework}\label{movingBoundaryFrameworkAPP_01sept2018} 
\dt{For completeness,} \dt{in the following} we will \dt{briefly} describe two \dt{key ingredients of the multiscale moving boundary framework introduced in \cite{Dumitru_et_al_2013}, namely: (1) the computationally feasible boundary tracking; and (2) the direction and magnitude induced by the micro-scale for the tumour boundary relocation at macro-scale.}
\subsection{The computationally feasible boundary tracking}
 \dt{Following a series of topological constraints as detailed in \cite{Dumitru_et_al_2013}, at any given time $t_{0}$,  for the maximal tissue cube} $Y \subset \mathbb{R}^N$ (which \dt{includes the growing tumour $\Omega(t_{0})$) we appropriately select the coarsest uniformly decomposed into a union of dyadic cubes $\epsilon Y$ with the property that
  from this dyadic decomposition of $Y$ taken together with all associated families of \emph{half-way shifted} dyadic cubes (as defined in \cite{Dumitru_et_al_2013}), a subfamily of overlapping cubes $\{\epsilon Y\}_{_{\epsilon Y}\in \P}$ can be extracted with the following two key characteristics:} 
 \begin{itemize}
 \item[(1)]\dt{each cube $\epsilon Y\in \P$ convey a neighbourhood for $\epsilon Y \cap \partial \Omega(t_{0})$ with both the part inside the tumour $\epsilon Y \cap \Omega(t_{0})$ and the part outside the tumour $\epsilon Y \setminus\Omega(t_{0})$ having their interior as connected sets;} 
 \item[(2)]\dt{$\{\epsilon Y\}_{_{\epsilon Y}\in \P}$ provides a complete covering of the boundary $\partial\Omega(t)$. }
\end{itemize}
\dt{On this bundle of covering boundary micro-domains $\{\epsilon Y\}_{_{\epsilon Y}\in \P}$, which is schematically illustrated in Figure \ref{fig:fullmodelintro}, in the presence of the MDEs source induced from the macro-scale (as detailed in \cite{Dumitru_et_al_2013} and briefly outlined in Section \ref{multMovingBoundaryPerspect}), the MDEs micro-dynamics is then explored.} 

\subsection{The microscopic movement of the tumour boundary} 
\label{micro_bound_move}

During the micro-dynamics, the MDEs present in the peritumoural region interact with the ECM captured by each $\epsilon Y$ cube on the boundary of the tumour. Within each micro-domain $\epsilon Y$, the degradation of the ECM is dependent on the pattern of the front of the advancing spatial distribution of MDEs in $\epsilon Y \setminus \Omega(t_{0})$ that have been secreted by the cancer cells. This pattern of the degradation defines a movement direction, $\eta_{\epsilon Y}$, and displacement magnitude, $\xi_{\epsilon Y}$, for the progression of the tumour boundary. This activity is the driving force behind the relocation of the boundary midpoints $x^*_{\epsilon Y}$ which are translated to the movement of the tumour boundary at macro-scale. 

\dt{Thus, proceeding as detailed in \cite{Dumitru_et_al_2013}, an appropriate uniform dyadic decomposition $\{\mathcal{D}_{j}\}_{j=1,2^{s}}$ is chosen for $\epsilon Y$. Since the pattern of degradation is ultimately given by the part of the level set of significant MDEs peaks that are transported at the largest possible distance from the tumour interface $\epsilon Y \cap \Omega(t_{0})$ within the peritumoural region $\epsilon Y \setminus \Omega(t_{0})$, as discussed in \cite{Dumitru_et_al_2013}, by denoting the barycentre of each $D_{j}$ by $y_{j}$,}  the movement direction $\eta_{\epsilon Y}$ and displacement magnitude \dt{$\xi_{\epsilon Y}$} of each boundary midpoint \dt{$x^{*}_{\epsilon Y}$} are represented mathematically as 
\[
\eta_{\epsilon Y} = x^{*}_{\epsilon Y} + \nu \sum_{l \in \mathcal{J}^{*}} \left( \int_{\mathcal{D}_j} m(y,\tau_f) \ dy \right) (y^{*}_{j} - x^{*}_{\epsilon Y})
\]
and 
\[
\xi_{\epsilon Y} = \sum_{l \in \mathcal{J}^{*}} \frac{\int_{\mathcal{D}_j} m(y,\tau_f) \ dy}{ \sum_{l \in \mathcal{I}^*_\delta}\int_{\mathcal{D}_j} m(y,\tau_f) \ dy} |\overrightarrow{x^*_{\epsilon Y} y_j^{*} }|. 
\]
where $\mathcal{J}^{*}$ is the family of indices of the dyadic cubes \dt{which track only the most advanced frontier formed by the significant peaks at the tip of the progressing front of the MDEs (i.e., situated at the furthest away distance from the $x^{*}_{\epsilon Y}$) within the micro-domain $\epsilon Y$.}
\vspace{-0.5cm}
\section{Table for the parameter set $\Sigma$}\label{paramSection}
Here we present a table for the parameter set $\Sigma$.
\begin{table}[h]
\centering
\begin{threeparttable}
\caption{The parameters in $\Sigma$}
\begin{tabular}{ c c c c}
  \hline		
  Parameter & Value & Description & Reference \\
  \hline
  $D_1$ & $10^{-4}$ & diffusion coeff. for cell population $c$ & \cite{Domschke_et_al_2014}\\
  $D_m$ & $10^{-3}$ & diffusion coeff. for MDEs & Estimated \\
  $\mu_1$ & $0.25$ & proliferation coeff. for cell population $c$ & \cite{Domschke_et_al_2014} \\
  $\gamma_{1}$ & 2 & non-fibrous ECM degradation coeff. & Estimated \\
  $\gamma_{2}$ & 1.5 & macroscopic fibre degradation coeff. & \cite{Peng2016} \\
  $\omega$ & 0 & non-fibrous ECM remodelling coeff. & \cite{Domschke_et_al_2014} \\
  $\alpha$ & $2.5 \times 10^{-3}$ & MDE secretion rate & \cite{Peng2016} \\
  $\textbf{S}_{max}$ & 0.5 & cell-cell adhesion coeff. & \cite{Domschke_et_al_2014} \\
  $\textbf{S}_{cF}$ & 0.1-0.2 & cell-fibre adhesion coeff. & Estimated \\
  $\textbf{S}_{cl}$ & 0.01 & cell-matrix adhesion coeff. & Estimated \\
  $R$ & 0.15 & sensing radius & Estimated \\
  $r$ & 0.0016 & width of micro-fibres & Estimated \\
  $f_{\text{max}}$ & 0.6360 & max. micro-density of fibres & Estimated \\
  $p$ & 0.2 & percentage of non-fibrous ECM & Estimated \\
  $h$ & 0.03125 & macro-scale spatial discretisation size & \cite{Dumitru_et_al_2013}\\
  $\epsilon$ & 0.0625 & size of micro-domain $\epsilon Y$ & \cite{Dumitru_et_al_2013} \\
  $\delta$ & 0.03125 & size of micro-domain $\delta Y$ & Estimated \\
   \hline 
  \label{table:parameters} 
\end{tabular}
\end{threeparttable}
\end{table}

\section*{Acknowledgment} RS and DT would like to acknowledge the support received through the EPSRC DTA Grant EP/M508019/1 on the project: \emph{Multiscale modelling of cancer invasion: the role of matrix-degrading enzymes and cell-adhesion in tumour progression}.   


\bibliography{ThesisReferences}

\end{document}